\documentclass[11pt,a4paper]{article}
\pdfoutput=1
\usepackage{jheppub}
\usepackage[utf8]{inputenc}
\usepackage{graphicx,fancybox,float,comment}
\usepackage{units}
\usepackage{subcaption}
\usepackage{multirow,array,arydshln}

\usepackage{amsmath,amssymb}

\usepackage{yfonts}
\usepackage{tensor}
\usepackage{paralist}

\usepackage{tensor,mathrsfs}

\title{Correlations far from equilibrium 
in charged strongly coupled 
fluids 
subjected to a strong magnetic field}
\author[]{Casey Cartwright, Matthias Kaminski}
\affiliation[]{Department of Physics and Astronomy, University of Alabama,\\ Tuscaloosa, AL 35487, USA}
\emailAdd{cccartwright@crimson.ua.edu}
\emailAdd{mski@ua.edu}
\begin{document}
\newcommand{\eref}[1]{Eq.~(\ref{#1})}
\newcommand{\exd}{\mathrm{d}}
\newcommand{\reals}{\mathbb{R}}
\newcommand{\mB}{\mathcal{B}}

\abstract{
Within a holographic model, we calculate the time evolution of 2-point and 1-point correlation functions (of selected operators) within a charged strongly coupled system of many particles. That system is thermalizing from an anisotropic initial charged state far from equilibrium towards equilibrium while subjected to a constant external magnetic field. One main result is that thermalization times for 2-point functions are significantly (approximately three times) larger than those of 1-point functions. Magnetic field and charge amplify this difference, generally increasing thermalization times. However, there is also a competition of scales between charge density, magnetic field, and initial anisotropy, which leads to an array of qualitative changes on the 2- and 1-point functions. There appears to be a strong effect of the medium on 2-point functions at early times, but approximately none at later times. At strong magnetic fields, an apparently universal thermalization time emerges, at which all 2-point functions appear to thermalize regardless of any other scale in the system. Hence, this time scale is referred to as saturation time scale. As extremality is approached in the purely charged case, 2- and 1-point functions appear to equilibrate at infinitely late time. We also compute 2-point functions of charged operators. Our results can be taken to model thermalization in heavy ion collisions, or thermalization in selected condensed matter systems.
}

\maketitle

\section{Introduction}
Fluids far from equilibrium are of current principal interest to physicists from various disciplines. A prime example for the high energy community is the Quark-Gluon Plasma (QGP)~\cite{Romatschke:2017ejr,Endrodi:2018ikq}. Such fluids are created at the Relativistic Heavy Ion Collider (RHIC) and the Large Hadron Collider (LHC) by colliding ultra relativistic heavy ions. Recently, questions have arisen concerning the magnetic fields generated during this process and their effects on the thermalization of the fluid and the correlation of particles traveling through this medium~\cite{Gursoy:2018yai,Ye:2018jwq,Gursoy:2014aka}. Given the history of the AdS/CFT correspondence~\cite{Maldacena:1997re} as a means to describe thermalizing degrees of freedom an application of this correspondence to the present question of magnetic field effects should be illuminating. Additionally, application to thermalizing condensed matter systems is possible. For example, consider a material allowing a fluid description, such as a material conducting electric current, which is driven far away from equilibrium, e.g. through an external magnetic field. Other obvious fields of application are cosmology and astrophysics, where charged fluids dynamically evolving in presence of magnetic fields are abundant (e.g.~\cite{Kaminski:2014jda}). For the sake of clarity, we will restrict ourselves in this work to draw analogies and provide conclusions for heavy ion applications and the QGP. 

The topic of thermalization in the AdS/CFT correspondence is nearly as old as the correspondence itself. The pioneering work of its founders~\cite{Witten:1998qj,Gubser:1998bc,Ferrara:1998jm} lead to a furry of activity focused on the interpretation of the various probes of CFT in terms of bulk AdS quantities and this often lead to discussion of thermalization of out of equilibrium quantities. An example of this would the use of the scale radius duality to the interpretation of the expansion of holographic bubbles in the dual CFT~\cite{KeskiVakkuri:1998nw}, (some further examples~\cite{PhysRevD.59.104021,Horowitz:1999jd,KalyanaRama:1999zj}). 

A principal example of the approach to equilibrium was provided in a series of papers by Keski-Vakkuria et al.~\cite{Danielsson:1999zt,Danielsson:1998wt,Danielsson:1999fa}. We find these to be the first instance of the use of a Vaidya like spacetime in the description of far from equilibrium CFT. In the same year Balasubramanian and Ross report the ability to detect point particles in the AdS$_3$ bulk via kinks in the corresponding Green functions~\cite{Balasubramanian:1999zv}. Their study also details the events of point particle collisions in the bulk and describes the formation of black holes during energetic particle collisions. However the most important aspect of their study for the present work is the explicit introduction of the geodesic approximation. 
 
It is the work of Aparicio, Abajo-Arrastia and Lopez~\cite{Aparicio:2011zy,AbajoArrastia:2010yt} which first makes use of the geodesic approximation in an AdS Vaiyda spacetime. This topic is covered in the same year in a complementary paper titled ``Holographic thermalization'' by Balasubramanian et al.~\cite{Balasubramanian:2011ur}. It is from this point on that the Vaiyda spacetime makes a full-fledged appearance as a community work horse of holographic thermalization. A significant portion of these previous works has been dedicated to understanding the thermalization of various quantum quenches, some examples are~\cite{Caceres:2012em,Ebrahim:2010ra,Keranen:2011xs,Camilo:2014npa,Hu:2016mym,Giordano:2014kya,Zhang:2014cga,Galante:2012pv,Dey:2015poa,Arefeva:2012jp,Zhang:2015dia,PhysRevD.84.066006,Ageev:2017wet,Andrade:2016rln,Grozdanov:2016zjj}. These examples cover a wide range of AdS Vaidya setups from the addition of charge (Reissner-Nordstrom) to modified gravity theories such as Gauss-Bonnet gravity. Yet they are all tuned to understanding the process of thermalization in strongly coupled CFTs. A reoccurring result in these works is top-down thermalization, UV degrees of freedom thermalize first. When examining non local probes this translates to the statement that short distance scales approach thermal equilibrium before long distance scales. This result is easily seen from the geodesic approximation in Vaiyda setups. 2-point functions (correlators) of small length separations correspond to geodesics which do not probe deep into the AdS bulk. As the matter shell falls further into the AdS bulk clearly geodesics which do not probe deeply into the bulk will be the first to not cross the infalling shell. As a result of their shallow trajectory they will see a black brane and hence report the value of a thermal correlator. It remains to be seen whether AdS Vaiyda setups are capable of reproducing bottom-up thermalization behavior of dual CFTs.

Before the flood of Vaidya papers many individuals indicated that colliding gravitational shock waves in AdS should be able to serve as a model of highly Lorentz contracted colliding nuclei~\cite{Grumiller:2008va,AlvarezGaume:2008fx,Lin:2009pn,Gubser:2009sx}. This motivated Chesler and Yaffe to begin the study of the adaptation of numerical solutions to Einstein equations to AdS spacetimes~\cite{Chesler:2008hg,Chesler:2009cy}. Their first work described the isotropization of a boost invariant far from equilibrium SYM fluid utilizing the characteristic formulation of Einstein's equations. Their work complemented previous studies of gravitational duals to boosted black brane solutions~\cite{Shuryak:2005ia,Janik:2005zt,Nakamura:2006ih,Janik:2006gp,Bak:2006dn,Sin:2006pv,Heller:2007qt,Heller:2008mb,Beuf:2009cx}. The combination of the previous studies on the hydrodynamics of boost invariant SYM fluid and the new numerical method to solve the Einstein equations provided a fully dynamic method of creating models to study the thermalization of strongly coupled far from equilibrium QFTs. This program has continued with a myriad of studies including the sought after gravitational wave collision in AdS~\cite{Chesler:2010bi,Heller:2012km,Casalderrey-Solana:2013aba,vanderSchee:2013pia,Chesler:2015fpa,Chesler:2015bba,Bellantuono:2015gda,Bellantuono:2015hxa,Casalderrey-Solana:2016xfq,Critelli:2018osu}. Although these systems are not QCD, they have provided practitioners of the AdS/CFT correspondence with a practical means to study far from equilibrium dynamics in a controlled setting.

Of central interest to our work was a generalization of the original anisotropic geometry due to Yaffe and Fuini~\cite{Fuini:2015hba}. The authors introduced an external magnetic field as well as a nonzero charge density, fully expecting large changes to the resulting evolution. However, it was a surprise for them to find very little difference at all! The largest deviations came from the introduction of the magnetic field. Analysis of the boundary pressure anisotropy found two time scales on which the thermalization occurred. The first is the time for the anistropic pulse to reflect off the boundary and subsequently fall into the horizon (this can be seen quite nicely by looking at the Kretschmann scalar as done for example in Fig. 3 in~\cite{Heller:2013oxa}). The second time scale is {the} time taken for the gravitational system to come to  equilibrium as a magnetic black brane geometry. Thus the study revealed that with static magnetic fields there exist potentially competing interests in the gravitational dual between an anisotropic pulse propagating to the AdS boundary and the subsequent collapse towards a magnetic black brane. These competing interests are brought to the forefront in our current work.  

Non-local probes of the dynamical systems such as correlation functions provide further insight to the propagation of information. The majority of the works above, in Vaidya spacetime, make use of the geodesic approximation to calculate 2-point functions (correlators). Some of these works also include the calculation of further non-local probes such as minimal surfaces which are dual to the entanglement entropy of the region $A\subset\mathbb{R}^{d+1}$ in the boundary CFT where $A$ is a boundary to the surface $S\subset AdS_{d+2}$ ~\cite{Ryu:2006bv}. In our work we focus on the geodesic approximation leaving the calculation of minimal surfaces for another time. Of the examples of CFT's dual to geometries calculated in the characteristic formulation two of those mentioned above have been studied further via non-local probes. Ecker et\ al.\ made first use of this method of investigating equal time 2-point correlations in a thermalizing anisotropic supersymmetric Yang-Mills fluid~\cite{Ecker:2015kna}. This study showed, among other things, even non-local probes of a strongly coupled system thermalize quickly. The second geometry used was a continuation of this study in a collaboration between Ecker et\ al.\ and van der Schee probing non-local observables of gravitational shock wave collisions~\cite{Ecker:2016thn}. 

Due to the current interest in the community concerning magnetic field effects~\cite{Gursoy:2018yai,Ye:2018jwq,Gursoy:2014aka} in (charged) strongly coupled fluids, we seek to fill the gap and provide a study of correlations in thermalizing anisotropic fluids in the presence of external magnetic fields and charge simultaneously. Furthermore, previous studies were concerned with computing the 2-point correlations of uncharged scalar operators. However, the presence of the (baryon) chemical potential and the external magnetic field provides us with the ability to study 2-point functions of charged scalar operators as well. In the bulk geometry this quantity is dual to the geodesic of a charged particle accounted for by the introduction of a Lorentz force term~\cite{Giordano:2014kya}.

\begin{figure}
  \centering
  \includegraphics[width=1.0\linewidth]{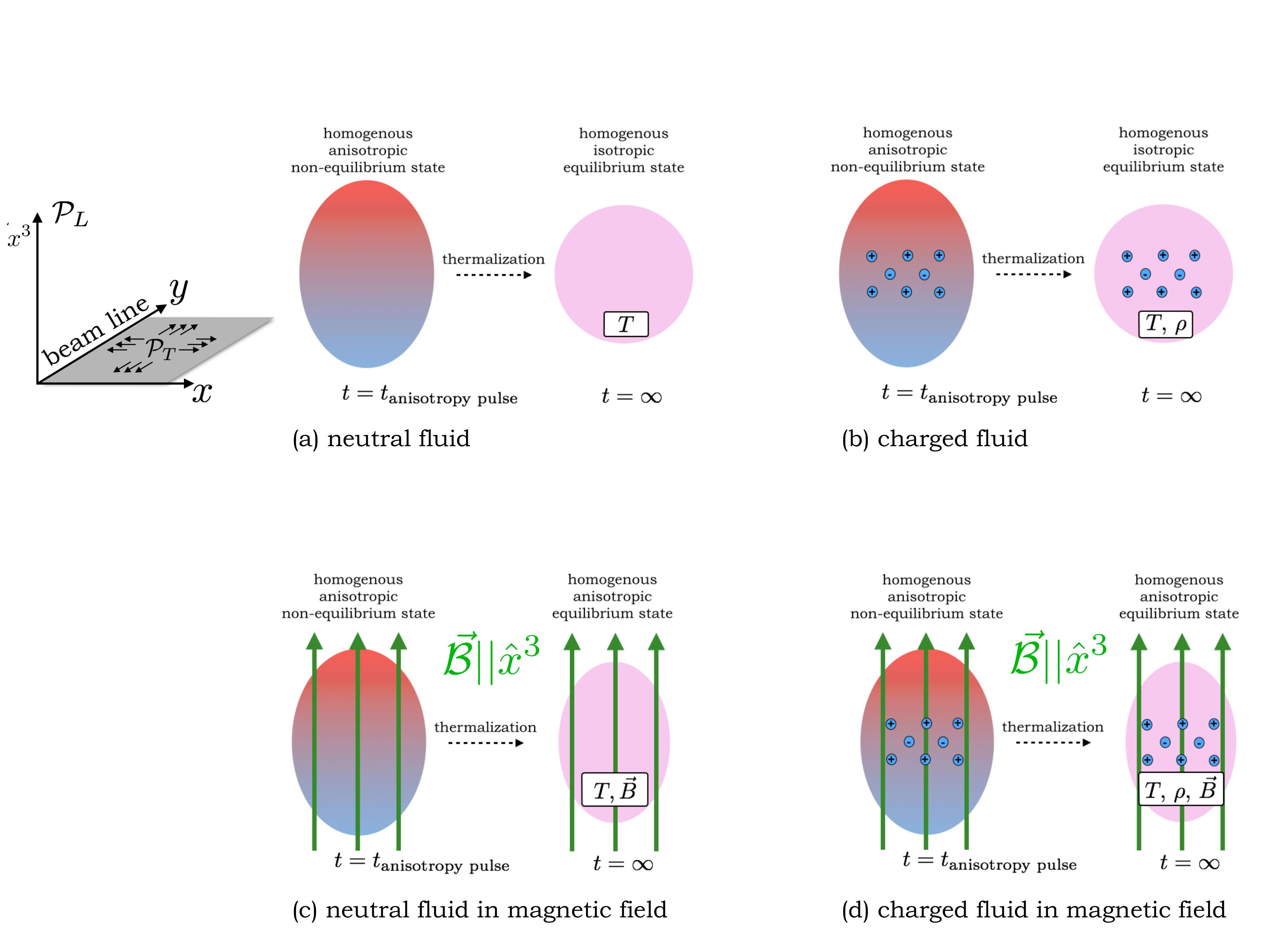}
\caption{\label{fig:introFigureSetup}
Schematic sketch of the four types of thermalization processes considered in this work. In each of the subfigures (a)--(d), a homogeneous anisotropic non-equilibrium state evolves into a homogeneous equilibrium state characterized by temperature $T$, charge density $\rho$, and magnetic field $\vec B$. The beam line can be chosen along the $y$-direction. We distinguish two sources of anisotropy, namely the initial anisotropy of the fluid and the anisotropy due to the presence of the magnetic field $\vec{\mathcal{B}}$. Both of these anisotropies distinguish the {$x^3$-direction} (longitudinal) from the $x$- and $y$-directions (transverse). Accordingly, a pressure anisotropy $\Delta \mathcal{P}=\mathcal{P}_T-\mathcal{P}_L$ is given by the difference between longitudinal pressure $\mathcal{P}_L$ (along the $x^3$-direction) and transverse pressure $\mathcal{P}_T$ (in the $x,y$-plane).}
\end{figure}
%
To be more precise, in this paper, we study the evolution of 2-point correlation functions (of scalar operators with large operator dimension) in $\mathcal{N}=4$ Super-Yang-Mills theory at strong coupling in a time-dependent (charged or uncharged) state in (presence or absence of) an external magnetic field. We consider our system to be in a state which throughout its whole time evolution can best be described as {\it fluid}, i.e. anything that can flow. A fluid could be a gas, a liquid, or a plasma. 
While we consider our system to be in a state which is best characterized as a fluid it is important to understand that this does not mean that this fluid is in a hydrodynamic regime. 

In particular, we study the four types of fluid states outlined in Fig.~\ref{fig:introFigureSetup}. Our convention is at odds with the heavy ion community. Typically it is the case that the ``longitudinal'' direction is chosen along the beam line which is the {$x^3$}-axis. The $x_1$ and $x_2$ coordinates are then the ``transverse'' coordinates. There is an anisotropy between the {$x^3$}-axis and the {$x_1, x_2$}-plane. And there is a further anisotropy between $x_1$ and $x_2$ in off central collisions which stems from the cigar shaped overlap of the colliding nuclei. The magnetic fields which appear due to the motion of the colliding nuclei are typically along the $x_1$ and $x_2$ direction~\cite{Pang:2016yuh}. In our work we use the coordinates $(v,x=x_1,y=x_2,{x^3},z)$ with $v$ a temporal direction and $z=1/r$ the fifth coordinate in the bulk AdS space. With our metric ansatz it is more natural for us to call the $x^3$ direction the longitudinal direction along which we align our magnetic field. This choice leaves us with a $\text{SO}(2)$ symmetric $x-y$ plane which we call transverse (see Fig.~\ref{fig:introFigureSetup}). 
Consequently, our setup can be thought to mimic the initial cigar-shaped overlap anisotropy in a heavy ion collision, and a magnetic field created by the colliding ions and parallel to the cigar-shape.
All of our initial states are prepared in an initial state far from equilibrium. This state then develops a pressure anisotropy $\Delta \mathcal{P}=\mathcal{P}_T-\mathcal{P}_L$ (defined as the difference between longitudinal pressure $\mathcal{P}_L$ along the third spatial direction,{ $x^3$}, and transverse pressure $\mathcal{P}_T$ in the $x,y$-plane):

\begin{itemize}
\item[(a)] Neutral fluid; note that the initial anisotropy is defined with the longitudinal pressure $\mathcal{P}_L$ along the {$x^3$}-direction, i.e. orthogonal to the beam line.
\item[(b)] Charged fluid; otherwise like (a).
\item[(c)] Neutral fluid in presence of a magnetic field along the {$x^3$}-direction, which generates a second source of anisotropy for the state (in addition to the initial pressure anisotropy); in contrast to the initial anisotropy, the anisotropy generated by the magnetic field is also present in an equilibrium state; otherwise like (a).
\item[(d)] Charged fluid, otherwise like (c).
\end{itemize}
We perform our calculations in the holographically dual setup defined by time-dependent metric and gauge field configurations which asymptote to (charged and/or magnetic) black brane solutions of Einstein-Maxwell theory.\footnote{Generically, a top-down construction also yields a Chern-Simons term. Our analysis here is undertaken at vanishing Chern-Simons coupling for simplicity. On the field theory side, this amounts to studying the electromagnetic $U(1)$, neglecting axial currents and the chiral anomaly associated with them. The anisotropic hydrodynamic regime of Einstein-Maxwell theory (including a Chern-Simons term) with a strong external magnetic field will be explored in~\cite{Ammon:2019}. In that reference, Kubo formulae are derived relating 2-point correlation functions to a multitude of transport coefficients, mentioned in part already in~\cite{Kovtun:2016lfw, Hernandez:2017mch,Kovtun:2018dvd,Ammon:2017ded}, see also~\cite{Bu:2018drd,Bu:2018psl,Bu:2016vum,Bu:2016oba}. It is important to note that~\cite{Ammon:2019} considers external non-dynamical electromagnetic fields and is hence different from modern magnetohydrodynamics with dynamical electromagnetic fields, as discussed e.g. in~\cite{Grozdanov:2017kyl,Grozdanov:2018fic,Armas:2018atq,Armas:2018zbe,Grozdanov:2016tdf,Hernandez:2017mch}.}

\paragraph{Main results. }
For a quick summary of our main results, we refer the impatient reader to our comparison of thermalization times of 2-point and 1-point functions provided in Fig.~\ref{fig:comparingThermalizationTimes}, Fig.~\ref{fig:comparingThermalizationTimes_Length_Dependence}, and table~\ref{tab:thermalizationTimes}. 
We find four main results: 
\begin{enumerate}
\item Thermalization times for 2-point functions, $t_{2pt}$ (probing nonzero length scales $l$ non-locally), are significantly larger than thermalization times for 1-point functions, $t_{1pt}$ (probing single points locally). Fig.~\ref{fig:comparingThermalizationTimes} shows examples in which $t_{2pt}\approx 3\, t_{1pt}$. 
All thermalization times 
generally increase with increasing charge density, as well as increasing length separation $l$ (between the two operators in the 2-point function). At first, with increasing magnetic field $\mathcal{B}\sim0,\dots,1$, thermalization times increase, and then decrease at larger magnetic field values $\mathcal{B} > 1$, as clearly seen from Fig.~\ref{fig:comparingThermalizationTimes}. 
\item At large magnetic fields, we observe a universal behavior in all 2-point functions despite distinct length separations $l$. All 2-point functions thermalize at approximately the same time, which we refer to as {\it saturation} time scale $t_{saturation}\approx 1.45$ (in units of the fixed energy density), approximately independent of the length separation $l$; see the $\mathcal{B}=3$ curves in  Fig.~\ref{fig:comparingThermalizationTimes_Length_Dependence}. Our definition of the thermalization time is given in Eq.~\eqref{eq:thermal}. A detailed comparison of thermalization times in all cases we studied, can be found in Sec.~\ref{sec:ComparisonofCases}. 
\item  We find a collection of effects which we attribute to a competition of the scales in the system, namely magnetic field, charge density, and the initial anisotropy (all in units of the fixed energy density). One example of scale competition effects is the change in thermalization time in charged and uncharged backgrounds. At magnetic field $\mathcal{B}=1$ the charge density has its largest effect on both the 1-point and 2-point functions, the difference between thermalization times of these cases is $\Delta t_{1pt}\approx 0.09$ and $\Delta t_{2pt}\approx 0.5$ for $l=1.5$ (see table~\ref{tab:thermalizationTimes}). As a comparison the thermalization time for the 1-point functions in the case of $\rho=0,\mathcal{B}=0$ is $t_{th}=0.687$. The difference in the two point functions is approximately $72\%$ of the time taken for the 1-point functions to thermalize in a background with $\rho=0,\mathcal{B}=0$. Yet the corresponding differences for $\mathcal{B}=3$ are $\Delta t_{1pt}\approx 0.007$ and $\Delta t_{2pt}=0.06$. The values of the dimensionless ratios $\mathcal{B}^2/\epsilon_B$ and $|\rho|^{4/3}/\epsilon_B$ display clearly the origin of this effect. For $\mathcal{B}=1,\rho=0.85,\epsilon_B=3/4$ we have $\mathcal{B}^2/\epsilon_B=1.33$ and $|\rho|^{4/3}|/\epsilon_B=1.07$, the magnetic field scale and charge density scale are close together. This can be contrasted with $\mathcal{B}=3,\rho=0.85,\epsilon_B=3.22$ for which we have $\mathcal{B}^2/\epsilon_B=2.79$ and $|\rho|^{4/3}/\epsilon_B=0.25$, in this regime the magnetic field is the dominant scale as our data in table~\ref{tab:thermalizationTimes} confirms\footnote{The equilibrium solutions for magnetic black branes can be described in terms of a single dimensionless parameter $\mathcal{B}^2/\epsilon_B$~\cite{Fuini:2015hba}. In our work we choose to work with a different renormalization scheme, the relation between our $\epsilon$ and $\epsilon_B$ is $\epsilon_B/\mathcal{B}^2=\epsilon/\mB^2 +\frac{1}{2}\log{|\mB|}$ where we have already set $L=1$. More on this topic is discussed in Sec.~\ref{sec:backgrounds} and in~\cite{Fuini:2015hba}.}.  
 \item  There appear to be strong (possibly non-local) medium effects in the fluid at early times, which seem to vanish at late times. We find nodes in equal-time correlators, see {e.g.} the two bottom graphs in Fig.~\ref{fig:BBUnchargedProbe}. The location of these nodes changes with charge density and magnetic field, as seen {e.g.} in the bottom graphs in Fig.~\ref{fig:ChargedUncharged}. As argued in Sec.~\ref{sec:results}, we interpret this to indicate the medium effects described above. 
\end{enumerate}
A comprehensive summary of our results is provided in the last section, Sec.~\ref{sec:discussion}. 
\section{Time-dependent gravitational backgrounds}\label{sec:backgrounds}
We follow the characteristic formulation of general relativity by Bondi and Sachs~\cite{Bondi:1960jsa,Sachs:1962wk}. We give only a brief explanation of this due to the well written reviews of the subject given in the papers~\cite{Chesler:2008hg,Chesler:2013lia}, by its first users in holographic settings. The backgrounds for cases (a) neutral, (b) charged, and (c) magnetic, have been extensively covered by Fuini and Yaffe~\cite{Fuini:2015hba}. In addition, a fourth case sparks our interest. Case (d), a charged fluid in the presence of a static magnetic field. 

All four fluid states described above are holographically described by four distinct background solutions to Einstein-Maxwell-Chern-Simons theory,
\begin{equation}
 S=\frac{1}{16\pi G}\int{\exd^4x \sqrt{-g} (R-2 \Lambda 
+F_{\mu\nu}F^{\mu\nu})+\gamma\epsilon^{\alpha\beta\gamma\delta\eta} 
A_{\alpha}F_{\beta\gamma}F_{\delta\eta}}.\label{CS-System}
\end{equation}
In the present work, we are focusing on vanishing Chern-Simons coupling which amounts to setting the chiral anomaly coefficient in the dual field theory to zero. The equations of motion resulting from Eq.~\eqref{CS-System} with $\gamma=0$ are,
\begin{align}
R_{\mu\nu}-\frac{1}{2}g_{\mu\nu}R+\Lambda 
g_{\mu\nu}&=2(F_{\mu\lambda}\tensor{F}{_{\nu}^{\lambda}}-\frac{1}{4}F_{\alpha\beta}F^{
\alpha\beta}g_{\mu\nu}), \\
\nabla_{\mu} F^{\mu\nu}&=0.
\end{align}
If we take as an ansatz for the metric,{$x_3$}
\begin{equation}
 \exd s^2=-A(v,r)\exd t^2+2\exd r\exd v +S(v,r)^2(e^{B(v,r)}(\exd x^2+\exd 
y^2)+e^{-2B(v,r)}(\exd x^3)^2), \label{eq:eddfinkmetric}
\end{equation}
we can find that the Maxwell equations are satisfied by taking a gauge field of the form,
\begin{equation}
\mathscr{A}(v,r)=(0,-\phi(v,r),-\frac{1}{2}y\mathcal{B},\frac{1}{2}x\mathcal{B},
0),\label{eq:gauge}
\end{equation}
with $\mathcal{B}$ a constant and the radial derivative of $\phi$ defined as,
 \begin{equation}
 -\partial_r\phi(v,r)=\mathcal{E}(v,r)=\frac{\rho}{S(v,r)^3}.\label{eq:maxwellsol}
 \end{equation}

In order to solve the Einstein equations we define new variables, corresponding to outgoing and in-falling null 
hypersurfaces. These new variables reduce the Einstein equations to a nested list of partial 
differential equations, that is the list of equations can be solved one by one in a sequential manner. The radial diffeomorphism symmetry of our metric ansatz allows us to shift the bulk radial coordinate by an arbitrary function of time. This motivates the introduction of a ``covariant'' derivative under this residual bulk radial shift and defines a derivative in the direction of outgoing bulk radial null geodesics,
\begin{equation}
 \dot{f}=\partial_v f+\frac{1}{2}A\partial_r f.\label{dotder}
\end{equation}
The radial derivative $\partial_r$ points in the direction of 
ingoing null geodesics. 

Using the definition of the gauge field given in \eref{eq:gauge} and 
\eref{eq:maxwellsol} we find the Einstein equations in the characteristic 
formulation to be as given below,
\begin{subequations}
\begin{align}
S''(v,r)&=-\frac{1}{2} B'(v,r)^2 S(v,r) \label{einS} \, ,\\
\dot{S}'(v,r)&=\frac{\mathcal{B}^2 e^{-2 B(v,r)}}{3 S(v,r)^3}-\frac{2 S'(v,r) 
\dot{S}(v,r)}{S(v,r)}+\frac{\rho ^2}{3 S(v,r)^5}+2 S(v,r)\label{einSd} \, ,\\
\dot{B}'(v,r)&= -\frac{3 \dot{B}(v,r) S'(v,r)}{2 S(v,r)}-\frac{3 B'(v,r) 
\dot{S}(v,r)}{2 S(v,r)}+\frac{2 \mathcal{B}^2 e^{-2 B(v,r)}}{3 S(v,r)^4}\label{einBdot}\, ,\\
A''(v,r)&= -3 B'(v,r) \dot{B}(v,r)-\frac{10 \mathcal{B}^2 e^{-2 B(v,r)}}{3 
S(v,r)^4}+\frac{12 S'(v,r) \dot{S}(v,r)}{S(v,r)^2}-\frac{14 \rho ^2}{3 
S(v,r)^6}-4\label{einA}\, ,\\
\ddot{S}(v,r)&= \frac{1}{2} A'(v,r) \dot{S}(v,r)-\frac{1}{2} \dot{B}(v,r)^2 
S(v,r).\label{einsdotdot}
\end{align}
\end{subequations}
We can see that the Eq.'s~\eqref{einS} to~\eqref{einA} can be solved in 
order for $S,\dot{S},\dot{B}, \text{ and } A$ given a profile for $B$ on the initial 
time slice. The data for the time derivative of $B$ on the initial slice can be 
obtained via $\partial_t B = \dot{B} -\frac{1}{2}A\partial_r B$. The code is 
pushed forward to the next time slice. We evolve the system with a 4th order Adams-Bashforth routine. The final Einstein equation is not required to propagate the initial data to the next time step and hence serves as a constraint. We calculate the residual of Eq.~\eqref{einsdotdot} on every time step to monitor the accuracy of our method. 

Each of the ordinary differential equations in Eq.'s~\eqref{einS} to \eqref{einA} is solved via a spectral method. We use a 
Chebyshev representation of the functions,
\begin{equation}
 f(r)\approx \sum_{i=0}^{N}T_{i}(r)a_i 
,\hspace{.5cm}C_j(r)=\frac{2}{Np_j}\sum^{N}_{m=0}\frac{1}{p_m}T_{m}(r_j)T_m(r).
\end{equation}
with $T_i$ the i\textsuperscript{th} Chebyshev polynomial, the $a_i$ are the 
expansion coefficients in the Chebyshev basis, and $C_j(r)$ are the cardinal functions. 
Each of the differential equations is written as a matrix 
equation with the use of the discrete derivative given in terms of derivatives of 
the cardinal functions~\cite{boyd},
\begin{equation}
 D_{ij}=\frac{\exd C_j(r)}{\exd r}\left|_{r=r_i}\right., \hspace{.5cm} D^2=D\circ 
D.
\end{equation}
The solution of the matrix equation is the value of the function at each of the 
$N$ Chebyshev grid points,
\begin{equation}
r_i=\frac{1}{2}(a+b) + \frac{1}{2}(a -b)\cos(i \pi/(N - 1))
\end{equation}

The additional bulk radial shift invariance $r\rightarrow r+\xi$ is used to fix 
the apparent horizon at the location of $z_h=1/r_h=1$ in terms of the inverted 
coordinate $z$, when appropriate.\footnote{Below we detail how the apparent horizon was fixed.} The residual shift invariance is apparent in an undetermined term 
when solving the Einstein equations order by order near the boundary. A near 
boundary expansion in powers of $r$ gives the following,
\begin{subequations}\label{eq:bdyExpansion}
\begin{align}
S(v,r)&=r+\xi(v)+\mathcal{O}(r^{-7}), \\
 B(v,r)&=\log (r) \left(-\frac{20 \mathcal{B}^2 \xi (v)^3}{3 r^7}+\frac{10 
\mathcal{B}^2 \xi (v)^2}{3 r^6}-\frac{4 \mathcal{B}^2 \xi (v)}{3 
r^5}+\frac{\mathcal{B}^2}{3 r^4}\right)+\frac{b_4(v)}{r^4}+\mathcal{O}(r^{-8}), 
\\
A(v,r)&= (r+\xi(v))^2-2\xi'(v)+\frac{a_4(v)}{r^2} \nonumber \\
&+\log (r) \left(\frac{8 
\mathcal{B}^2 \xi (v)^3}{3 r^5}-\frac{2 \mathcal{B}^2 \xi (v)^2}{r^4}+\frac{4 
\mathcal{B}^2 \xi (v)}{3 r^3}-\frac{2 \mathcal{B}^2}{3 
r^2}\right)+\mathcal{O}(r^{-6}).
\end{align}
\end{subequations}

Spectral methods can approximate non-singular functions very well. Since the functions in Eq.~\eqref{eq:bdyExpansion} contain terms of order $\mathcal{O}(r^n)$ for $n>0$ (and logarithms at subleading orders $n<0$), we choose to work with ``subtracted'' functions $S_s,\, B_s,\, A_s$,
\begin{align}
 S(v,r)&=\frac{1}{r^4}S_s(v,r)+r+\xi(v), \label{subS}\\
 B(v,r)&=\frac{1}{r^4}B_s(v,r)+\log (r) \left(-\frac{20\mathcal{B}^2\xi(v)^7}{3r^7}+\frac{10 \mathcal{B}^2 \xi (v)^2}{3 r^6}-\frac{4 
\mathcal{B}^2 \xi (v)}{3 r^5}+\frac{\mathcal{B}^2}{3 
r^4}\right), \\
A(v,r)&= \frac{1}{r^2}A_s(v,r)+(r+\xi(v))^2-2\xi'(v)\nonumber\\
&+\log (r) \left(-\frac{10\mathcal{B}^2\xi(v)^4}{3r^6}+\frac{8 \mathcal{B}^2 \xi (v)^3}{3 
r^5}-\frac{2 \mathcal{B}^2 \xi (v)^2}{r^4}+\frac{4 \mathcal{B}^2 \xi (v)}{3 
r^3}-\frac{2 \mathcal{B}^2}{3 r^2}\right)\label{subA}.
\end{align}
Our code solves the Einstein equations for the subtracted functions ($B_s$, $S_s$, $A_s$) 
rather then the full functions themselves. At anytime the full functions can be 
restored if necessary using~\eref{subS} to \eref{subA}.

The holographic renormalization procedure yields the dual 
energy momentum tensor of the field theory~\cite{Fuini:2015hba}, the non-zero components are,
\begin{subequations}
\begin{align}
 \langle T_{00}\rangle &=-\frac{3}{4} a_4 +\frac{1}{2}\mathcal{B}^2 \ln \mu L \label{eq:energyMomentumTensorT00}\\
 \langle T_{11}\rangle =\langle T_{22}\rangle&=-\frac{1}{4}a_4+b_4-\frac{1}{4}\mathcal{B}^2 +\frac{1}{2}\mathcal{B}^2 \ln \mu L\label{eq:energyMomentumTensorTxx}\\
\langle T_{33}\rangle&=-\frac{1}{4}a_4-2b_4-\frac{1}{2}\mathcal{B}^2 \ln \mu L  \label{eq:energyMomentumTensorTzz}
\end{align}
\end{subequations}
By introducing a magnetic field we have broken scale invariance. This can be seen 
when computing the trace of the boundary stress energy-momentum tensor 
\cite{Fuini:2015hba},
\begin{equation}
 T_\alpha{}^\alpha=-\frac{1}{2}\kappa \mathcal{B}^2.
\end{equation}
This conformal anomaly and the renormalization scheme dependence of this setup~\cite{Fuini:2015hba} necessitates a careful study of the renormalization point 
dependence of the boundary energy momentum tensor which we will not 
repeat here; the interested reader is referred again to~\cite{Fuini:2015hba}; see also the detailed discussion in the context of dynamical electromagnetic fields~\cite{Grozdanov:2017kyl}. We 
take the same convention as~\cite{Fuini:2015hba}, namely  
we choose the renormalization scale $\mu=1/L=1$, and define the energy density $\epsilon=\langle T_{00}\rangle$. Note however, that this energy $\epsilon$ defined at a scale $\mu=1/L=1$ is not invariant under the scaling symmetries discussed in Sec.~2 of~\cite{Fuini:2015hba}. As a result, $\epsilon$ is not a unique label for charged magnetic black brane solutions. Hence, we also introduce the scale-invariant \begin{equation}
 \epsilon_{\mathcal{B}}  = \epsilon +\frac{1}{2} \mathcal{B}^2 \ln{|\mathcal{B}|} \, , 
\end{equation}
which can be thought of as a unique parameter labeling magnetic black brane solutions. This parameter, $\epsilon_B$, in addition to the scale invariant parameter $\rho^{4/3}/\mB^2$ is required to uniquely identify charged magnetic black brane solutions as can be seen in Fig.~\ref{fig:Physically_Distinct_Solutions}.

In order to provide a meaningful comparison to previous work on non-local probes of an isotropizing fluid, we keep the energy density fixed to 
$\epsilon=3/4$ for every case. This choice of energy density corresponds to the choice of $a_4=-1$. When computing the solution for isotropization towards a RN black brane we need to fix the charge density of the final equilibrium solution. As given in~\cite{Fuini:2015hba} the blacking factor can be put into the form,
\begin{equation}
U(r)=r^2-\frac{m}{r^2}+\frac{1}{3r^4}\rho^2
\end{equation}
where we have chosen to already set the AdS curvature $L=1$. The charge density $\rho$ has an upper bound given by $\rho_e=\sqrt{2}m^{3/4}/3^{1/4}$. With our choice of energy density and mass our upper bound is $\rho_e=\sqrt{2}/3^{1/4}$. We choose to write our charge density as a fraction of this extremal charge density and find computing a solution with $\rho=0.78\rho_e$ to be sufficient to capture the effects of the addition of charge\footnote{More precisely, we use $\rho=0.7896\rho_e$.}. There has been interest expressed to us in isotropization to near extremal equilibrium solutions. We can confirm that it is difficult to construct consistent initial data as the charge density approaches the extremality bound~\cite{Fuini:2015hba}. For us to consider the initial data to be consistent we require an apparent horizon is formed at a finite location for which $r_h\rightarrow r_h'=r_h+\xi=1$. With this condition we have two choices, either 1) we use the residual diffeomorphism symmetry to place the horizon at a convient location (for us this is at $z_h=1$) or 2) we can run the calculation without  fixing the horizon. When we do not fix the horizon location we set the radial shift $\xi$ associated with the residual diffeomorphism symmetry to zero ($\xi(v)=0$). Our method relies on the fact that we work in a domain of fixed size. If we are to simulate a region of size $z\in [0,1]$ then we must be sure that all of the dynamics we wish to capture is within this region leaving all of the causally disconnected regions of spacetime to be in the region $[1,\infty)$. We use the residual diffeomorphism symmetry to place the horizon at $z_h=1$ so that the causally connected portion of the spacetime lies within our domain. The initial value of the anisotropy can be so large that we cannot form an apparent horizon at the required location of $z=1$. In the case of the RN black brane isotropization the radial shift required to hide the horizon outside of our computational domain diverges to infinity as we approach extremality. As a result of the diverging shift all initial data is inconsistent in terms of our first condition. This leaves us with the second condition. Suppose that we know the location of the apparent horizon throughout the whole evolution of our system $z_h(v)$. Suppose further that $b\in\reals$ and that,
\begin{equation}
    z_h(v)>b\hspace{.4cm} \forall\hspace{.4cm} v\in\reals_+.
\end{equation} 
Provided that,
\begin{equation}
    z_h(v)-b << 1\hspace{.4cm}\forall\hspace{.4cm} v\in\reals_+,\label{eq:Noshift}
\end{equation}
we can work in a domain $[0,b]$. If it is the case that~\eref{eq:Noshift} is not satisfied our numerics can suffer from instabilities. The long range nature of the Chebyshev polynomials is to blame for these instabilities. If a portion of the total spacetime is not captured by the domain the accuracy of the resulting spectral approximation suffers. In the RN case the diverging value of the radial shift indicates that the grid size continues to grow as we approach extremality and as a result we are unable to render the grid we need to work in finite. In the process of searching for initial data which would allow either of the above situations to occur we can further confirm that as we decrease the initial anisotropy we can begin to approach the 
extremal 
charge density.

On the first time step, a guess for $\xi(v)$ is made, and that guess is iteratively improved in order to satisfy the horizon condition $\dot S=0$ at the shifted $r=1$. In each further time step $\xi(v)$ is found via solving a first order ODE. We extract this ODE from the horizon stationarity condition,
\begin{equation}
 \left.   \partial_v \dot{S}(v,r)\right|_{r=r_h}=0.
\end{equation}
We can rewrite the time derivative using \eref{dotder} and then use the Einstein Eq.'s \eqref{einsdotdot} and \eqref{einSd} with $\dot{S}(v,r_h)=0$ to find,\footnote{Our expression for the horizon stationarity condition seems to contain additional terms compared with that of~\cite{Fuini:2015hba}.}
\begin{equation}
 \left. A(v,r) + \dot{B}(v,r)^2 \frac{3S(v,r)^6}{(6 S(v,r)^6 - \rho^2 - 
     e^{-2 B(v,r)} S(v,r)^2 \mathcal{B}^2)}
 \right|_{r=r_h}=0. \label{eq:xicon} 
\end{equation}
If we rewrite \eref{eq:xicon} in terms of the subtracted functions then we find a first order ODE for $\xi(v)$. We use this equation to propagate the value we found on the first time step to every other time step. Although we fix the horizon to $z=1$ on the initial time step and subsequently evolve the shift needed to place the horizon to $z=1$ on each time step we do find mild horizon drifting as reported in~\cite{wilkethesis} on the order of $10^{-6}$.  

\begin{figure}[h]
 \begin{center}
  \includegraphics[width=3.5in]{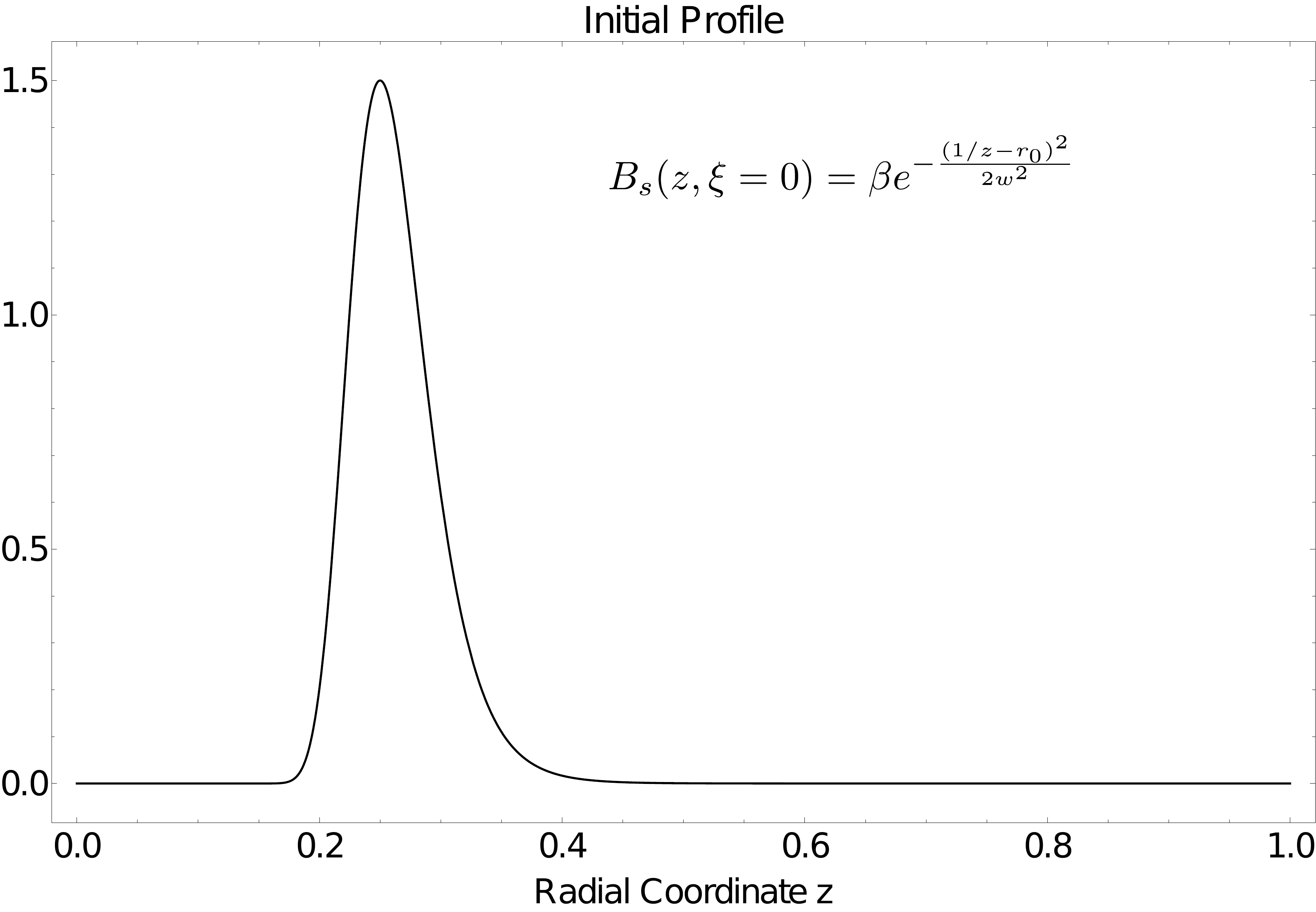}
 \end{center}
\caption{Initial Gaussian ``Pulse'' $B_s(z)$ is displayed in terms of the
inverted coordinate $z=1/r$. Parameters take the values $w=0.5$, $r_0=4$ and $\beta=1.5$.   \label{fig:initprofile}}
\end{figure}

We also keep fixed the choice of initial anisotropy function for every type of background,
\begin{equation}
 B_s(v,r)=\beta e^{-\frac{\left(\frac{1}{r}-\xi(v)-r_0\right)^2}{2 w^2}},\label{eq:init_Profile}
\end{equation}
with $w=0.5$, $r_0=4$ throughout all our numerical calculations. However we find that varying the value of the initial amplitude of the anisotropy function can lead to reduced constraint violation. We find that the principal contribution to the violation to be the accuracy with which we know the location of the apparent horizon. For $\mathcal{O}(1)$ values of the magnetic field we find initial anisotropy with amplitude of $\mathcal{O}(1)$ lends itself to the most accurate picture of the apparent horizon and hence minimal constraint violation. When the magnetic field and charge is absent we find that for amplitudes of $\mathcal{O}(10^{-1})$ we find minimal constraint violation when fixing the apparent horizon. However in order to compare the two point functions generated from these backgrounds we find it simpler to run each background at a fixed amplitude of $\beta=1.5$. 
The initial Gaussian with $w=0.5$, $r_0=4$ and $\beta=1.5$ is displayed in Fig.~\ref{fig:initprofile}.

\begin{figure}[h]
  \begin{subfigure}[b]{.5\linewidth}
  \begin{center}
\includegraphics[width=2.9 in]{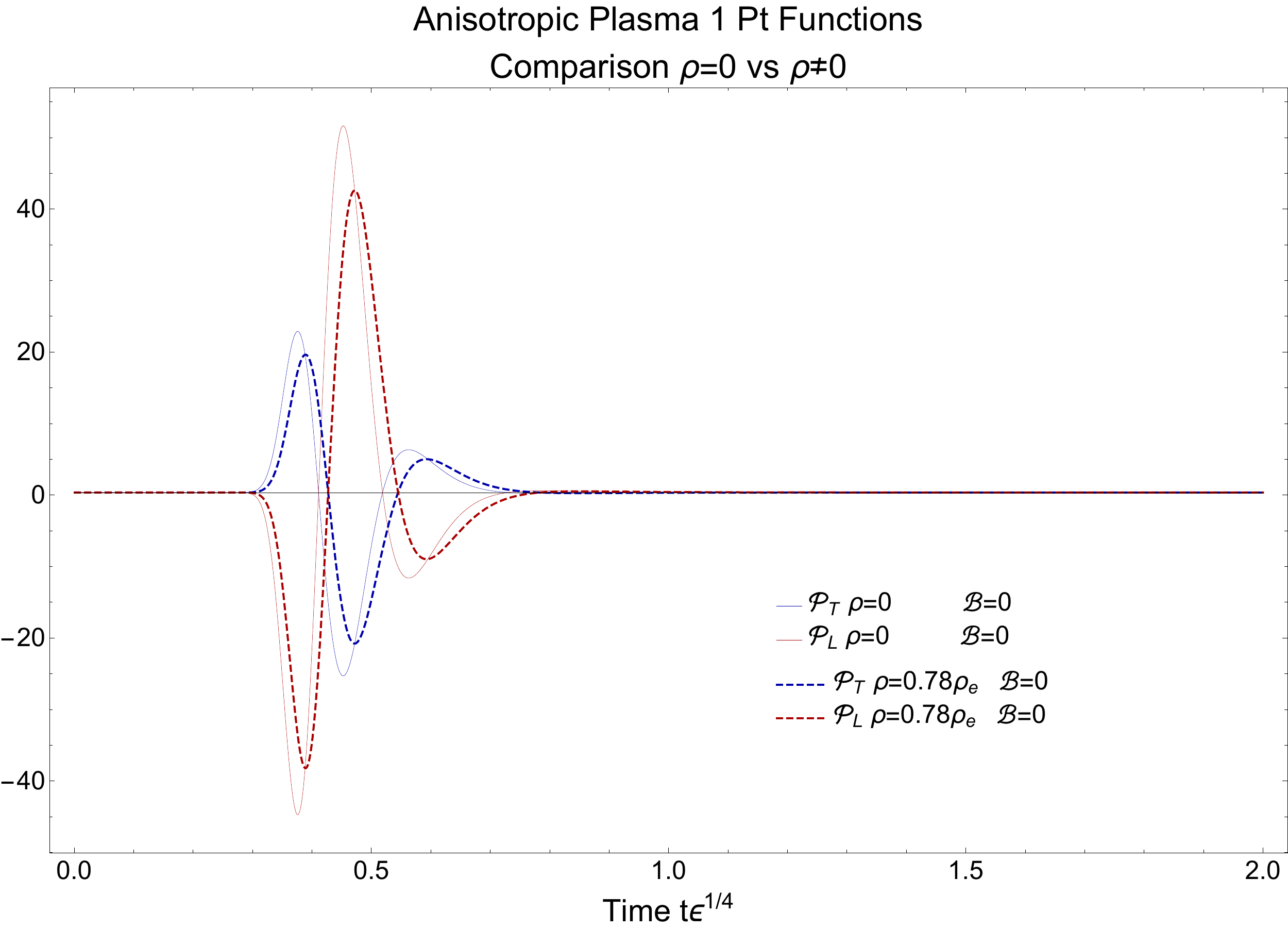}
\end{center}
\end{subfigure}
 \begin{subfigure}[b]{.5\linewidth}
  \begin{center}   
\includegraphics[width=2.9 in]{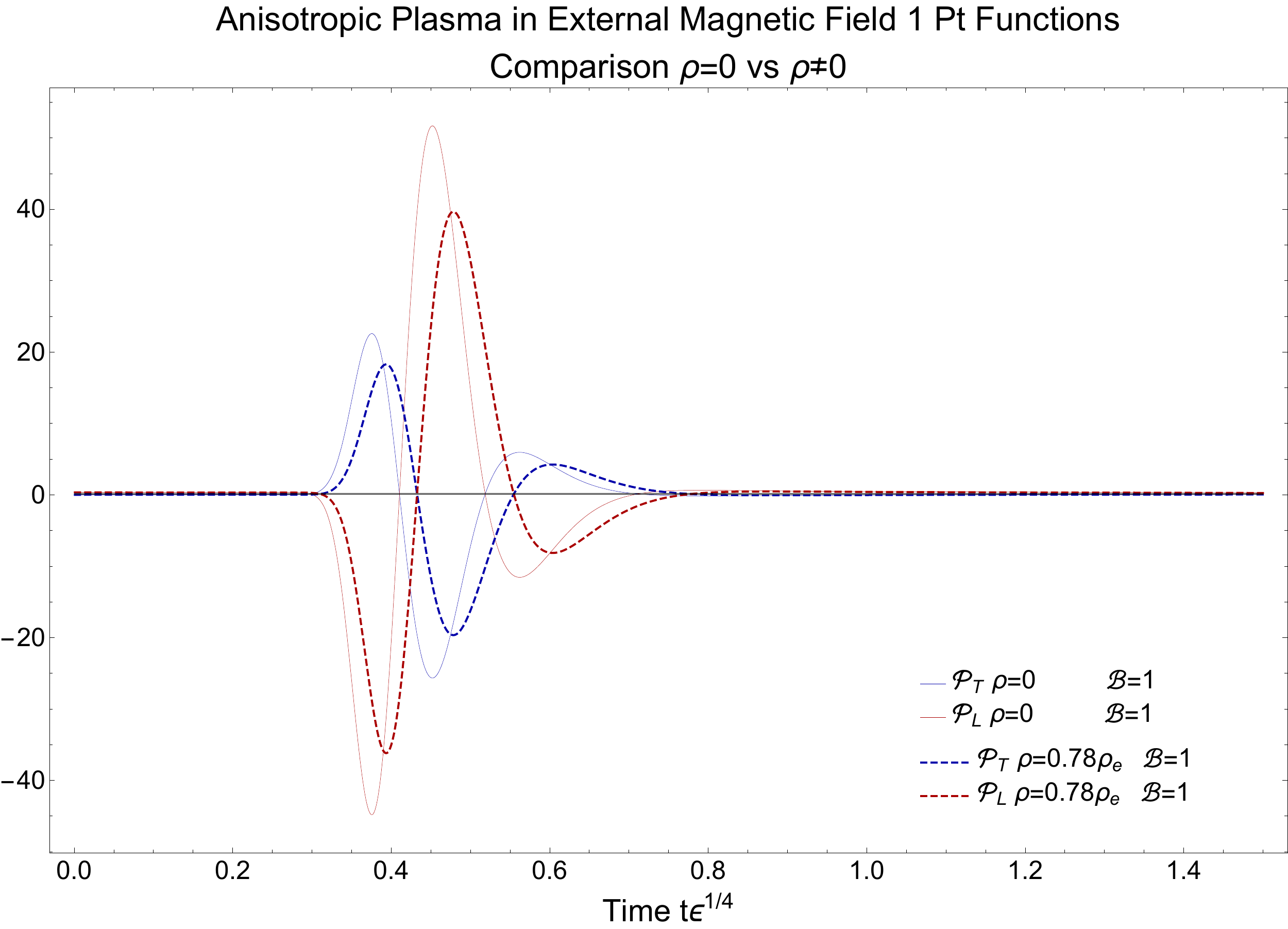}
  \end{center}
\end{subfigure}
 \begin{subfigure}[b]{.5\linewidth}
  \begin{center}   
\includegraphics[width=2.9 in]{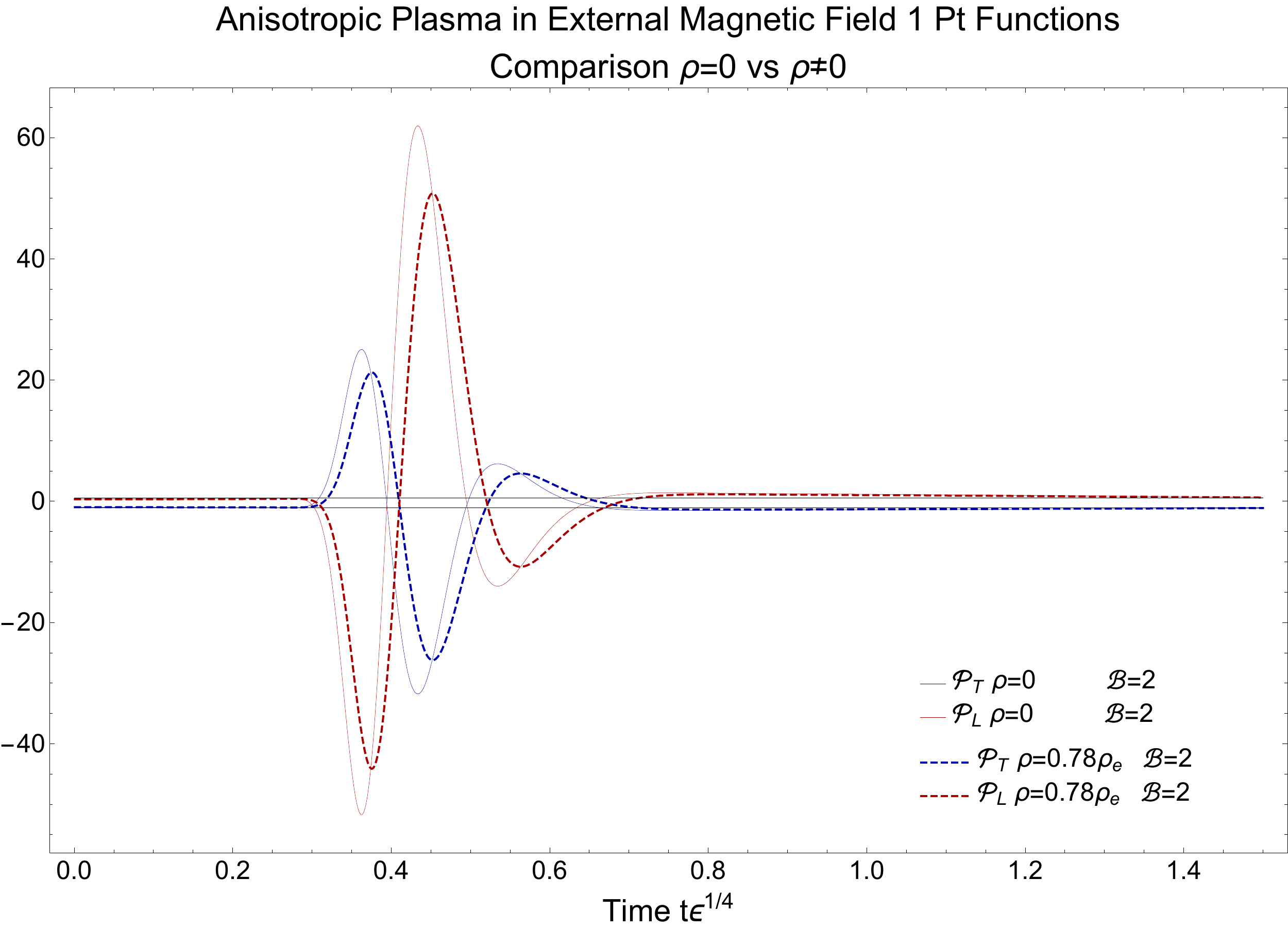}
  \end{center}
\end{subfigure}
 \begin{subfigure}[b]{.5\linewidth}
  \begin{center}   
\includegraphics[width=2.9 in]{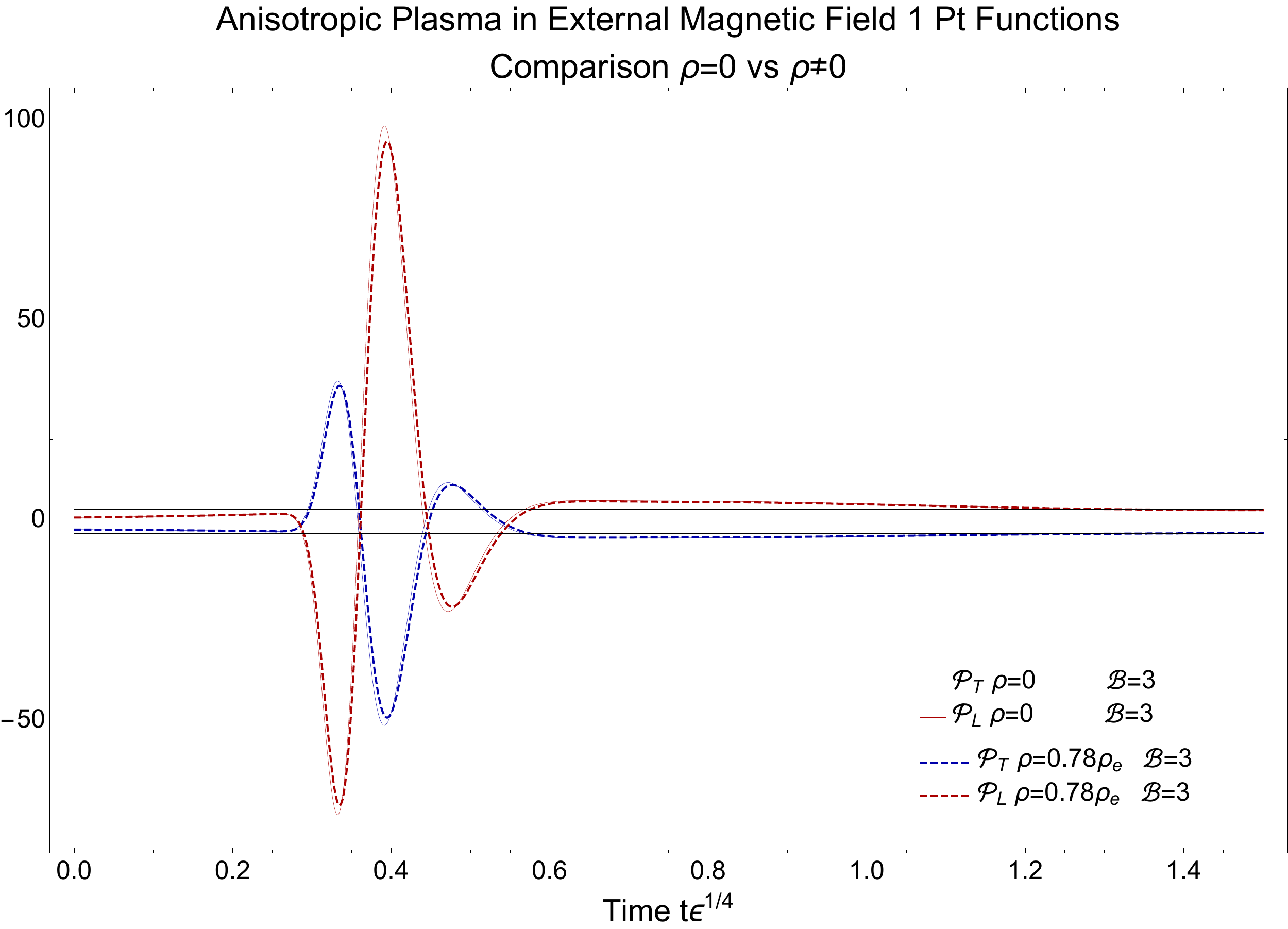}
  \end{center}
\end{subfigure}
\caption{\label{fig:ChargedUncharged}
Four plots are displayed, each showing the differences in the 1-point functions $\mathcal{P}_T$ and $\mathcal{P}_L$ obtained during the isotropization process due to the inclusion of a non zero charge density $\rho=0.78 \rho_e$. \textit{Top left:} No external magnetic field.
\textit{Top right:} External magnetic field $\mathcal{B}=1$.
\textit{Bottom left:} External magnetic $\mathcal{B}=2$.
\textit{Bottom right:} External magnetic $\mathcal{B}=3$.}
\end{figure}

\begin{figure}[h]
    \centering
    \includegraphics[width=.85\textwidth]{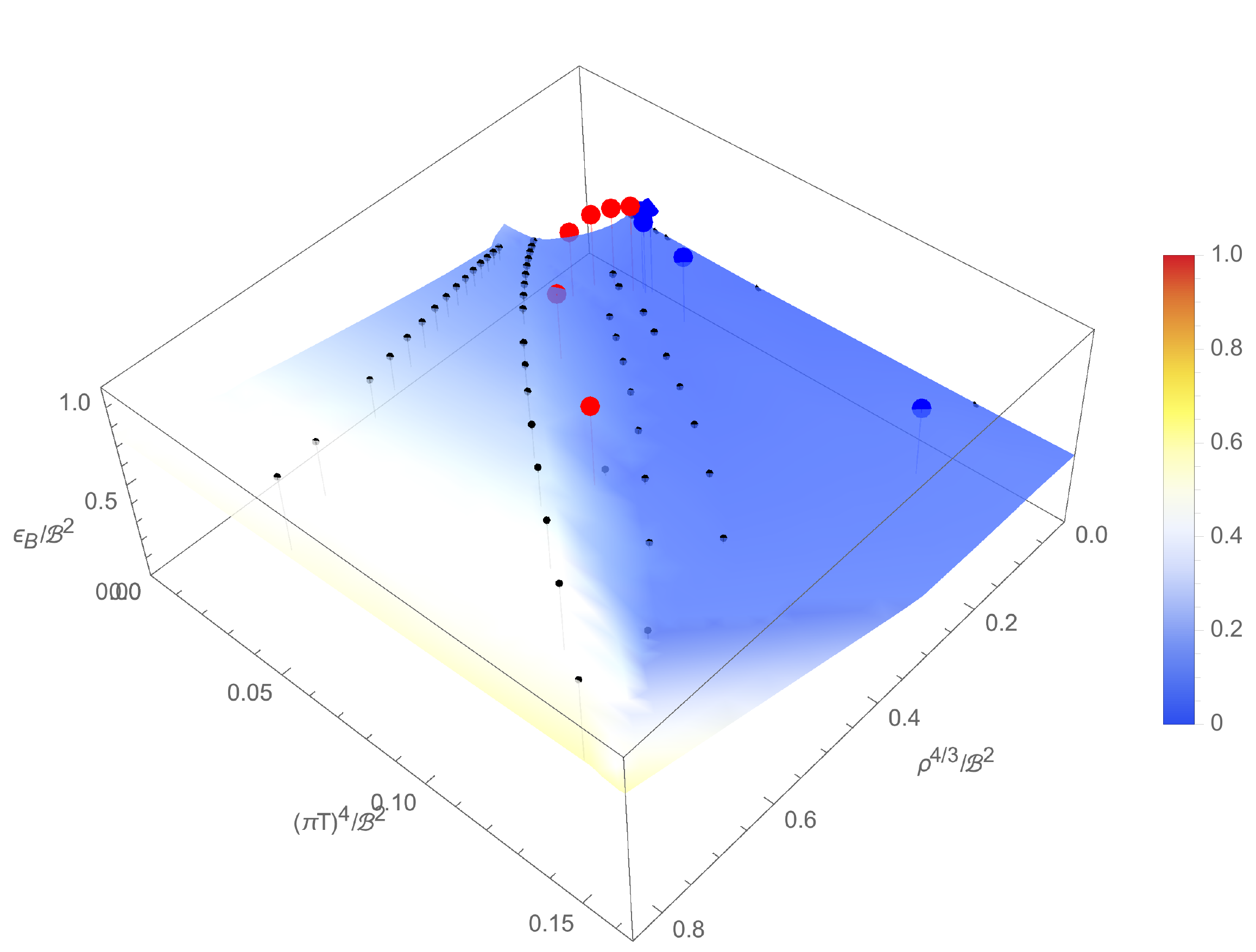}
    \caption{Physically distinct charged magnetic black branes are uniquely labeled by dimensionless ratios, $\epsilon_\mathcal{B}/\mB^2$, $(\pi T)^4/\mB^2$ and $\rho^{4/3}/\mB^2$~\cite{Fuini:2015hba}. 
    This figure shows a 3D-plot of that scale invariant thermodynamic data. Black small dots are equilibrium data generated with a shooting method (charged magnetic black brane code), the surface indicates the (rough) interpolation of this equilibrium data. Large red ($\rho=78\% \rho_e$, $\mB=0.75,\, 1,\, 1.5,\, 2,\, 2.5,\, 3$) and blue ($\rho=30\% \rho_e$, $\mB=1,\, 1.5,\, 2,\, 2.5,\, 3$) dots indicate data extracted from late time behavior of our fully dynamical setup. 
    \label{fig:Physically_Distinct_Solutions}}
\end{figure}

\begin{figure}[h]
    \centering
    \includegraphics[width=.75\textwidth]{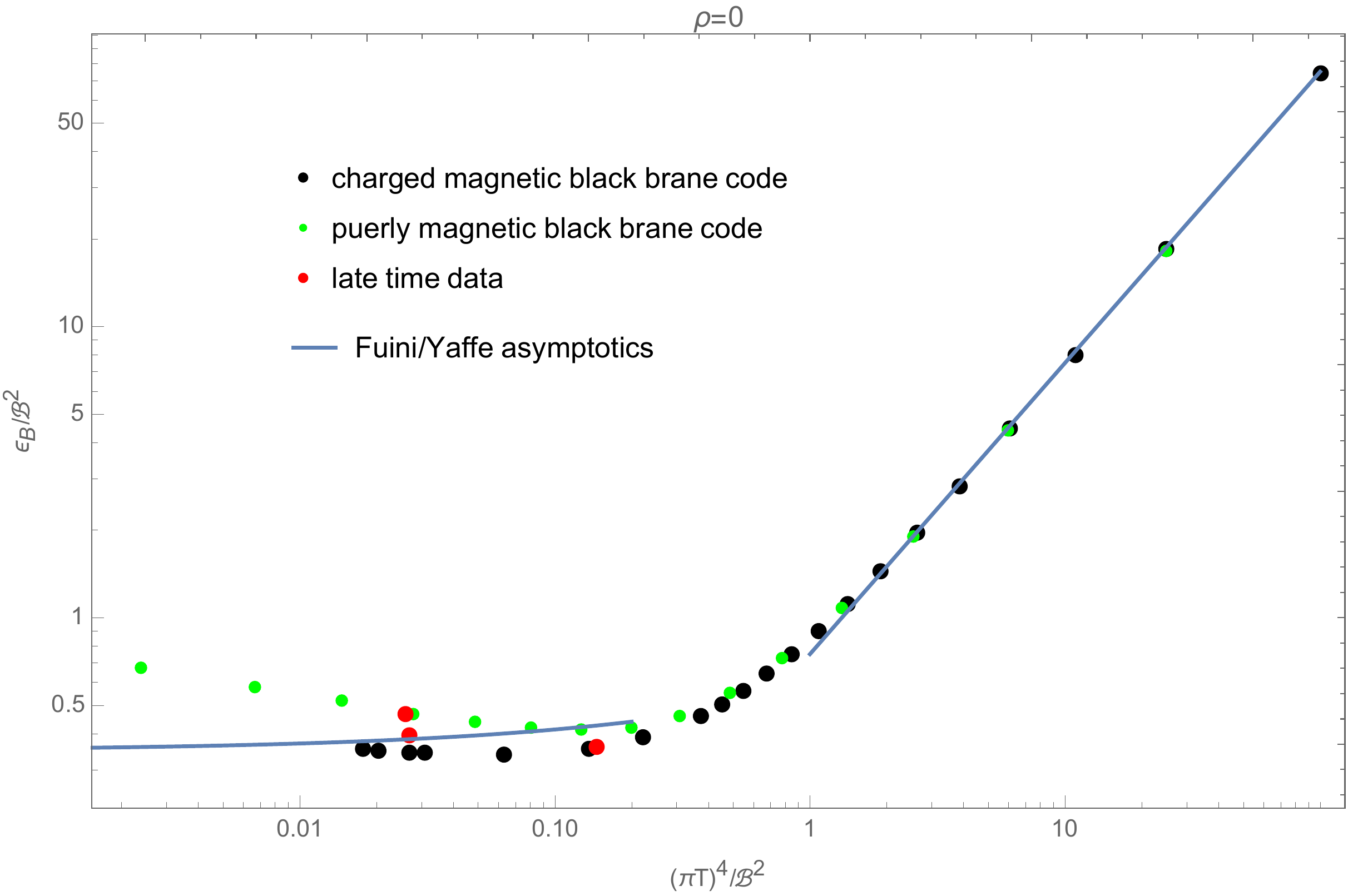}
    \caption{Physically distinct (uncharged) magnetic black branes are uniquely labeled by dimensionless ratios, $\epsilon_\mathcal{B}/\mB^2$, $(\pi T)^4/\mB^2$~\cite{Fuini:2015hba}. This plot is a slice of Fig.~\ref{fig:Physically_Distinct_Solutions} at $\rho^{4/3}/\mB^2=0$.
    In this figure, at vanishing charge density, the energy density and temperature are displayed normalized to the magnetic field. There is one unique curve on which all (uncharged) magnetic black brane solutions lie. Deviations of the data from this one curve is due to numerical errors, which increase with $\mathcal{B}$. Red dots (late time data from our fully dynamical evolution) are at $\mathcal{B}=1,\,2,\, 3$ increasing from right to left. Black and green dots indicate data from equilibrium calculations, both using the shooting method, however, the two codes use slightly different horizon/boundary cutoffs, different numerical working precision and different number of orders in the near-horizon expansion.
    \label{fig:Physically_Distinct_SolutionsRho0}}
\end{figure}

Previous work was concerned with the form of the 1-point functions of the energy momentum tensor obtained via the asymptotic coefficient $b_4$. With our choice of subtraction and scaling this coefficient can be obtained as $b_4(t)=B_s(0,t)$. Here we write the coefficient as a function of $t$ rather then $v$ since on the boundary $t=v$. We demonstrate that our method here reproduces the earlier work of Chesler and Yaffe~\cite{Chesler:2008hg}, the corresponding curves are shown in Fig.~\ref{fig:ChargedUncharged}.
We have checked that our late time thermodynamic data --up to numerical deviations-- agrees with the known static solutions in the Schwarzschild case (up to $7.01\times 10^{-4}$\%), the (charged) Reissner-Nordstr\"om case (up to $1.26\times 10^{-8}\%$), and the magnetic black brane case~\cite{Janiszewski:2015ura, D'Hoker:2009mm} (up to $0.503\%$ for $\mathcal{B}=1$, $0.750\%$ for $\mathcal{B}=2$ and $7.780\%$ for $\mathcal{B}=3$). See table~\ref{tab:lateTimeVsEquilibrium}. As a new equilibrium result, we compute the static charged magnetic black brane solutions to Einstein-Maxwell theory\footnote{The charged magnetic black brane solutions studied by~\cite{Ammon:2016fru,Ammon:2016szz,Ammon:2017ded} are solutions to Einstein-Maxwell-Chern-Simons theory, i.e. with a nonzero Chern-Simons coupling.}; see table~\ref{tab:lateTimeVsEquilibrium2} and Fig.~\ref{fig:Physically_Distinct_Solutions}. If our late time data is consistent with the equilibrium solutions, all the data needs to lie on the same surface in this 3D-plot of Fig.~\ref{fig:Physically_Distinct_Solutions}. Black small dots are equilibrium data generated with a shooting method, the surface indicates the interpolation of this equilibrium data. Large red ($\rho=78\% \rho_e$) and blue ($\rho=30\% \rho_e$) dots indicated data extracted from late time behavior of our fully dynamical setup. There is good agreement at $\mathcal{B}=1$ and $2$, at worst the late time data deviates by $12.2\%$ ($\mathcal{B}=2,\, \rho=30\% \rho_e$). At $\mathcal{B}=3$, there is a deviations of 29.9\% and 28.7\%. Fig.~\ref{fig:Physically_Distinct_SolutionsRho0} illustrates that numerical errors of this order ($\sim 30\%$) are already present in the equilibrium data when calculated with two different codes at vanishing charge density. This indicates that at large magnetic fields $\mathcal{B}\ge 3$ the numerical problem becomes increasingly challenging for the shooting method which we are using to find the equilibrium data. This makes our equilibrium data more and more unreliable at $\mathcal{B}\ge 3$ and it is more difficult to judge if the far-from equilibrium setup has reached an equilibrium at late time. Our convergence checks indicate that at late time the thermodynamic data (energy density, pressure anisotropy) extracted from our non-equilibrium setup asymptotes to constant values. We will assume that these are close to equilibrium values of this setup, even at $\mathcal{B}=3$, and leave the detailed investigation of this region $\mathcal{B}\ge 3$ to later work. The asymptotic lines are taken from Fuini/Yaffe~\cite{Fuini:2015hba}.  
See table~\ref{tab:lateTimeVsEquilibrium2} for values.
\begin{table}[h]
    \centering
\begin{tabular}{c|c|c|c|c|}
 $\mathcal{B}$ & 0 & 1 & 2 & 3  \\
 \hline
$\rho=0$ & $7.01\times 10^{-4} \%$ & $0.503\%$ & $0.750\%$ & $7.780\%$ \\
 \hline
$\rho=0.848528$   
& $1.26\times 10^{-8}\%$ & \multicolumn{3}{c|}{see table~\ref{tab:lateTimeVsEquilibrium2}}  \\
\hline
\end{tabular}
 \caption{\label{tab:lateTimeVsEquilibrium}
 Comparison to equilibrium data: Deviation (in percent) between our late time thermodynamic data (i.e. pressures $\mathcal{P}_{L,T}$) and the equilibrium solutions, i.e. the Schwarzschild, Reissner-Nordstr\"om, and purely magnetic black brane solutions. 
 }
\end{table}

\begin{table}[h]
    \centering
\begin{tabular}{c|c|c|c|c|}
 $\mathcal{B}$&
 $\frac{(\pi T)^4}{\mathcal{B}^2}$   &
 $\frac{\rho^{4/3}}{\mathcal{B}^2}$  &
 $\epsilon_\mathcal{B}/\mathcal{B}^2$&
 $2(\epsilon_\mathcal{B}-\epsilon^{equilibrium}_\mathcal{B})/(\epsilon_\mathcal{B}+\epsilon^{equilibrium}_\mathcal{B})$\\
 \hline
 \multicolumn{5}{c|}{red dots (30\% $\times \rho_{e}$ )}\\
 \hline
 1 & 0.0552605 & 0.126033 & 0.360787 & 0.70\%  \\
 \hline
 2 & 0.0138151 & 0.0242842 & 0.393449 &  12.2\% \\
\hline
 3 &  0.00614005 & 0.0255086 & 0.468773  & 29.9\% \\
\hline
 \multicolumn{5}{c|}{blue dots (78\% $\times \rho_{e}$ )}\\
 \hline
 1 & 0.197566 & 0.0247171 & 0.360787 & -5.40 \%  \\
 \hline
 2 & 0.0493915 & 0.0108224 & 0.393449 &  -7.90\% \\
\hline
 3 &  0.0219518 & 0.0210068 & 0.468773  & 28.7\% \\
\hline
\end{tabular}
 \caption{\label{tab:lateTimeVsEquilibrium2}
 Comparison to equilibrium data: last column shows deviation (in percent) between our late time energy density $\epsilon_\mathcal{B}/\mathcal{B}^2$ and that of the equilibrium solutions, i.e. the charged magnetic black brane solutions. These data points are displayed in Fig.~\ref{fig:Physically_Distinct_Solutions} as red and blue large dots.
 }
\end{table}

\begin{figure}[h]
\begin{subfigure}[b]{.48\linewidth}
\includegraphics[width=2.8in]{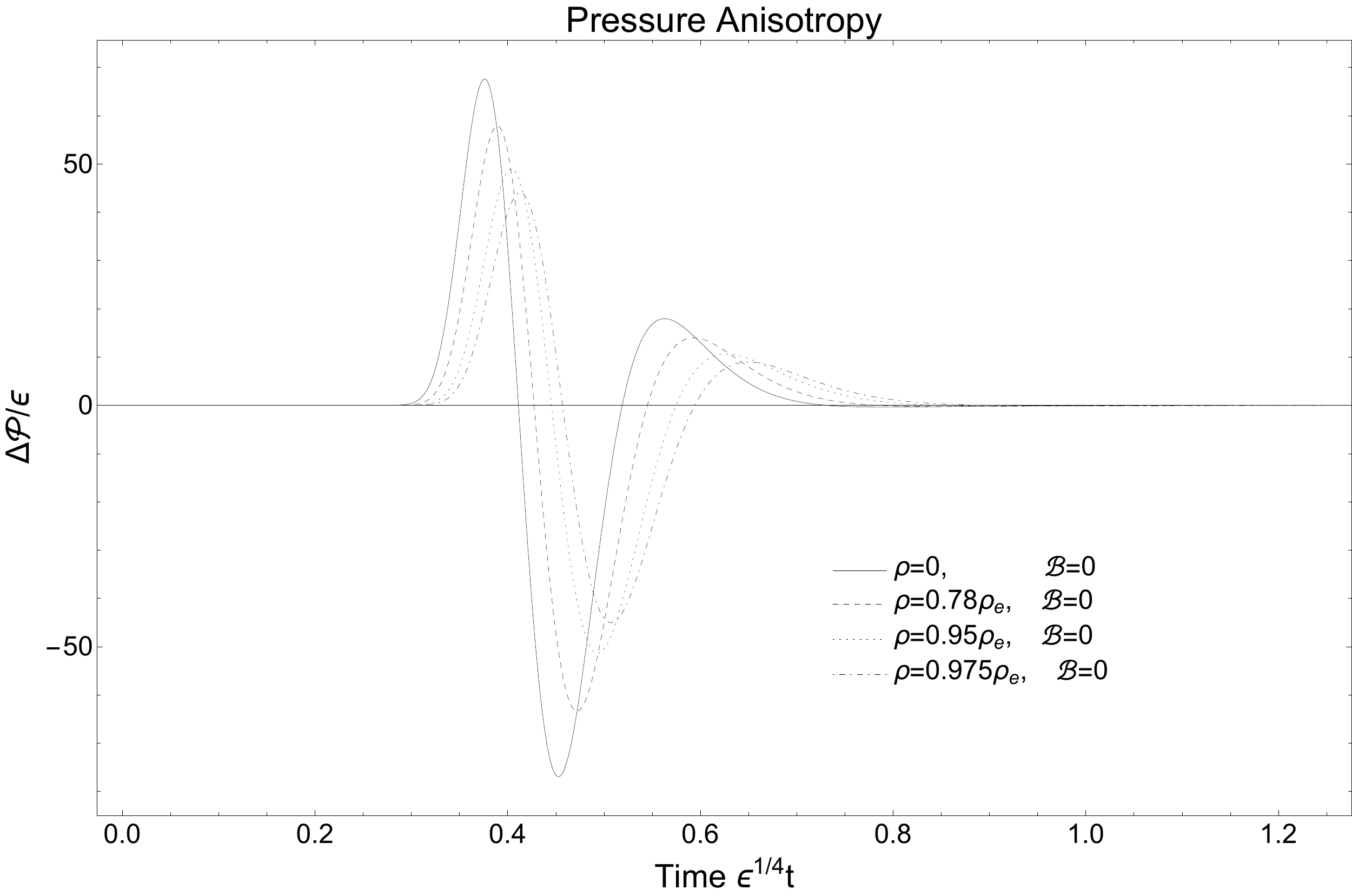}
\end{subfigure}
\begin{subfigure}[b]{.48\linewidth}
\includegraphics[width=2.8in]{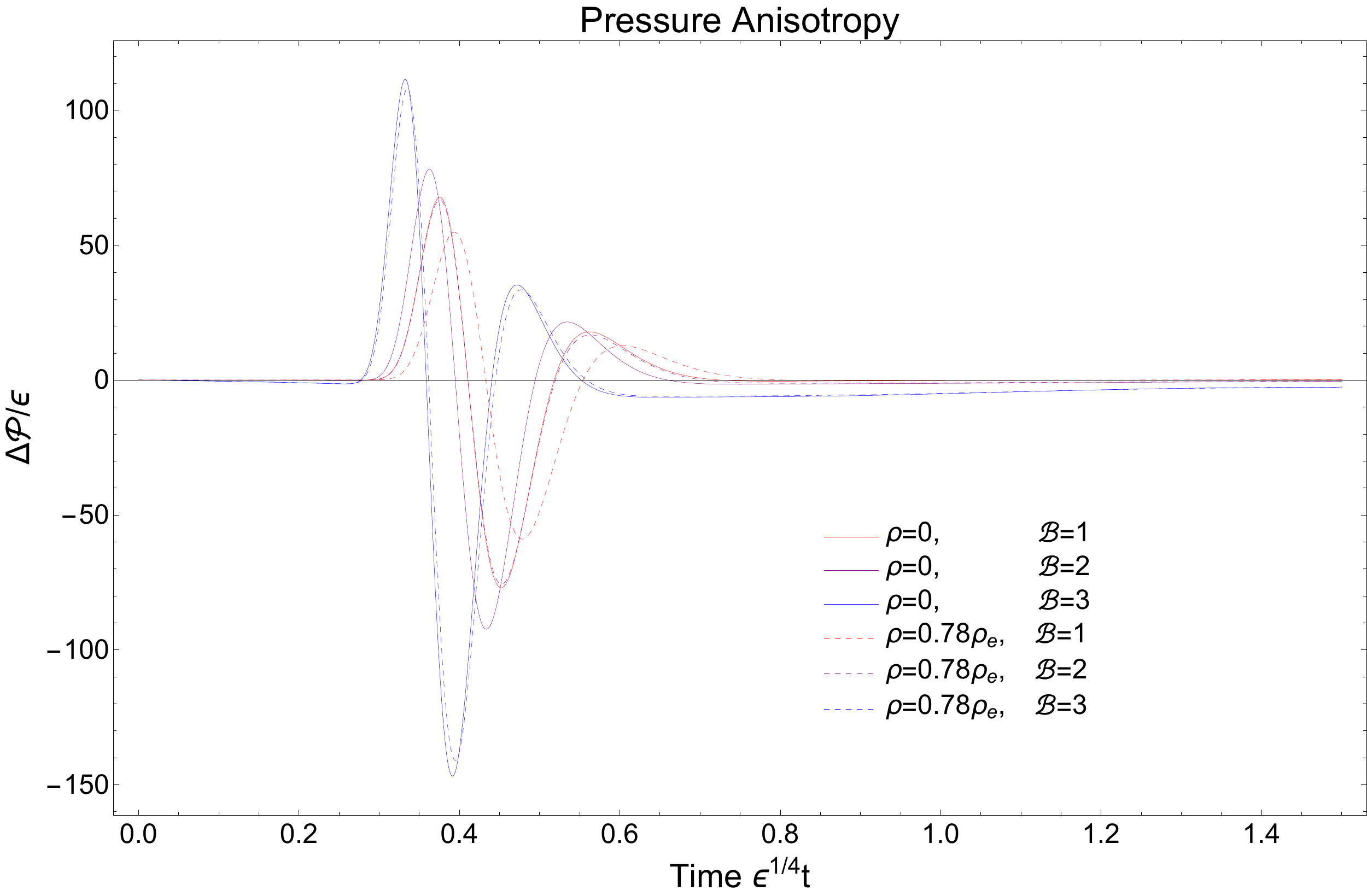}
\end{subfigure}
\caption{\label{fig:Everything} The pressure anisotropy $\Delta \mathcal{P}=\mathcal{P}_T -\mathcal{P}_L+\frac{1}{4}\mathcal{B}^2$ in units of energy density $\epsilon$ for every case considered in this work displayed together. }
\end{figure}

It should be noted that the pressures oscillate throughout the evolution, and pass through their eventual equilibrium values several times. These times will lead to distinct features in the 2-point functions, and we refer to these times as {\it equilibrium times in the background}, $t_{eq, n}$ (with $n=1,\, 2,\,3,\dots $). 
In Fig.~\ref{fig:Everything}, for every system under consideration we also display the pressure anisotropy 
\begin{equation}\label{eq:PAnisotropy}
\Delta\mathcal{P}=\mathcal{P}_T -\mathcal{P}_L +\frac{1}{4}\mathcal{B}^2
= \langle T^{xx}\rangle - \langle T^{33}\rangle
+\frac{1}{4}\mathcal{B}^2\, ,
\end{equation}
where $\langle T^{xx}\rangle$ is the transverse component of the energy momentum tensor 1-point function, and $\langle T^{33}\rangle$ is the longitudinal one. When magnetic fields are present there is a static contribution to the pressure anisotropy due to the magnetic field of $\mathcal{P}_{Static}=-\frac{1}{4}\mathcal{B}^2$.
This can be seen from Eq.~\eqref{eq:energyMomentumTensorTxx} and Eq.~\eqref{eq:energyMomentumTensorTzz}. For ease of comparison among cases we do not display this static contribution, we choose only to display the dynamic contribution calculated from the asymptotic expansion of the bulk metric.  

Only subtle differences between the systems are apparent. These differences although small have a large effect on the behavior of the two point correlation functions of scalar operators with large conformal operator dimension. The deviation between 2-point correlations is especially evident in the case of the isotropization of the magnetic black brane. While the effects of the charge density mirror what has been seen previously for the 1-point functions. 
In Sec.~\ref{sec:results}, we define a measure for thermalization time of our 1-point function of the energy momentum tensor, and discuss the effect of charge, magnetic field, and initial anisotropy on this thermalization time.
We normalize our timelike axis to the energy density of the corresponding field theory. We find this to be the most useful normalization in the presence of a magnetic field.  

\section{Geodesics}\label{sec:geodesics}
Readers interested in the results only may want to skip this section.
We employ the geodesic approximation as first shown in 
\cite{Balasubramanian:1999zv} and used first in asymptotically AdS Vaidya spacetime in~\cite{AbajoArrastia:2010yt,Aparicio:2011zy} and subsequently in 
\cite{Ecker:2015kna,Caceres:2012em,Camilo:2014npa,Hu:2016mym,Giordano:2014kya} 
etc. The Wightman functions are approximated via,
\begin{equation}
\langle\mathscr{O}(t,\vec{x}_1)\mathscr{O}(t,\vec{x}_2)\rangle=\int{\mathcal{DP}
e^{i\Delta \mathcal{L}(\mathcal{P})}}\approx\sum_{\text{geodesics}}e^{-\Delta 
L}\approx e^{-\Delta L}\, ,\label{eq:geoapprox}
\end{equation}
where the sum is taken over all geodesics. $L$ is the proper length of the 
geodesic and is given as,
\begin{equation}
 L=\int{\exd\lambda \sqrt{g_{\mu\nu}\dot{x}^{\mu}\dot{x}^{\nu}}}\, ,
\end{equation}
with the derivative $\dot{x}^{\mu}\equiv \frac{\exd x^{\mu}}{\exd \lambda}$, where $\lambda$ is an affine parameter. In order to utilize \eref{eq:geoapprox} we first calculate the variation of 
the action and solve the resulting equations of motion which is nothing other 
than solving the geodesic equations. Having found a solution to the geodesic equations 
we then use these solutions in the action and explicitly compute the integral to 
find the proper length of the geodesic. In practice this requires us to solve 
the geodesic equations in a numerically generated background. To do so we,
\begin{enumerate}
 \item generate the dynamic background,
 \item generate interpolations of the metric functions,
 \item discretize the geodesic equations using a relaxation scheme as in 
\cite{Ecker:2015kna},
 \item approximate the proper length using a Riemann sum.
\end{enumerate}
 The backgrounds we investigate are anisotropic so we follow along with 
\cite{Ecker:2015kna} and note we can separate the geodesics through the bulk into two classes with separations in the transverse ($x$-$y$) plane and longitudinal ($x^3$ axis) direction. We denote by $l$ the separation in the boundary 
theory and by $v_0$ the initial time at which the Wightman functions are calculated. Since the computational domain for time is fixed as $[0,v_{comp}]$ there is a critical time $v_c$ which bounds the initial time of our geodesic at fixed length from below $v_c<v_0$. The bound for each length is different, and the bound itself is due to 
the behavior of the Eddington-Finkelstein coordinate $v$ as we reach into the 
bulk~\cite{Ecker:2015kna}. 

We choose to use a relaxation scheme to solve the discrete geodesic equations  as shown in~\cite{Ecker:2015kna}. For the scheme to work we require an initial guess for the geodesic. We take as our initial guess the geodesic in an empty AdS space,
\begin{equation}
 ds^2=\frac{1}{z^2}(-\exd v^2-2\exd v\exd z +\exd x^2) \, .
\end{equation}

Analytic forms for spacelike geodesics can be computed via first integrals of 
motion as is done in~\cite{Ecker:2015kna,Ecker:2016thn} so we will not repeat 
the procedure of obtaining these solutions. They are given in terms of a 
non-affine parameter $\sigma$ as,
\begin{equation}
 z(\sigma)=\frac{l}{2}(1-\sigma^2),\hspace{1cm} 
x(\sigma)=\frac{l}{2}(\sigma\sqrt{2-\sigma^2}),\hspace{1cm} 
v(\sigma)=v_0-z(\sigma) \, .
\end{equation}
Using a non-affine $\sigma$, the relevant equation is the nonaffinely 
parameterized geodesic equation,
\begin{equation}
 \frac{\exd^2 x^{\mu}}{\exd 
\sigma^2}+\tensor{\Gamma}{^{\mu}_{\alpha\beta}}\frac{\exd x^{\alpha}}{\exd 
\sigma}\frac{\exd x^{\beta}}{\exd \sigma}=\frac{\frac{\exd^2 \lambda}{\exd 
\sigma^2}}{\frac{\exd \lambda}{\exd \sigma}}\frac{\exd x^{\mu}}{\exd\sigma}\label{eq:nonAffine}
\end{equation}

The set of spacelike solutions of the geodesic equation~\cite{Matschull:1999} required for this work also include timelike separated points in the boundary theory. Such trajectories through the bulk of empty AdS spacetime then correspond to non-equal time Wightman functions on the boundary and their solutions are given as,
\begin{subequations}
\begin{align}
  z(\sigma)&=\frac{\sqrt{l^2-\Delta t^2}}{2}(1-\sigma^2),\\
x(\sigma)&=\frac{l}{2}(\sigma\sqrt{2-\sigma^2}),\\ 
v(\sigma)&=\frac{\Delta t}{2}\sigma\sqrt{2-\sigma^2}+v_0-z(\sigma).
\end{align}
\end{subequations}
The $l$ again denotes the spatial separation but the $\Delta t$ is new and denotes the temporal separation. The correlation is then calculated between spacetime points $X_1=(v_0-\Delta t/2,-l/2)$ and $X_2=(v_0+\Delta t/2,l/2)$. To arrive at these solutions please refer to Ecker et.\ al.\ ~\cite{Ecker:2015kna} without setting the constant of motion to zero, i.e. keep $E\neq 0$.  

The relaxation method we use is exactly that which is described in 
\cite{numericalrecipes}. We first generate the geodesic equations using 
Mathematica and write $\exd x^{\mu}/\exd\sigma\equiv p^{\mu}$ to arrive at a set of 
six first order equations for six fields, summarized as $V^{\mu}=(v,z,x,p_{v},p_{z},p_{x})$, note this 
is not a vector quantity. We then discretize by making the replacements,
\begin{subequations}
 \begin{align}
 f(v(\sigma),z(\sigma))&\rightarrow 
\frac{1}{2}(f\left(v_{k-1},z_{k-1}\right)+f\left(v_{k},z_{k}\right)) \\
 f'(\sigma)&\rightarrow \frac{f_{k}-f_{k-1}}{\sigma_{k}-\sigma_{k-1}} \\
  f(\sigma)&\rightarrow \frac{1}{2}(f_{k}-f_{k-1}) \\
  J f(\sigma)&\rightarrow \frac{1}{2}(J_{k-1}f_{k-1}+J_{k}f_k),\hspace{1cm} 
J=\frac{\exd^2 \lambda}{\exd \sigma^2}/\frac{\exd \lambda}{\exd \sigma}
\end{align}\label{eq:discrete}
\end{subequations}
The goal is to then create a matrix which represents the values of the equations 
at the grid points and minimize the correction needed to each grid point with 
$V^{\mu}_{n+1}=V^{\mu}_{n}+\Delta V^{\mu}$. We work with,
\begin{equation}
 E^{\mu}\equiv\begin{cases} \frac{\exd x^{\mu}}{\exd\sigma} - p^{\mu} 
\hspace{.2cm} \text{if} \hspace{.2cm}\mu=1,2,3 \\
 \frac{\exd p^{\mu}}{\exd 
\sigma}+\tensor{\Gamma}{^{\mu}_{\alpha\beta}}p^{\alpha}p^{\beta}
 -\frac{\frac{\exd^2 \lambda}{\exd \sigma^2}}{\frac{\exd \lambda}{\exd 
\sigma}}p^{\mu}. \hspace{.2cm}\text{if} \hspace{.2cm}\mu=4,5,6
\end{cases}
\end{equation}
To calculate the corrections $\Delta V^{\mu}$  we first apply the discretization 
scheme shown in \eref{eq:discrete} to the residual functions $E^{\mu}$ and 
expand $E^{\mu}_n$ in a Taylor series up to first order and insist this Taylor 
series vanish as in~\cite{numericalrecipes}. If we set $M$ as the number of grid points and $N$ as the
number of variables $\mu_e$, defined as
\begin{equation}
 \mu_e=\frac{1}{M N}\sum^{M}_{i=1}{\sum^{6}_{\mu=1}{\Delta V^{\mu}_i}}<10^{-6},
\end{equation}
serves as our measure of success. The relaxation procedure is continued until the average correction to each of the variables is less 
than $10^{-6}$. At each step we evaluate the measure $\mu_e$ to decide how much 
of the correction we should add back to the coordinates. Our relaxation scheme is a first 
order method and is likely not the best judge of distance to the solution until 
$\mu_e$ is relatively small therefore we use,
\begin{equation}
 \alpha =\begin{cases}
          .05 & \text{if}\hspace{.35cm} \mu_e>.05 \\
          1 & \text{if}\hspace{.35cm} \mu_e<.05
         \end{cases}.
\end{equation}
At each step of the relaxation we compute $\Delta V^{\mu}$ and update the 
coordinates as $V^{\mu}_{n+1}=V^{\mu}_{n}+\alpha\Delta V^{\mu}$. 

The boundary conditions we employ are: 
\begin{enumerate}
 \item Equal time correlators with AdS$_3$ slices: \begin{align} z_0&=z_{N-1}=z_{UV},\\ v_0&=v_{N-1}=v_i,\\x_0&=-x_{N-1}=-l/2\,.\end{align}
 \item Non-equal time correlators AdS$_3$ slices: \begin{align} z_0&=z_{N-1}=z_{UV},\\ v_0&=v_i-\frac{\Delta t}{2}\, ,\quad v_{N-1}=v_i+\frac{\Delta t}{2},\\x_0&=-x_{N-1}=-l/2\,.\end{align}
\end{enumerate}

\begin{figure}
\begin{center}
\includegraphics[width=4in]{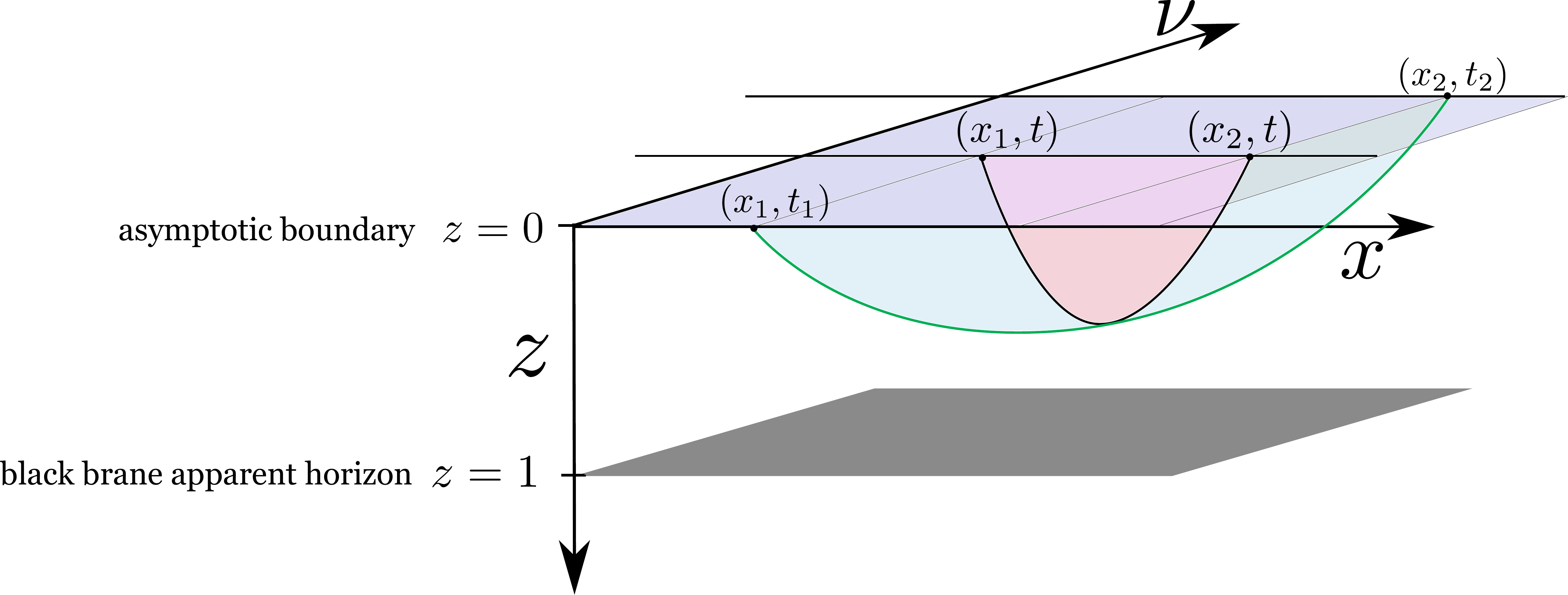}
\end{center}
\caption{\label{fig:BCs}A schematic representation of the boundary conditions used in this work. }
\end{figure}
Each geodesic takes about 3 iterations of our relaxation code. 
Since we generate geodesics for a large set of times at a fixed length we use 
the previously computed geodesic as the initial guess for the next time step 
with each set of geodesics at fixed length $l$ separated in time by $\delta 
t=0.0125$. At the critical time $v_c$ the 
coordinate time $v$ will reach beyond our numerical data. This critical 
time is different for each length and each type of background. To ensure 
we calculate the proper length for the longest time possible before we reach the 
unknown $v_c$ we first generate a set of geodesics at a comfortable position 
with $v_0>v_c$. Afterwards we run our code in reverse and subtract an 
increment as $v_{n+1}=v_n-\delta t$ until we hit $v_c$.

The approximation of the proper length is done via a simple Riemann sum. We use 
a trapezoidal rule and write the proper length as,
\begin{align}
L&=\int_{\sigma=\sigma_0}^{\sigma=\sigma_{M-1}}{\exd\sigma\mathscr{L}(A,B,S,V^{
\mu})}\nonumber\\
 &\approx 
\sum_{k=1}^{M-1}{(\sigma_{k}-\sigma_{k-1})\frac{1}{2}\left(\mathscr{L}(A_k,B_k,
S_k,V_k^{\mu})+\mathscr{L}(A_{k-1},B_{k-1},S_{k-1},V_{k-1}^{\mu})\right)}.\label
{eq:proplen}
\end{align}
We display the results of this process in the figures contained in 
the next sections. To display this data we calculate,
\begin{equation}\label{eq:<OO>}
 \langle O(t_2,x_2)O(t_1,x_1)\rangle =e^{-(L-L_{th})}\, .
 \end{equation}
The thermal value $L_{th}$ is calculated as the geodesic length in the equilibrium 
geometry. In practice this means we run our geometric evolution long enough so 
that the final data of geometry is in/near equilibrium. We then use the final 
geodesic length we compute as $L_{th}$. The geodesics diverge near the boundary so we employ a cutoff $z_{uv}=0.05$ when calculating the geodesics. All boundary conditions are imposed on the cutoff surface.   

\begin{figure}[h]
    \centering
    \includegraphics[width=4.0in]{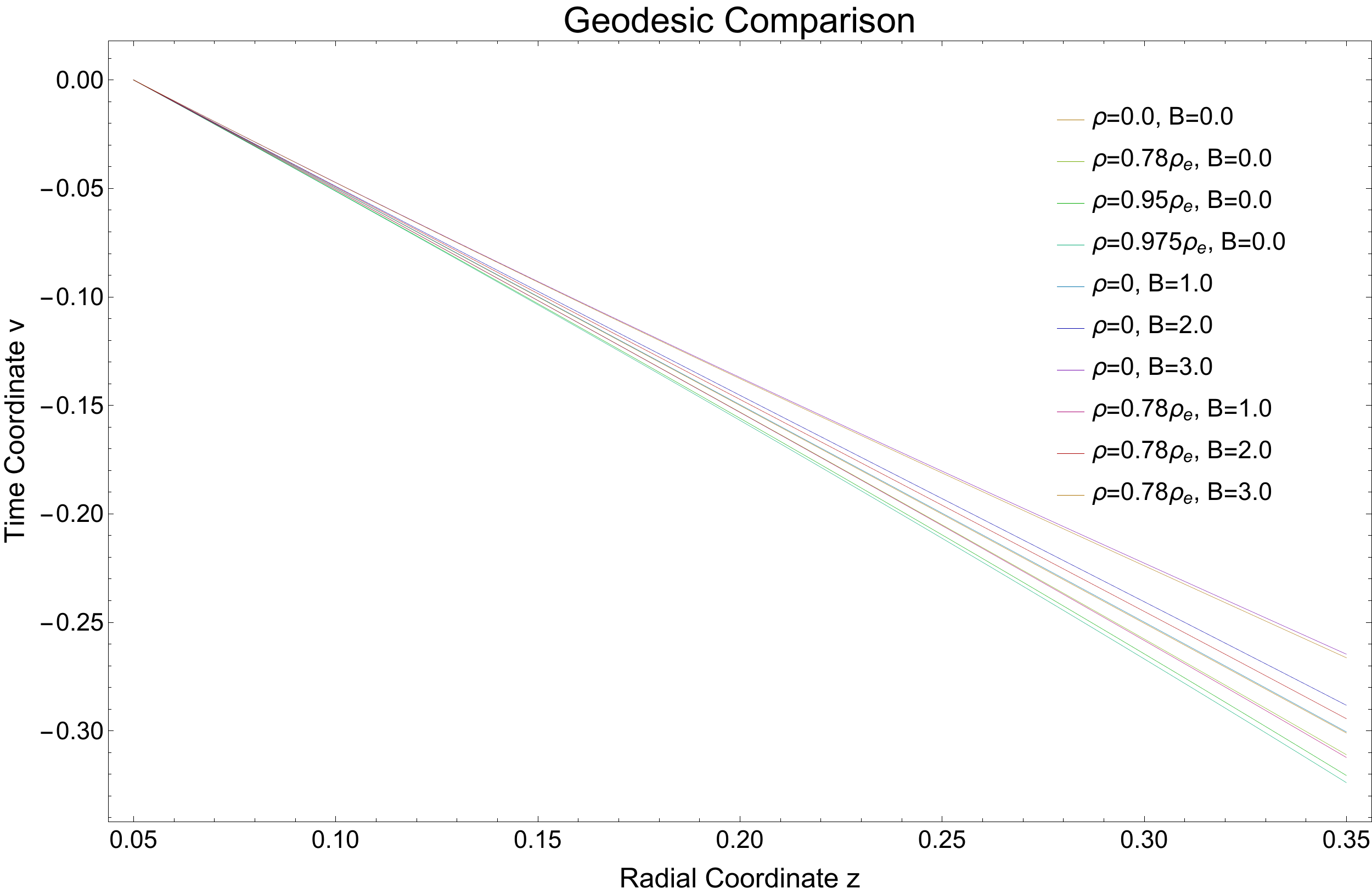}
    \caption{Comparison of the bulk time $v$ as a function of radial coordinate $z$ for the late time geodesics computed for all cases (a)-(d). All cases are plotted as $v(u)-v(0.05)$. Each case exhibits a different angle $\theta=\arctan{(v/z)}$ backwards through time into the bulk.
    \label{fig:RelativeAngle}}
\end{figure}
\section{Results: Correlators in thermalizing fluid \& thermalization times} \label{sec:results}
We recall that we call the {$x^3$}-direction longitudinal (aligned with the initial anisotropy and the magnetic field) and the $x$- and $y$-direction transverse (as these are transverse to both, initial anisotropy and magnetic field); one can think of the $y$-direction as aligned with the beam line in a heavy ion collision, see Fig.~\ref{fig:introFigureSetup}. Additionally, recall that we are computing 2-point functions of scalar operators with large operator dimension in this paper (via the geodesic approximation). Whenever we compare to 1-point functions, we compare to 1-point functions of the energy-momentum tensor, which has an operator dimension of 4. The reader should bear in mind that 4 is not necessarily large and that 2-tensor n-point functions can differ largely from scalar n-point functions.
\subsection{Case (a): Isotropization towards a Schwarzschild black brane}\label{sec:case(a)}
\begin{figure}[H]
\begin{subfigure}[b]{.5\linewidth}
 \begin{center}
  \includegraphics[width=2.9in]{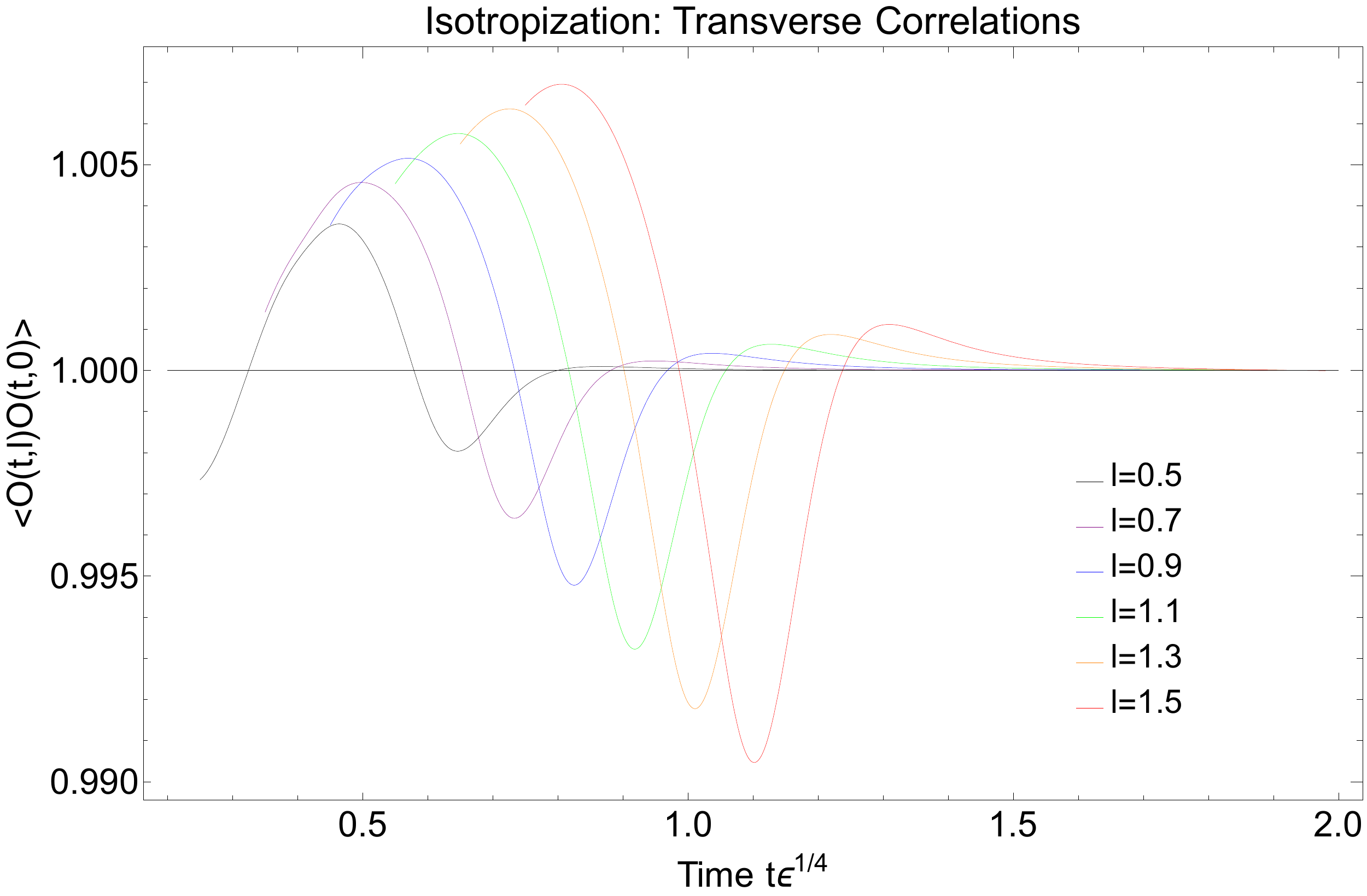}
 \end{center}
\end{subfigure}
\begin{subfigure}[b]{.5\linewidth}
 \begin{center}
  \includegraphics[width=2.9in]{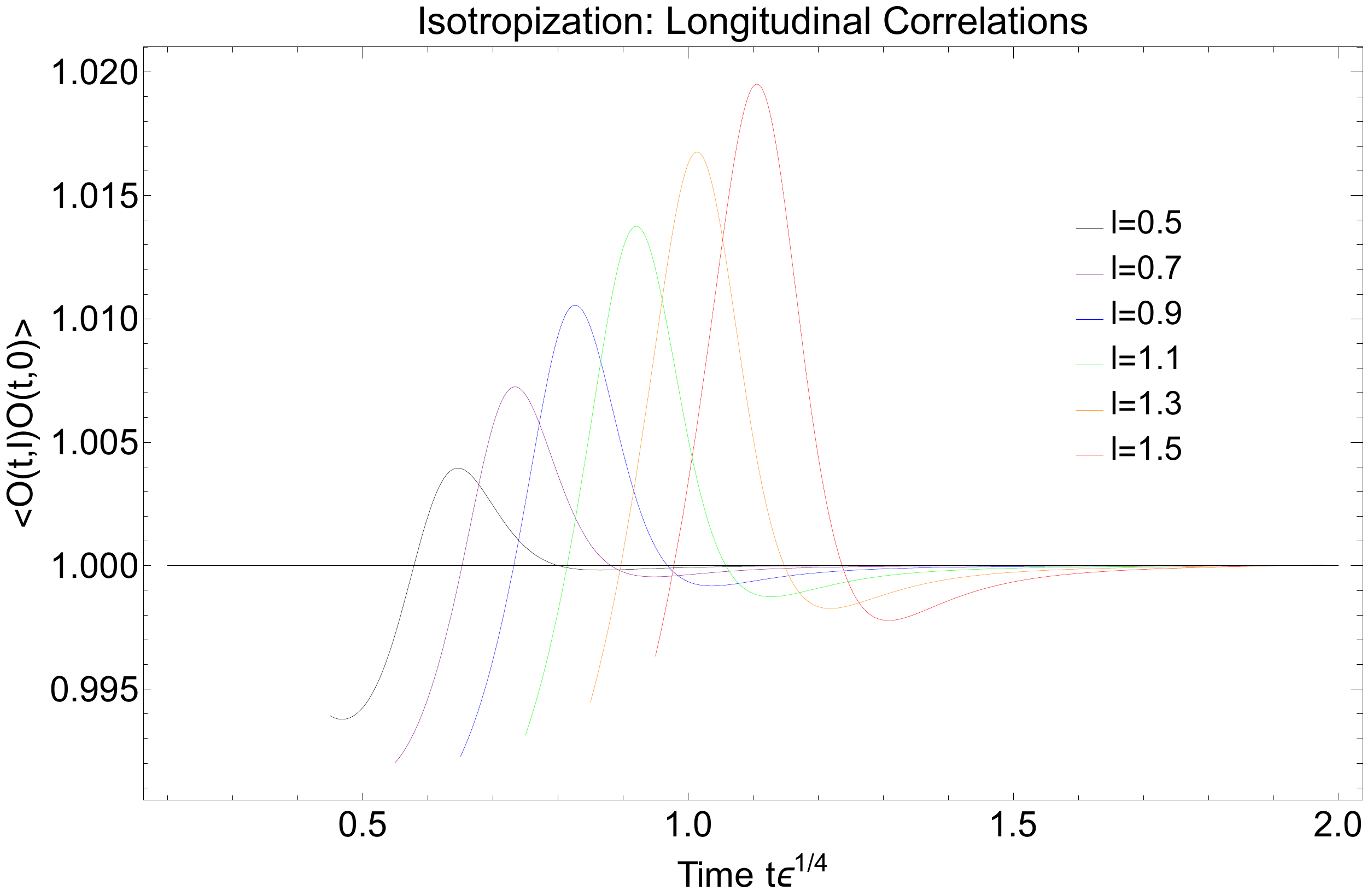}
 \end{center}
\end{subfigure}
\begin{subfigure}[b]{.5\linewidth}
\begin{center}
\includegraphics[width=2.9in]{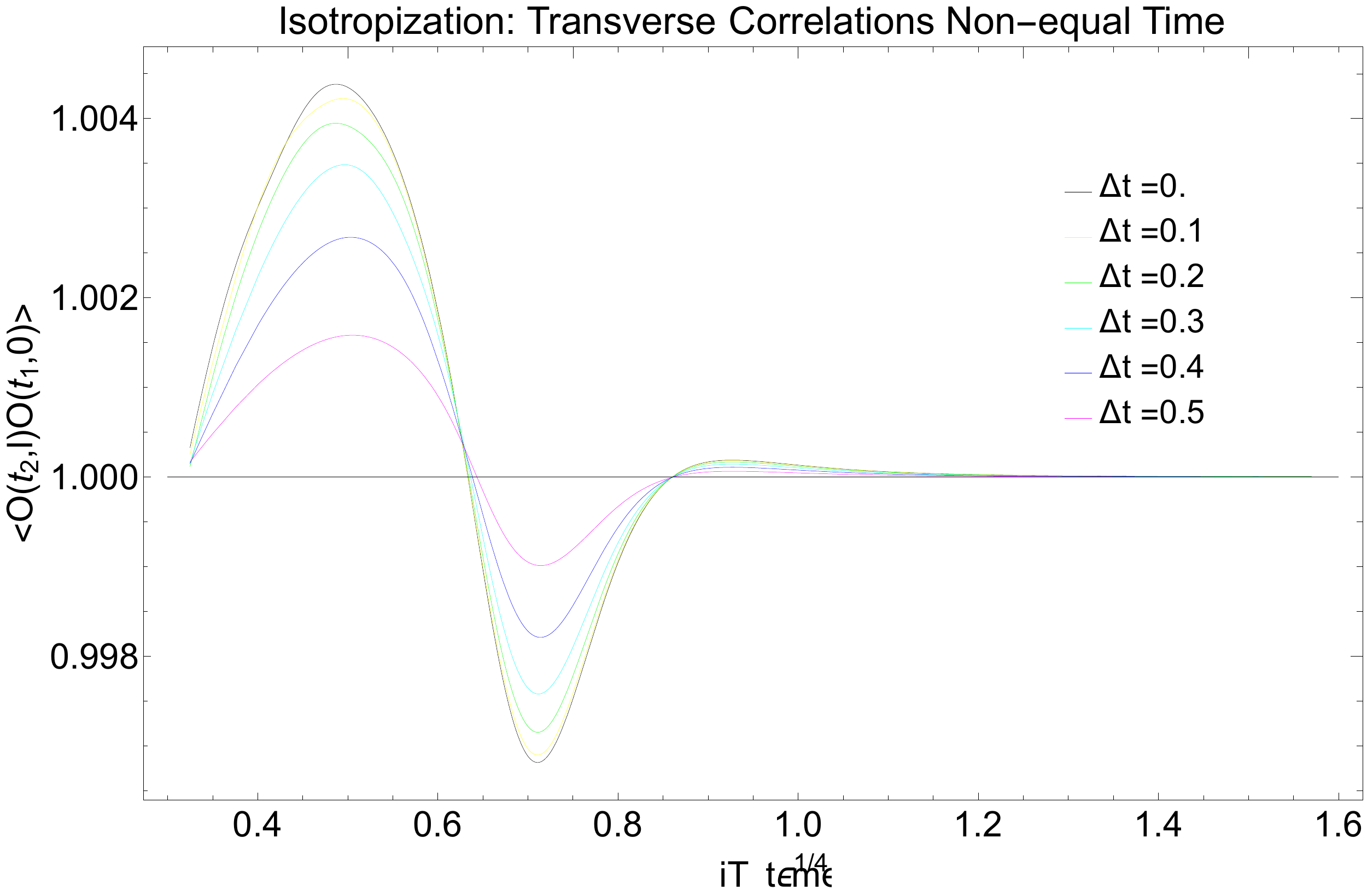}
\end{center}
\end{subfigure}
\begin{subfigure}[b]{.5\linewidth}
  \begin{center}   
\includegraphics[width=2.9in]{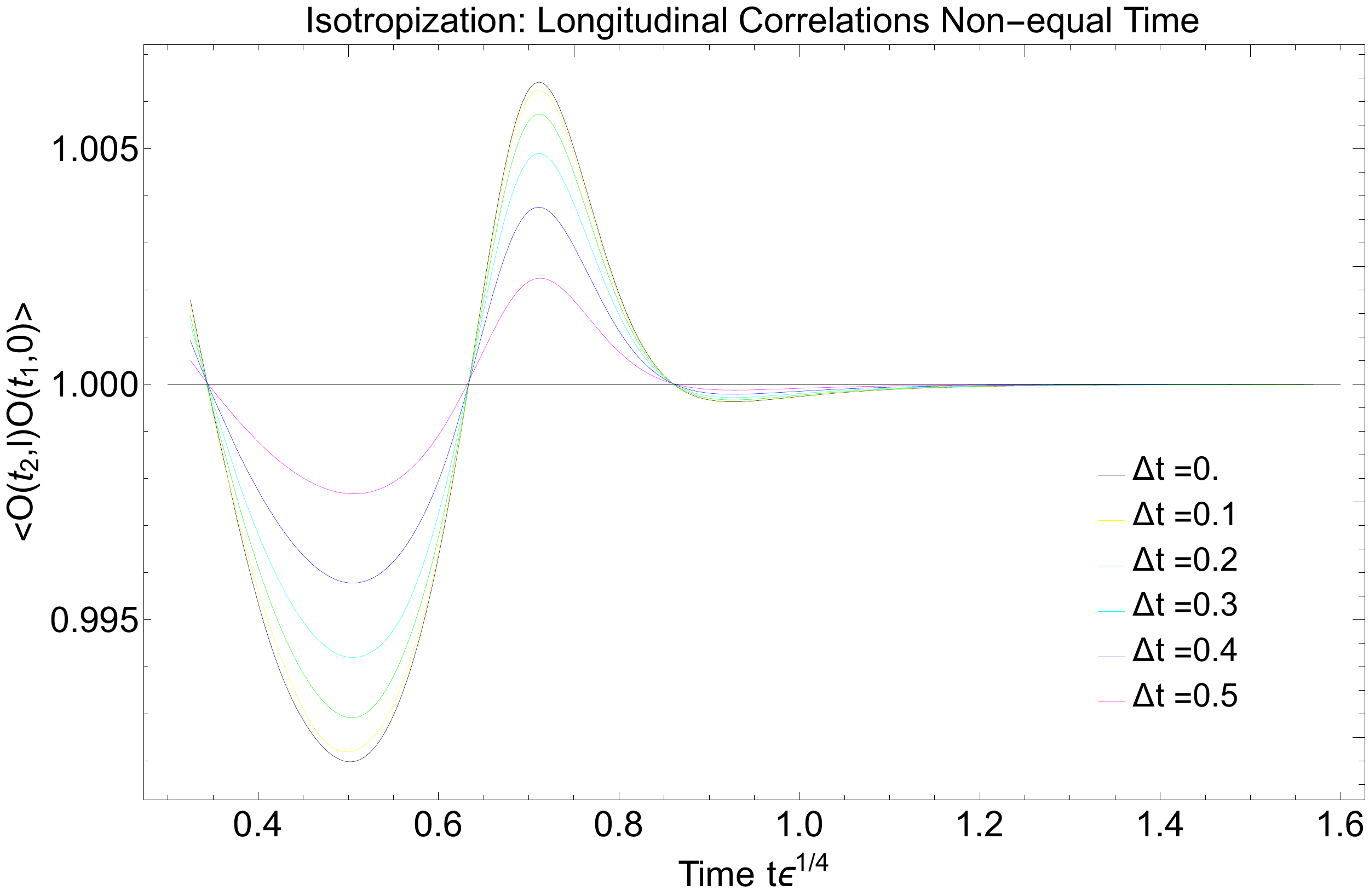}
  \end{center}
\end{subfigure}
\caption{\label{fig:BBUnchargedProbe}
2-point correlation functions of an (uncharged) scalar operator in a thermalizing fluid (dual to evolution towards a Schwarzschild black brane geometry), $\rho=0$, $\mathcal{B}=0$, case (a) in Fig.~\ref{fig:introFigureSetup}.
\textit{Top Left:} Lengths in the transverse direction at $\Delta t=0$. 
\textit{Top right:} Lengths in the longitudinal direction at $\Delta t=0$. 
\textit{Bottom left:} Transverse length separation by $l=0.65$ at various time separations $\Delta t$. 
\textit{Bottom Right:} Longitudinal length separation by $l=0.65$ at various times. }
\end{figure}
In Fig.~\ref{fig:BBUnchargedProbe} we display the results of our calculations of 2-point functions of two scalar operators in an initially anisotropic, uncharged thermal fluid (without magnetic field), which relaxes towards its isotropic equilibrium state\footnote{Similar results in this system have been previously reported by Ecker et.\ al~\cite{Ecker:2015kna}, though for different initial anisotropy parameter values. 
}, 

{i.e.}~case (a) (as sketched in Fig.~\ref{fig:introFigureSetup}).\footnote{This is dual to the computation of geodesics in the time-dependent (neutral) Schwarzschild black brane background.}  
In the top panels colors correspond to increasing length $l$ of the separation in the boundary theory in units of the energy density. In the bottom panels the colors denote the variation of $\Delta t$ separating the two operator insertions. The correlators are normalized to their late time (equilibrium) values. Hence, all curves asymptote to 1 (equilibrium) at late times.
We note in the top two graphics of Fig.~\ref{fig:BBUnchargedProbe} that the time it takes for the correlation to come to equilibrium is a function of the length of separation $l$ in both cases of longitudinal and transverse separation. 
Note the shift of the first visible peak to later times with increasing $l$, and the curves merging onto the value 1 at later times. 
Thus relaxation time for the 2-point functions increases with increasing length separations. As a working definition for thermalization time, we calculate the peak to peak amplitude, 
\begin{equation}
   A= \max(\Delta\langle\mathcal{O}(t,l)\mathcal{O}(t,0)\rangle)-\min(\Delta\langle\mathcal{O}(t,l)\mathcal{O}(t,0)\rangle),
\end{equation} 
of the correlator and take the time after which the correlator's deviation from equilibrium stays within a percentage of the peak to peak amplitude. That is, we calculate
\begin{equation}
\left|
\Delta\langle\mathcal{O}(t,l)\mathcal{O}(t,0)\rangle
-
\Delta\langle\mathcal{O}(t=\infty,l)\mathcal{O}(t=\infty,0)\rangle\right|\leq 0.01 A,\label{eq:thermal}
\end{equation}
where $\Delta\langle\mathcal{O}(t,l)\mathcal{O}(t,0)\rangle= \left(\langle\mathcal{O}(t,l)\mathcal{O}(t,0)\rangle_T - \langle\mathcal{O}(t,l)\mathcal{O}(t,0)\rangle_L\right)$,
and we define the 2-point function thermalization time, $t_{2pt.}$, {to be the time at which Eq.~\eqref{eq:thermal} is satisfied for all future times.}
This measure for the difference between a transverse 2-point function and its longitudinal counterpart is defined in analogy to the anisotropy measure for the 1-point functions, $\mathcal{P}_T-\mathcal{P}_L$, defined more carefully in Eq.~\eqref{eq:PAnisotropy}. We define the 1-point function thermalization time, $t_{1pt.}$, by replacing each 2-point function in Eq.~\eqref{eq:thermal} by a 1-point function. 
This measure~\eqref{eq:thermal} ``normalizes'' each thermalization time by taking into account the amplitude, $A$, of the corresponding correlation. This way, the single criterion~\eqref{eq:thermal} can be used for all of the 2-point and 1-point functions in the parameter region presented in this work. This yields thermalization times based on a unique criterion, which can therefore be meaningfully compared to each other. 
\footnote{We have also considered and discarded an alternative measure for thermalization time, details can be found in appendix~\ref{sec:altMeasure}.}

Furthermore, the curves in the top two graphics of Fig.~\ref{fig:BBUnchargedProbe} take on the value 1 at various times $t_{cor,n}$, which we will refer to as {\it equilibrium times in the correlators}. At these times $t_{cor,n}$, we observe $L-L_{th}=0$ according to Eq.~\eqref{eq:<OO>}, i.e. the geodesic length $L$ takes its equilibrium value $L_{th}$. This happens a certain time after the 1-point function goes through an equilibrium value at time $t_{eq, n}$. Also, we observe that for each equilibrium time $t_{eq, n}$ in the background there is one equilibrium time $t_{cor,n}$ in the correlator. There is an important distinction to be made here. These equilibrium times that we describe for the background field theory are strictly times at which the 1-point functions of the energy momentum tensor, $\mathcal{P}_T$ and $\mathcal{P}_L$ take on the values they would in equilibrium. However this does not mean that the background is in an equilibrium state. A quick analysis of the metric functions at these equilibrium times  show that the background is still highly dynamic and has not settled to an equilibrium configuration. As an example in the Schwarzschild isotropization although $\mathcal{P}_T=(1/2)(<T_{xx}>+<T_{yy}>)=0$, this does not mean $\partial_t\mathcal{P}_T$=0. We  discuss these equilibrium times of the background throughout the remainder of this work and as we do so please keep this distinction in mind.

\begin{figure}
\begin{center}
\includegraphics[width=3.5in]{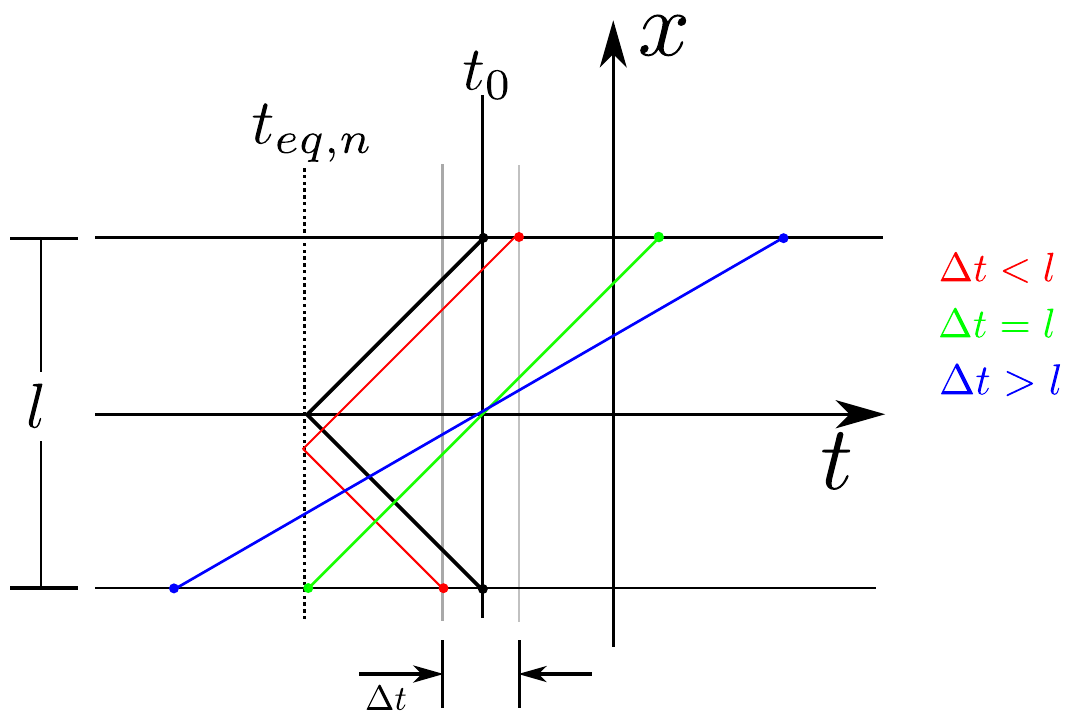}
\end{center}
\caption{\label{fig:causal}Here we display the causal structure behind the nodes displayed in the bottom two panels of Fig.~\ref{fig:BBUnchargedProbe}. The black wedge corresponds to the light cone which connects the points separated by a distance $l$. The red wedge does the same for points now also separated in time. Clearly they both touch the surface $t_{1pt.}$, a time for which the background passed though an equilibrium point. This indicates that for correlators separated in length and displaced in time around a central time $t$ will also take their thermal value.  }
\end{figure}

We find it to be interesting that the locations in time when $\Delta\mathcal{P}=\Delta\mathcal{P}_{eq}$ and $\partial_t\Delta\mathcal{P}\neq 0$ can be used to predict the times at which the 2-point functions take on thermal values. If we consider the causal structure of our field theory, as sketched in Fig.~\ref{fig:causal}, we can get some understanding of why it is instructive to study the times $t_{eq,n}$ in the backgrounds evolution. The diagram shows the horizontal field theory time axis labeled $t$, while the vertical axis labeled $x$ represents a spatial coordinate in the field theory~{($x,\,y$, or $x^3$)}.\footnote{From the gravity perspective, this figure is a sketch of the four-dimensional boundary of our five-dimensional gravity spacetime defined by $u=0$.} Consider the dotted line labeled $t_{eq,n}$ as a time at which the background 1-point functions pass through their equilibrium values, as seen {e.g.}~in Fig.~\ref{fig:ChargedUncharged}.\footnote{As mentioned in Sec.~\ref{sec:backgrounds} this generally happens at multiple times, so we keep the general label $n$ in Fig.~\ref{fig:causal}.} Let two operators be correlated at the point in the middle of that line at $(t=t_{eq,n},x=0)$. Let us, for the sake of the argument, assume that this local correlation is the only source for all future correlations of the two operators. The black wedge indicates the causal future of this point. Then, at the later time $t_0$, we measure the correlation between an operator located at $x=l/2$ and one at $x=-l/2$. These operators were last in causal contact at the point $(t=t_{eq,n},x=0)$, at which they both were immersed in an equilibrium background. After the two operators lose causal contact, their correlation appears to be simply propagated forward along the light-cone until it is measured at the time $t_0$. Hence, their correlator attains its equilibrium value at a time $t_{cor,n} = t_{eq,n}+l/2$. We can only check this in our data for 2 of the 4 equilibrium intersections of the 1-point functions, our analysis is summarized in table~\ref{tab:EQPureAnis}. If the operators take on an equilibrium value at exactly $t_{cor,n}=t_{eq,n}+l/2$ then we call this process \textit{free streaming}. That is the two operators we insert are not interacting with the medium, the correlation is not affected by interactions with the medium.
\begin{table}
\begin{center}
\begin{tabular}{|c|c|c|c|c|c|}
\multicolumn{6}{c}{Transverse separation }\\
\hline
$n$   & $t_{eq,n}$ & $l\epsilon^{1/4}$ & Predicted $t_{cor,n}$ & Numerical $t_{cor,n}$ & Relative percent error   \\
    \hline
\multirow{6}{*}{2 } &\multirow{6}{*}{0.411 }&0.5  & 0.661 & 0.579 &13.314 \\
  
   & &0.7& 0.761  & 0.653 & 15.342\\
   
   & & 0.9& 0.861 &0.733 &16.059 \\

   & & 1.1& 0.961 &0.817 & 16.233\\
   
   & &1.3& 1.061  &0.902 & 16.238\\
   
   & &1.5& 1.161 &0.986 &16.277 \\
   \hline
\multirow{6}{*}{3 } &\multirow{6}{*}{0.519 }&0.5  & 0.769 & 0.8 &4.016 \\
  
   & &0.7& 0.869  & 0.881 & 1.437\\
   
   & & 0.9& 0.969 & 0.969 &0.0282 \\

   & & 1.1& 1.069& 1.058 & 0.991\\
   
   & &1.3& 1.169  & 1.148 & 1.76\\
   
   & &1.5& 1.269 & 1.238 & 2.488 \\
   \hline
\end{tabular}
\end{center}
\caption{\label{tab:EQPureAnist}
Example equilibrium times of the 1-point functions, $t_{eq,n}$, and 2-point functions, $t_{cor,n}$, in a thermalizing fluid (dual to the evolution towards a Schwarzschild black brane geometry).}
\end{table}
\begin{table}
\begin{center}
\begin{tabular}{|c|c|c|c|c|c|}
\multicolumn{6}{c}{Longitudinal separation }\\
\hline
$n$   & $t_{eq,n}$ & $l\epsilon^{1/4}$& Predicted $t_{cor,n}$ & Numerical $t_{cor,n}$ & Relative percent error   \\
    \hline
\multirow{6}{*}{2 } &\multirow{6}{*}{0.411 }&0.5  & 0.661 & 0.578 &13.301 \\
  
   & &0.7&0.761  & 0.653 & 15.342\\
   
   & & 0.9&0.861 &0.732 &16.188 \\

   & & 1.1&0.961 &0.814 & 16.553\\
   
   & &1.3&1.061  &0.897 & 16.785\\
   
   & &1.5&1.161 &0.978 &17.063 \\
   \hline
\multirow{6}{*}{3 } &\multirow{6}{*}{0.519 }&0.5  & 0.769 & 0.8 & 4.026 \\
  
   & &0.7&0.869  & 0.881 & 1.43\\
   
   & & 0.9 & 0.969 & 0.968 &0.042 \\

   & & 1.1&1.069 & 1.0581 & 1.013\\
   
   & &1.3&1.169  & 1.1481 & 1.797\\
   
   & &1.5&1.269 & 1.237 & 2.553 \\
   \hline
\end{tabular}
\end{center}
\caption{\label{tab:EQPureAnis}
Example equilibrium times of the 1-point functions, $t_{eq,n}$, and 2-point functions, $t_{cor,n}$, in a thermalizing fluid (dual to the evolution towards a Schwarzschild black brane geometry).}
\end{table}

We find, at early times in the isotropization process, an average relative percent difference of $15.58\%$ between the expected numerical value and the predicted value. This deviation must stem from non-local correlations or rather from the interaction of the two operators with the strongly correlated fluid, which do not stem from their correlation at the point $(t=t_{eq,2},x=0)$. This gives a quantitative measure for the non-local correlations and/or interaction with the fluid early in the process of isotropization. This observation leads us to drop our previous assumption that the only source for correlations is the background equilibration point $(t=t_{eq,2},x=0)$. This is reminiscent of the behavior of geodesics observed in~\cite{AbajoArrastia:2010yt, Aparicio:2011zy}, where the authors consider the geodesic approximation applied to an infinitely thin infalling shell of null dust. 
They find that in order for a region of size $l$ to appear thermalized a time of at least $l/2$ must pass. 
At later times ($n=3$), we find the relative percent difference to be only $1.79\%$ in stark contrast with the early ($n=2$) time $15.58\%$ deviation mentioned above. Furthermore we find that the relative percent difference between the predicted and numerical values at later times appears to have a approximately parabolic relationship, i.e. there is a minimum at a certain length. This indicates there is an optimum length scale at which our hypothesis applies best. For $t_{eq,3}$ this length is approximately $l=0.9$ for both longitudinal and transverse separations (see table~\ref{tab:EQPureAnist} and table~\ref{tab:EQPureAnis}). In summary due to the small relative error in the prediction of the later equilibrium point correlations at this time can be explained by local correlations generated at the point $(t=t_{eq,3},x=0)$ and subsequent free streaming. 

We have also calculated non-equal time correlators shown in the bottom two graphics of Fig.~\ref{fig:BBUnchargedProbe}. 
These geodesics are centered around the same time which is used to compute the reference curve with $\Delta t=0$. We use as boundary conditions $t_i=t_0-\Delta t/2$ and $t_f=t_0+\Delta t/2$. The main effect of increasing the separation in time $\Delta t$ is to bring the 2-point function closer to its equilibrium value. As the time separation $\Delta t$ approaches the value $l$ (that is as the spacelike path connecting the points in boundary approaches a lightlike trajectory),  
we note that the deviation of the non-equilibrium correlation from the equilibrium correlation vanishes. 
From the gravity perspective, this vanishing deviation from equilibrium appears plausible. For $\Delta t=l$, the geodesic does not enter the AdS bulk anymore but remains on the boundary. In other words, for two boundary points which are separated by spatial length $l$ and by time $\Delta t=l$ the shortest connection between them is a straight line along the AdS boundary, and there is no ``shortcut'' through the bulk (like there was for $\Delta<l$). The geometry of the boundary is Minkowski at all times, so this geodesic has no information about the time-dependent AdS bulk geometry. Within the geodesic approximation the lightlike separated points on the AdS boundary can not display any out of equilibrium correlations. It is here, when the geodesic path lies only in the boundary that the geodesic approximation breaks down. 

The time separated correlations come to their thermal value in approximately the same time as the correlation with no time separation. We do note however that there appears to be a small decrease in the thermalization time as the time separation increases. The correlators with a time separation comparable to length separation appear to thermalize in the shortest amount of time. This behavior is displayed in both longitudinal and transverse correlations.

As a distinct feature in all non-equal time correlators with the same length separation $l$, we observe nodes, see the bottom two graphics in Fig.~\ref{fig:BBUnchargedProbe}. Those nodes are at times at which all the correlators take an identical equilibrium value, independent of the time separation $\Delta t$. This can again be understood by considering the causal structure of the field theory sketched in Fig.~\ref{fig:causal}. Previously, we considered the black causal wedge with its  origin at $(t=t_{eq,n},x=0)$. For that wedge $\Delta t=0$. Now consider the red wedge centered slightly below the black one in space, but still originating at the equilibrium time for the background $t_{eq}$. For that red wedge the time separation of the two operators as we measure them is $\Delta t$, because we measure one operator at $(t=t_0-\Delta t/2,x=-l/2)$, and the other one at $(t=t_0+\Delta t/2,x=l/2)$. So, again, the last point of causal contact between these two operators when we measure them has been at time $t_{eq}$ at which they were immersed in an equilibrium background. Hence, the correlation again takes on its equilibrium value for the family of points parameterized by $\Delta t\ge 0$. 
Deviations from this prediction should be attributed to non-local correlations, or interactions with the fluid. 
If we consider the nodes shown in the bottom left panel of Fig~\ref{fig:BBUnchargedProbe} we calculate their locations to be $t=0.736$ and $t=0.843$. Checking our hypothesis that the correlator's equilibrium point occurs at $t_{cor,n}=t_{eq,n}+l/2$ we find a relative percent error of $14.41\%$ and $2\%$ for the two nodes. We have stated that each of the correlation functions for time separated points should take on an equilibrium value at the same point. We can quantify this statement by calculating the standard deviation of times at which this node occurs for each of the geodesics at different time separations from our numerical data. We find a standard deviation of $\sigma=0.00396$ for $t_{eq,2}$ and $\sigma=0.00058$ for $t_{eq,3}$. We conclude that our data confirms that the correlation functions take on the same equilibrium value for the family of points parameterized by $\Delta t\ge 0$. It is interesting that although the first node appears to experience a larger degree of non-local correlations or interaction effects with the fluid, the equilibrium point is shared by the whole family of correlations parameterized by distinct values of $\Delta t\ge 0$.

The reasoning of the previous paragraph applies to spacelike separated operators ($\Delta t<l$). Increasing $\Delta t$, we shift the causal wedge in Fig.~\ref{fig:causal} further down until we reach the case of lightlike separated operators $\Delta t =l$. For larger timelike time separations ($\Delta t>l$) the two operators are causally connected over the whole interval $\Delta t$, so we would expect a qualitative change in the correlation functions.
Naively applying to this case the geodesic approximation as described in Sec.~\ref{sec:geodesics} leads to complex-valued correlators, which we chose not to display here. A phase factor discrepancy was found between correlators computed from a conformal field theory and from the geodesic approximation for timelike separated operators~\cite{Aparicio:2011zy}. It has yet to be seen if a (modified) geodesic approximation can correctly reproduce this phase factor. 

\subsection{Case (b): Isotropization towards Reissner-Nordstr\"om black brane} \label{sec:RNBlackBraneBG}
\begin{figure}[H]
\begin{subfigure}[b]{.5\linewidth}
 \begin{center}
  \includegraphics[width=2.9in]{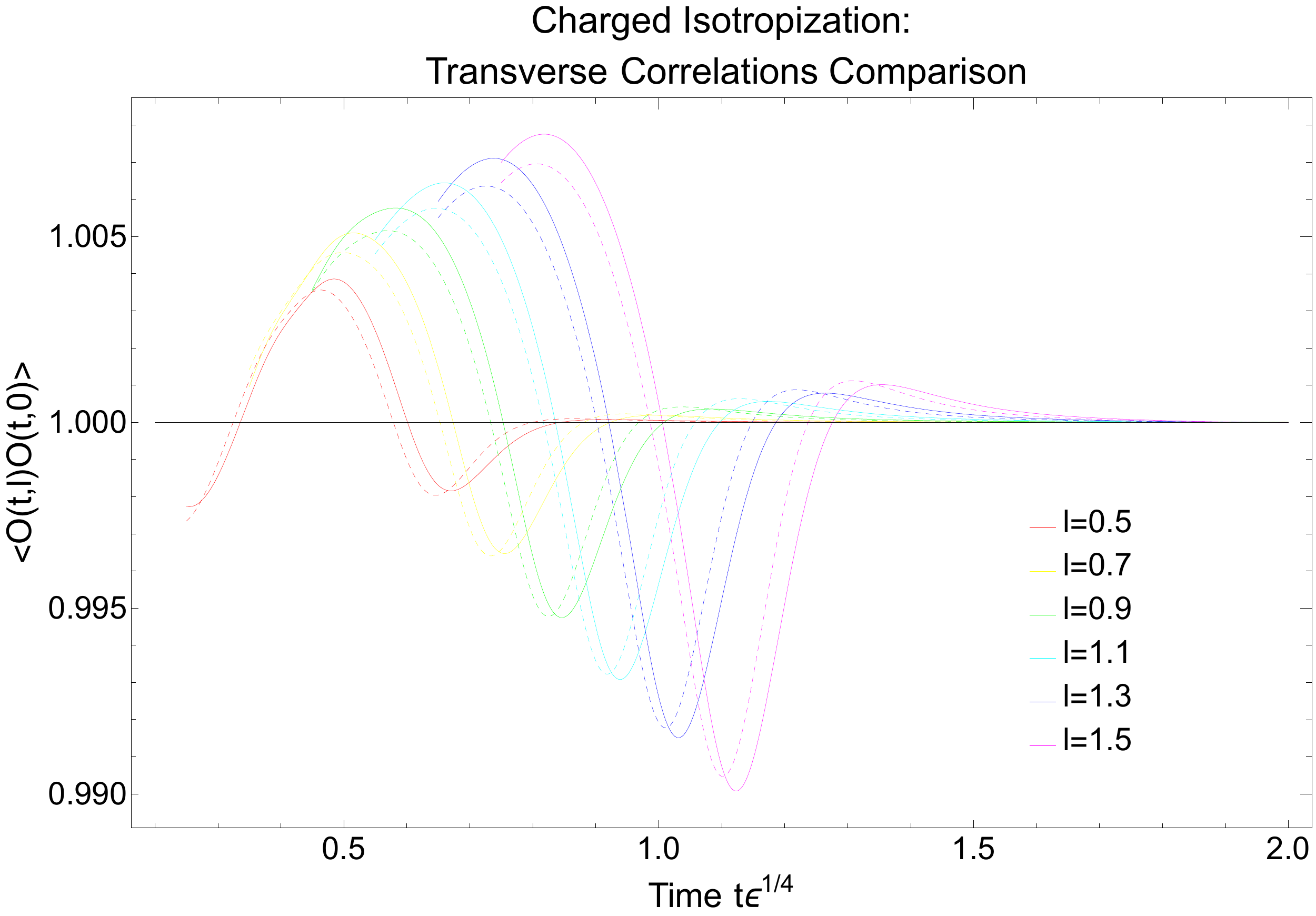}
 \end{center}
\end{subfigure}
\begin{subfigure}[b]{.5\linewidth}
 \begin{center}
  \includegraphics[width=2.9in]{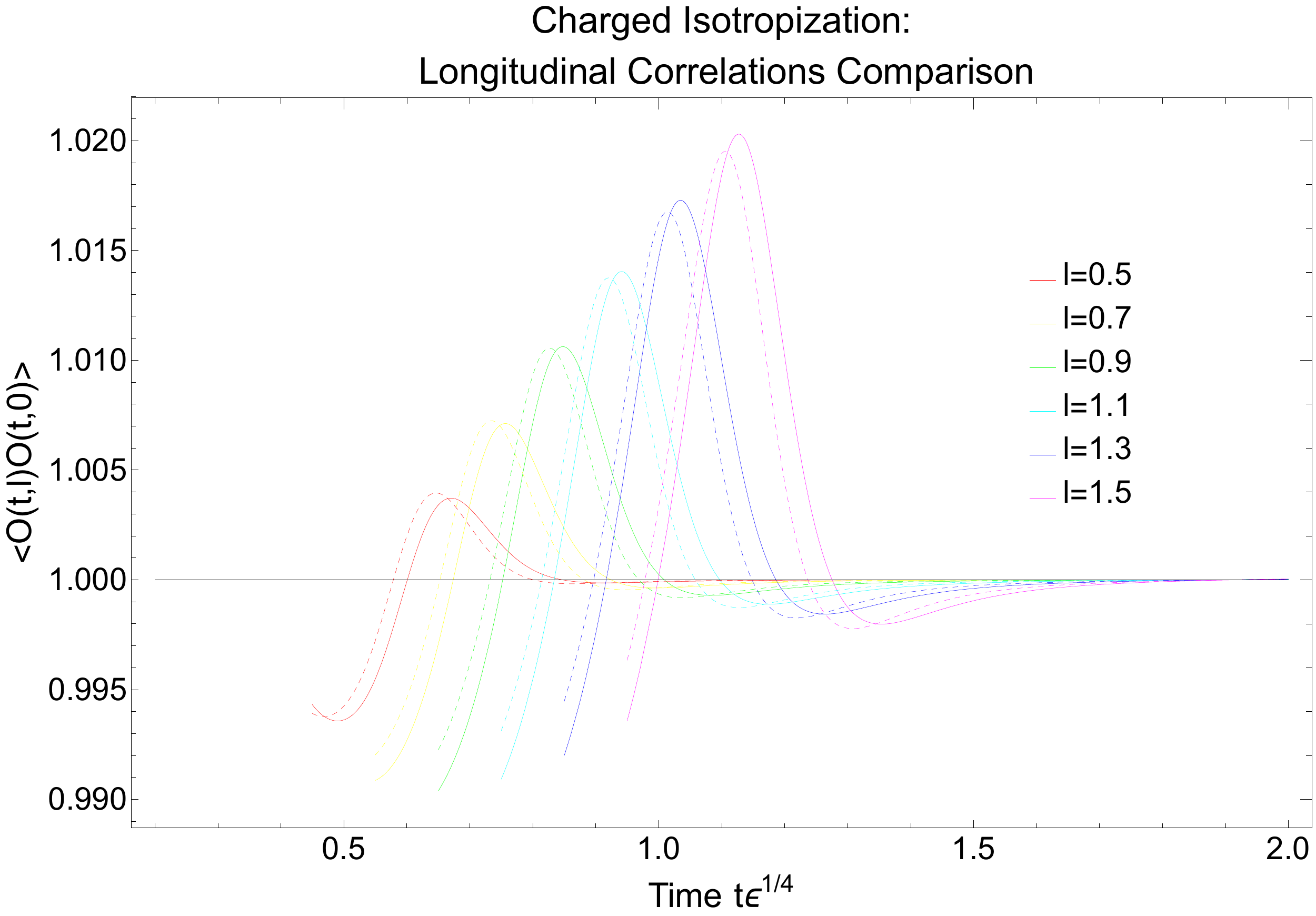}
 \end{center}
\end{subfigure}
\begin{subfigure}[b]{.5\linewidth}
\begin{center}
\includegraphics[width=2.9in]{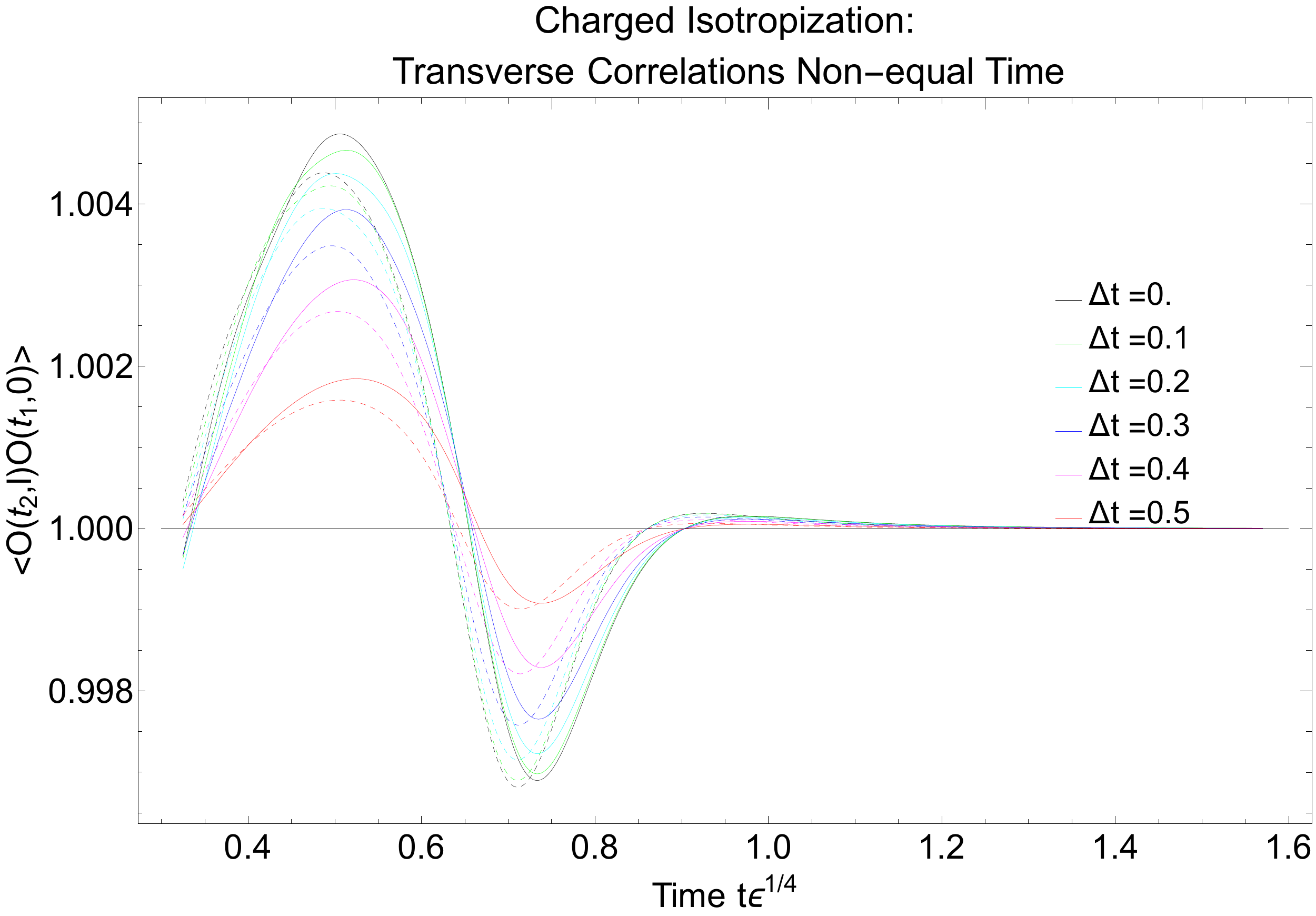}
\end{center}
\end{subfigure}
\begin{subfigure}[b]{.5\linewidth}
  \begin{center}   
\includegraphics[width=2.9in]{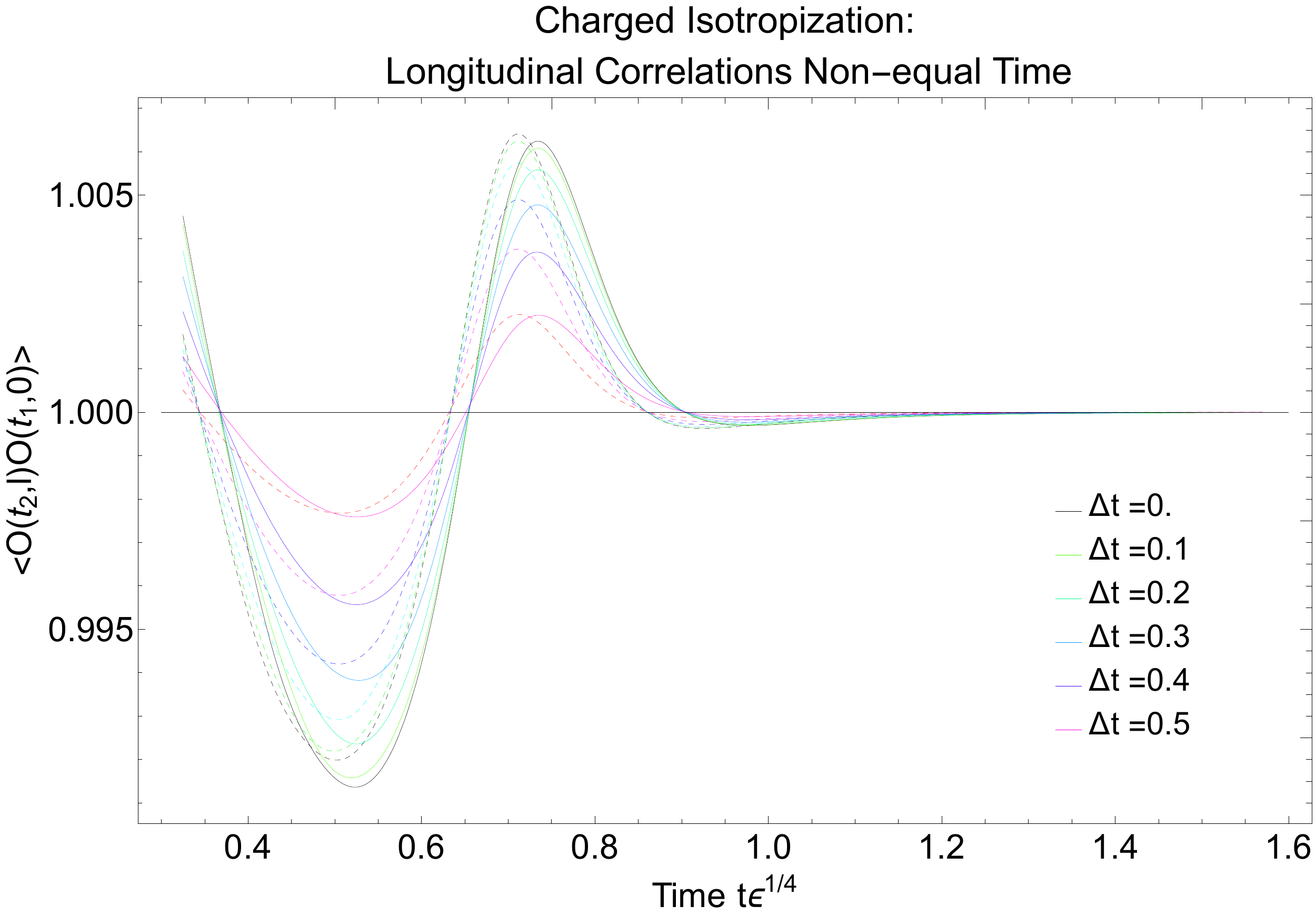}
  \end{center}
\end{subfigure}
\caption{\label{fig:RNUnchargedprobe}
2-point correlation functions for (uncharged) scalar operator in a charged thermalizing fluid (RN black brane geometry), $\rho=0.78\rho_{e}, \, \mathcal{B}=0$, case (b) in Fig.~\ref{fig:introFigureSetup}.  
\textit{Top Left:} Various length separations in the transverse direction at $\Delta t=0$. 
\textit{Top right:} Various length separations in the longitudinal direction at $\Delta t=0$. 
\textit{Bottom left:} Various time separations $\Delta t$, fixed transverse length separation $l=0.65$.
\textit{Bottom Right:} Various time separations $\Delta t$, fixed longitudinal length separation $\Delta l=0.65$. 
In all plots dashed lines are calculated in the neutral anisotropic case (a) shown for comparison, and solid lines correspond to the charged anisotropic case (b).}
\end{figure}
In Fig.~\ref{fig:RNUnchargedprobe} we display the results of our calculation for the 2-point functions in an initially anisotropic charged thermal fluid (without a magnetic field), which relaxes towards its isotropic equilibrium state, {i.e.}~case (b) (as sketched in Fig.~\ref{fig:introFigureSetup}).\footnote{This is dual to the computation of geodesics in the time-dependent charged black brane background.}
In the top panels colors correspond to increasing length of the separation in the boundary theory in units of the energy density. In the bottom panels the colors denote the variation of $\Delta t$ separating the two operator insertions. 
The overall observation is that the presence of a considerable amount of charge changes the geodesics, and hence the correlators, only very little compared to the uncharged case. This can be seen comparing the dashed (uncharged) to the solid (charged) curves in Fig.~\ref{fig:RNUnchargedprobe}, which have been generated for a background with $78\%$ of the extremal charge density, $\rho_{e}$, of the thermalized RN black brane. 
\begin{figure}
    \begin{subfigure}[b]{.5\linewidth}
    \includegraphics[width=2.9in]{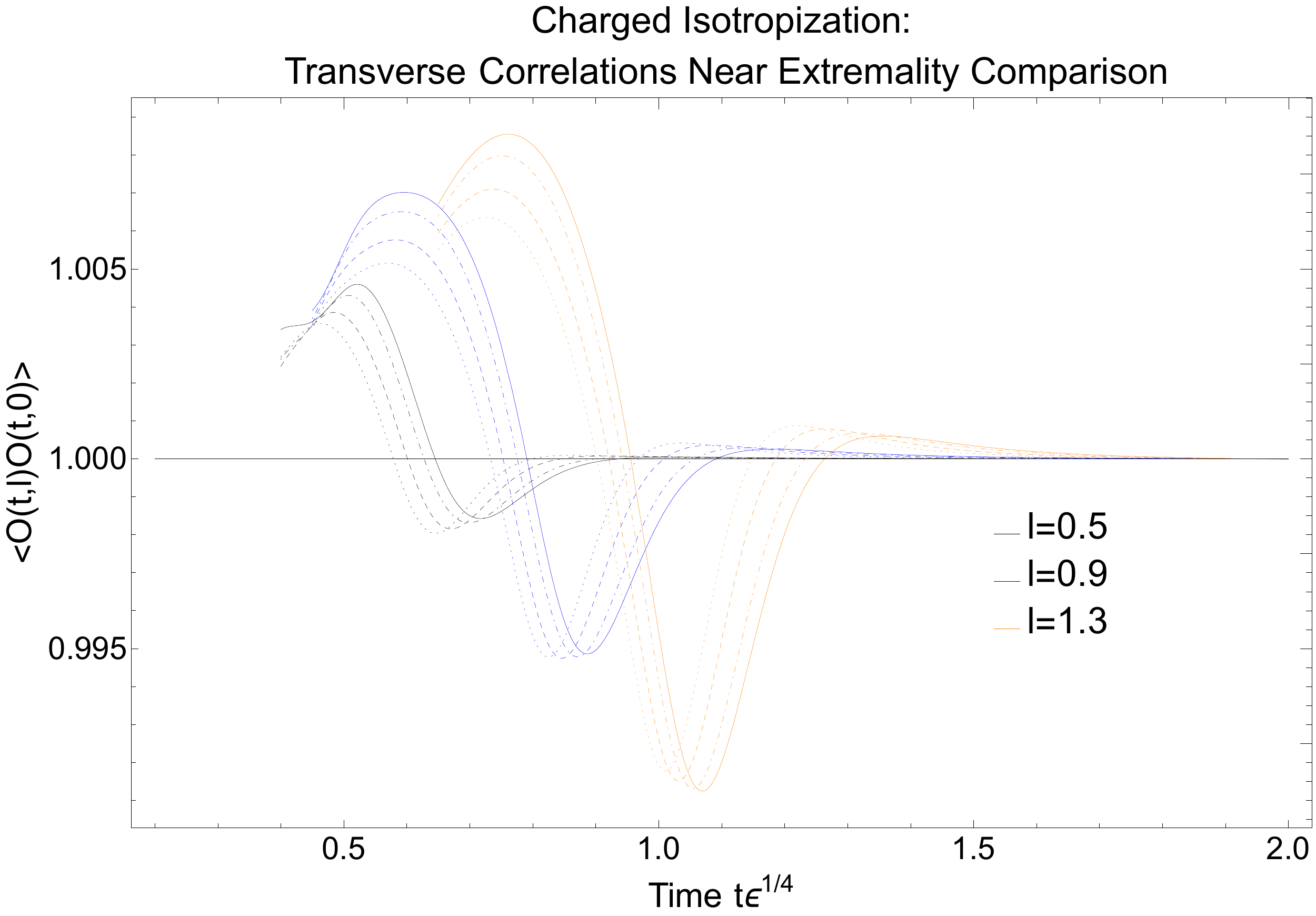}
    \end{subfigure}
     \begin{subfigure}[b]{.5\linewidth}
    \includegraphics[width=2.9in]{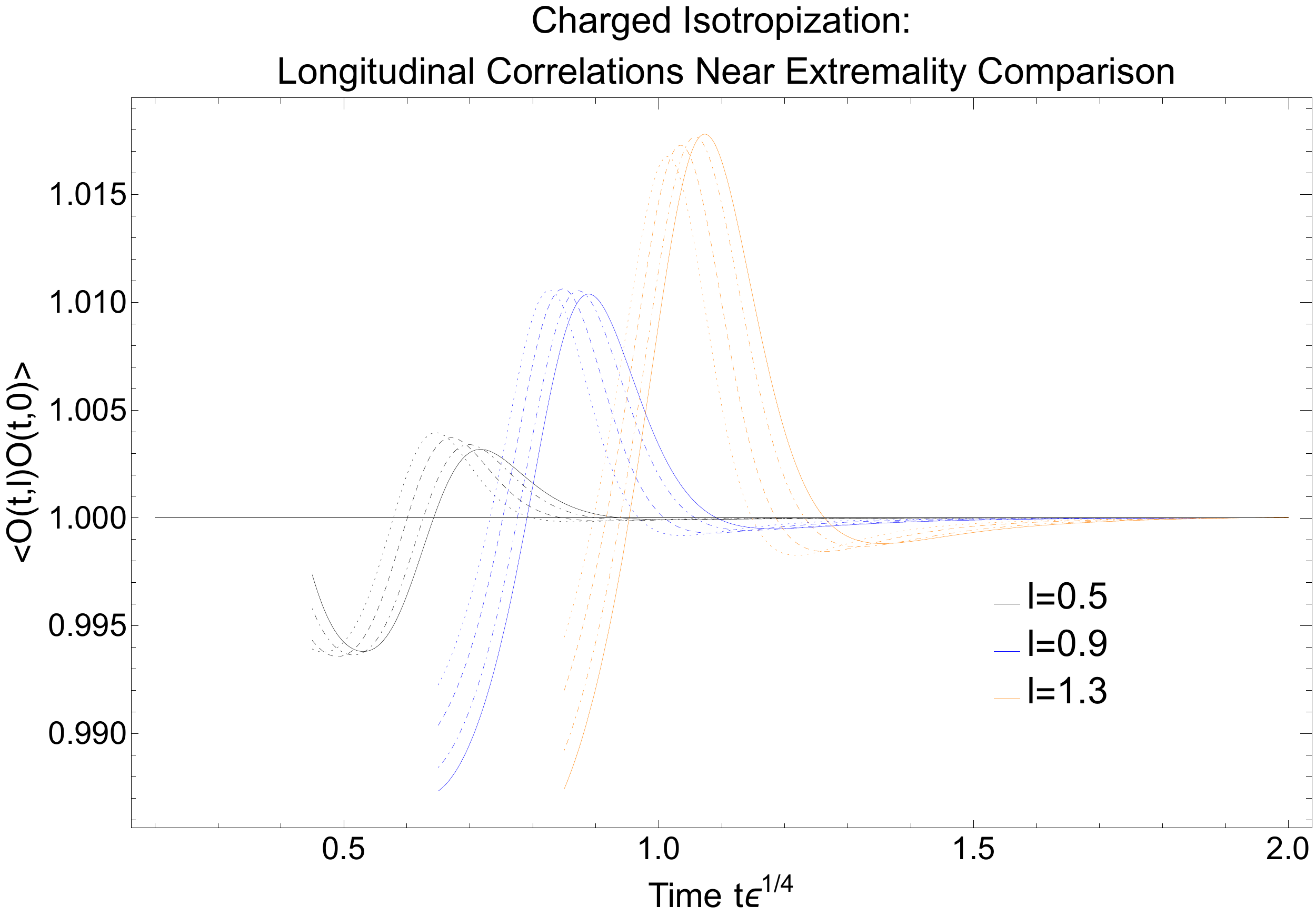}
    \end{subfigure}
    \caption{2-point correlation functions calculated for a scalar operator in a charged thermalzating fluid (RN black brane geometry).  \textit{Left:} Various length separations in the transverse direction. \textit{Right:} Various length separations in the longitudinal direction. In all plots dotted lines represent $\rho=0$, dashed lines represent $\rho=0.78\rho_e$, dot-dashed lines represent $\rho=0.95\rho_{e}$ and solid lines represent $\rho=0.975\rho_{e}$.
    \label{fig:nearExtremalLS}}
\end{figure}
It is interesting to note that for the 1-point correlation functions displayed in Fig.~\ref{fig:Everything} the main difference was a decrease in amplitude of the pressure anisotropy. We note the case of 2-point functions, see~Fig.~\ref{fig:RNUnchargedprobe}, we also see a small change in the amplitude of the 2-point functions. As said in Sec.~\ref{sec:case(a)} our two point functions are normalized to the late time equilibrium values. We can therefore say the larger the deviation from horizontal line at one represents a larger departure from equilibrium.
We note that the amplitude of the correlation function is increased for two points transversely separated. 
The longitudinally separated correlation functions display slightly different behavior. We note that at short distance scales the amplitude of the response is smaller than the corresponding response of the uncharged fluid. However as we increase the length scale the response of the correlation function has an amplitude larger than the corresponding two point function of the uncharged fluid. This indicates further from equilibrium behavior. This is interesting since we may expect that at short distance scales the 2-point functions should more closely resemble the 1-point functions. In the case of longitudinal separation at short distance scales we behavior similar to the 1-point function, a decrease in amplitude. We also note that the increase in amplitude to be smaller than the amplitude increases found for the correlation functions in the transverse direction indicating larger departures from equilibrium in the transverse direction. 

Additionally we find a shift in the response, {e.g.}~peaks and other features are shifted to later times for all lengths (relative to the energy density scale). In particular, the thermalization time, $t_{2pt.}$ after which the 2-point function satisfies Eq.~\eqref{eq:thermal} occurs slightly later, see table~\ref{tab:thermalizationTimes}. The 2-point correlation of a scalar probe of this charged SYM fluid takes a slightly longer time to thermalize than it does in the previous case of an uncharged fluid. Already the authors of ~\cite{Fuini:2015hba} noted that the difference between the 1-point functions found for the charged versus the neutral SYM fluid is surprisingly small, with little effect on the thermalization time. Here we confirm that observation, and add that although there is a difference in the 2-point correlations between charged and neutral fluids, that difference is also small.

The addition of charge has a dramatically small effect on the correlations compared to what one could have naively expected. Still, near extremal charge densities might lead to departures from this behavior. When fixing the charge density to $\rho=0.95\rho_{e}$, as displayed in Fig.~\ref{fig:nearExtremalLS}, we find little qualitative difference in the behavior of the correlations. We do find that as we increase the value of the charge density the features of the plot e.\ g.\ peaks and valleys shift in the positive $t\epsilon^{1/4}$ direction. When increasing from $\rho=0$ (dotted line) to $\rho=0.78\rho_e$ (dashed line) Fig.~\ref{fig:nearExtremalLS} shows a shift of about $\delta t\epsilon^{1/4}\approx 0.03$. As we increase to $\rho=0.95\rho_e$ we again see a shift of $\delta t\epsilon^{1/4}\approx 0.03$. This trend is continued during the increase to $\rho=0.975\rho_e$. Despite the increasingly smaller increases to the charge density the shift of the features of the 2-point function is approximately the same. It appears that in the approach to extremality the response time of the 2-point function is shifted to infinity. The 1-point function thermalization times appear to follow the same trend. Meanwhile, the 1-point functions decrease in amplitude whereas the 2-point functions increase as we approach extremality.

Asides from the curious behavior of the response time the overall trend seen in Fig.~\ref{fig:RNUnchargedprobe} is continued in Fig.~\ref{fig:nearExtremalLS}. The correlation functions of two points separated in the transverse direction see positive peak amplitude increases. The longer length scales see increasingly larger positive peak amplitudes. However small length scales in the longitudinal direction see increased amplitudes of the 2-point function as compared to the case of $\rho=0.78\rho_e$ where we find a decrease at short lengths. The longer length scales in the longitudinal case again see larger positive peak amplitudes as compared to the correlation functions of the uncharged fluid.
At all length scales we probed we find an increase in the thermalization time. One may suspect that varying the amplitude of the initial anisotropy pulse could increase or decrease the relative effect of the charge on the 2-point functions. A comparison of our data at initial pulse amplitudes of $\beta=1.5$ versus $0.15$ does indicate only minor quantitative changes. 

In the non-equal time correlators, displayed in the bottom two graphs of Fig.~\ref{fig:RNUnchargedprobe}, we again observe the same type of nodes which we found in the neutral case. The effect of the charge is to shift these nodes to later times, and --somewhat surprisingly--  off the equilibrium line. Previously, in case (a), we stated the hypothesis that these nodes are the result of the operators being causally disconnected after feeling an equilibrium background. The earlier node being shifted considerably far away from the equilibrium value seems to make that explanation mute. Despite that, we repeat the analysis performed in Sec.~\ref{sec:case(a)} and display the results in table~\ref{tab:EQRNt} and table~\ref{tab:EQRN}. As is the case in the uncharged fluid we find for the later time equilibrium point the relative difference between the predicted and numerical value to follow a near parabolic relationship. Again this indicates there is an optimum length scale at which the hypothesis of free streaming works best. In both the transverse and longitudinal separations this length is approximately $l=1.1$ (see table~\ref{tab:EQRNt} and table~\ref{tab:EQRN}). Interestingly this length scale is longer then what was found in the uncharged case in Sec.~\ref{sec:case(a)} where a free streaming hypothesis works best near $l=0.9$.

Additionally we also perform a fit to find that our numerical data for the location of the equilibrium time in the 2-point function as a function of the length satisfies the following formulas. For the second node we find
\begin{equation}
  t_{eq,2}(l)= 0.421+0.343 l+0.0285 l^2,\label{eq:RN_Fit_node_2}
\end{equation}
with an adjusted R squared value of $0.999997$. For the third node we find our data is fitted by,
\begin{equation}
  t_{eq,3}(l)= 0.654+0.365 l+0.033 l^2,\label{eq:RN_Fit_node_3}
\end{equation}
with an adjusted R squared value of $0.999996$. As is the case with neutral thermalizing fluid the best fit to data describing the location of the nodes is a quadratic one. This provides further evidence of a deviation from our hypothesis of free streaming. We do note that the fits given by Eq.~\eqref{eq:RN_Fit_node_2} and Eq.~\eqref{eq:RN_Fit_node_3} do bear some resemblance to the work of~\cite{Caceres:2012em,Galante:2012pv}. The authors of~\cite{Caceres:2012em} investigate a Vaidya metric for the AdS-RN background, i.e. an infalling mass-shell, in various dimensions with non-local probes including geodesics. 
There, the authors also find a quadratic relationship for the dependence of their thermalization measure 
on the length separation. Here we do not have an infalling shell but we do have a pulse of anisotropy bouncing back and forth between AdS-boundary and the apparent horizon. As a result our 2-point functions pass through equilibrium multiple times on their path to thermalization. 

\begin{table}
\begin{center}
\begin{tabular}{|c|c|c|c|c|c|}
\multicolumn{6}{c}{Transverse separation }\\
\hline
$n$   & $t_{eq,n}$ & $l\epsilon^{1/4}$& Predicted $t_{cor,n}$ & Numerical $t_{cor,n}$ & Relative percent error   \\
    \hline
\multirow{6}{*}{2 } &\multirow{6}{*}{0.4281 }&0.5  & 0.677 & 0.601 &11.937\\
  
   & & 0.7&0.777  & 0.673 & 14.318\\
   
   & & 0.9&0.877 &0.753 &15.26 \\

 & &1.1&0.977 &0.836 &15.586 \\
   
   & &1.3 & 1.077  &0.921 & 15.671\\
   
   & &1.5& 1.177 &1.000 &15.706 \\
   \hline
\multirow{6}{*}{3 } &\multirow{6}{*}{0.5461}&0.5  & 0.795 & 0.846 &6.204 \\
  
   & &0.7&0.895  & 0.924 &3.184 \\
   
   & & 0.9&0.995 & 1.009 & 1.384 \\

   & & 1.1&1.095 & 1.097 &0.191\\
   
   & &1.3&1.195  & 1.187 &0.711 \\
   
   & &1.5&1.295 1 & 1.276 & 1.482 \\
   \hline
\end{tabular}
\end{center}
\caption{\label{tab:EQRNt}
Example equilibrium times of the 1-point functions, $t_{eq,n}$, and 2-point functions, $t_{cor,n}$, in a thermalizing fluid (dual to the evolution towards a RN black brane geometry).}
\end{table}

\begin{table}
\begin{center}
\begin{tabular}{|c|c|c|c|c|c|}
\multicolumn{6}{c}{Longitudinal separation }\\
\hline
$n$   & $t_{eq,n}$ & $l\epsilon^{1/4}$ & Predicted $t_{cor,n}$ & Numerical $t_{cor,n}$ & Relative percent error   \\
    \hline
\multirow{6}{*}{2 } &\multirow{6}{*}{0.4281 } &0.5 & 0.677 & 0.601 & 11.97 \\
  
   & &0.7 &0.777  & 0.674 & 14.286\\
   
   & & 0.9&0.877 &0.752 &15.329 \\

   & & 1.1&0.977 &0.834 & 15.818\\
   
   & &1.3&1.077  &0.917 & 16.1055\\
   
   & &1.5&1.177 &0.999 &16.366 \\
   \hline
\multirow{6}{*}{3 } &\multirow{6}{*}{0.5461 } &0.5 & 0.795 & 0.846 & 6.237 \\
  
   & &0.7&0.895  & 0.924 & 3.183\\
   
   & & 0.9&0.995 & 1.009 &1.372 \\

   & & 1.1&1.095 & 1.097 & 0.171\\
   
   & &1.3&1.195  & 1.186 & 0.744\\
   
   & &1.5&1.295 & 1.275 & 1.534 \\
   \hline
\end{tabular}
\end{center}
\caption{\label{tab:EQRN}
Example equilibrium times of the 1-point functions, $t_{eq,n}$, and 2-point functions, $t_{cor,n}$, in a thermalizing fluid (dual to the evolution towards a RN black brane geometry).}
\end{table}

In principle, we can also probe the charged fluid with charged operators and calculate their correlation functions. The geodesic approximation can be extended in order to relate such 2-point functions of a charged scalar operator to the geodesic of a charged probe particle in our asymptotically $AdS$ spacetime as done in~\cite{Giordano:2014kya}. We consider some further details in appendix~\ref{sec:chargedGeodesics} and in Sec.~\ref{sec:magneticBlackBraneResults} where we compute the geodesics of charged particles in a isotropizing magnetic black brane background. However, for the RN black brane, the charge density in the field theory is uniformly distributed over the fluid. In other words, in a uniformly charged fluid every direction looks the same to a charged probe particle. Hence, the geodesic of a charged probe particle in the dual gravitational theory is not affected electromagnetically as there is no potential gradient arising from the electromagnetic sector of the theory. In order to see an effect on charged operator 2-point functions, we would have to introduce charge gradients or electromagnetic fields in the field theory fluid.

\subsection{Case (c): Isotropization towards a magnetic black brane}

\label{sec:magneticBlackBraneResults}
 \begin{figure}[H]
  \begin{subfigure}[b]{.5\linewidth}
  \begin{center}   
\includegraphics[width=2.9in]{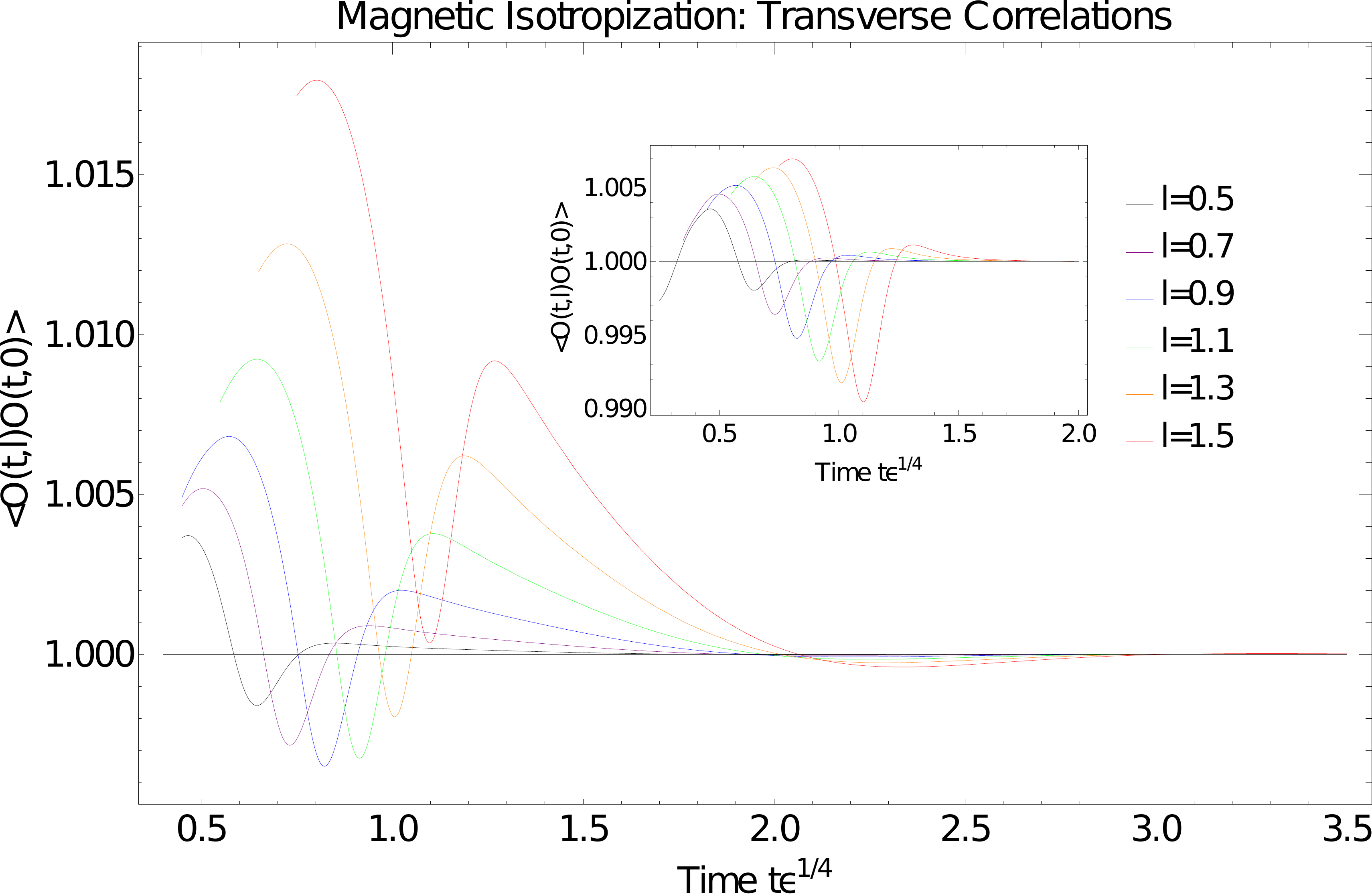}
  \end{center}
\end{subfigure}
 \begin{subfigure}[b]{.5\linewidth}
  \begin{center}   
\includegraphics[width=2.9in]{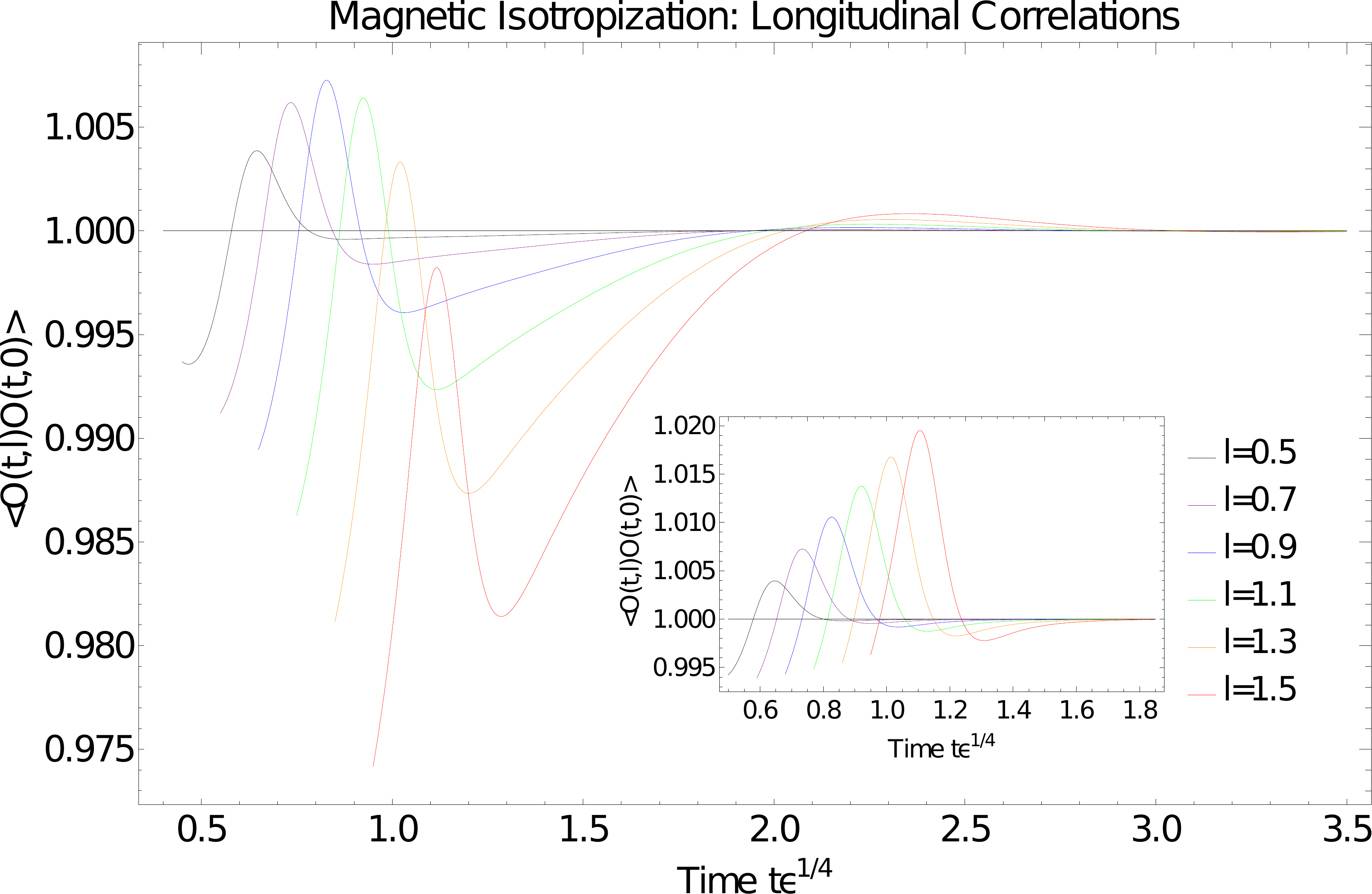}
  \end{center}
\end{subfigure}
 \begin{subfigure}[b]{.5\linewidth}
  \begin{center}   
\includegraphics[width=2.9in]{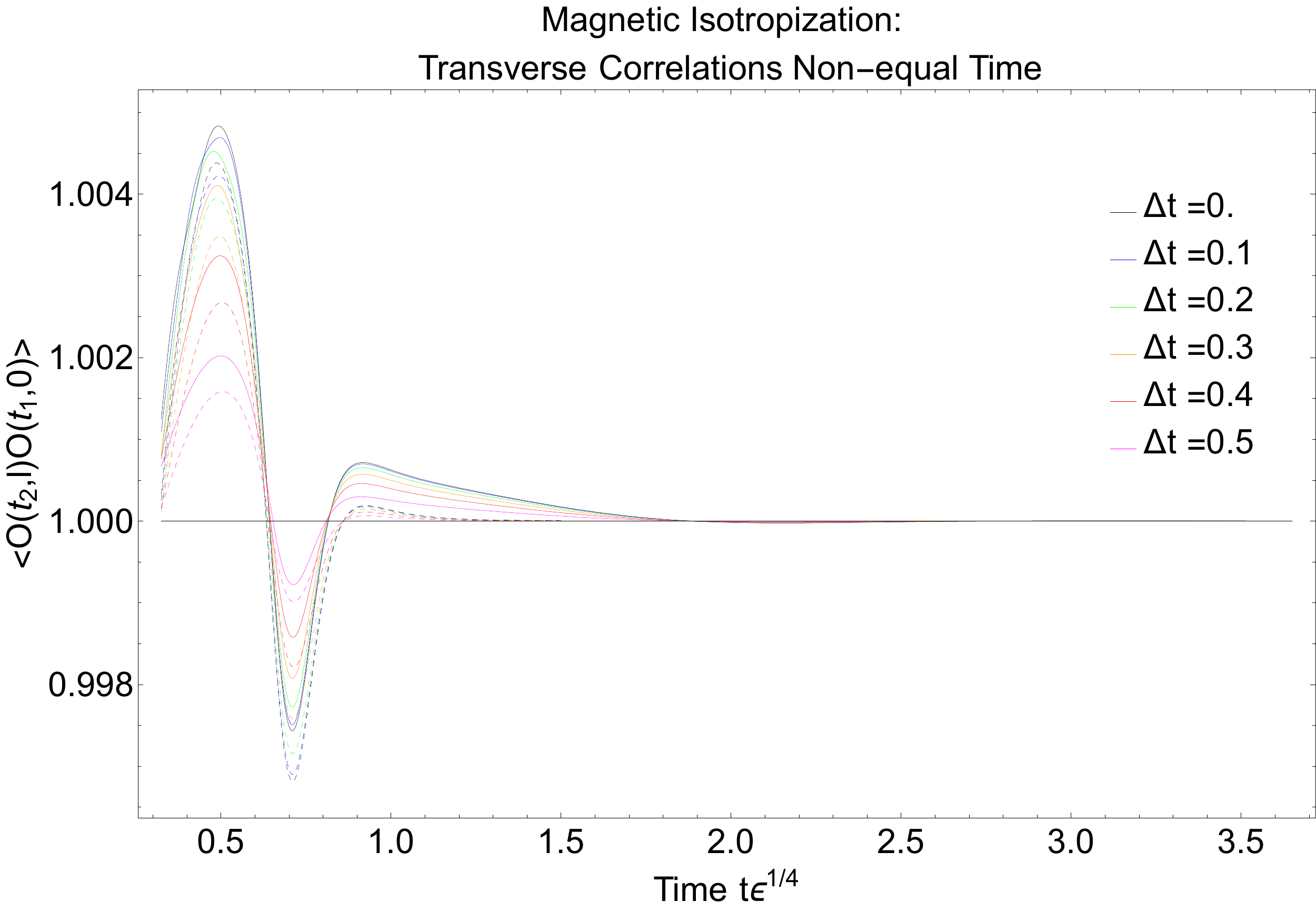}
  \end{center}
\end{subfigure}
 \begin{subfigure}[b]{.5\linewidth}
  \begin{center}   
\includegraphics[width=2.9in]{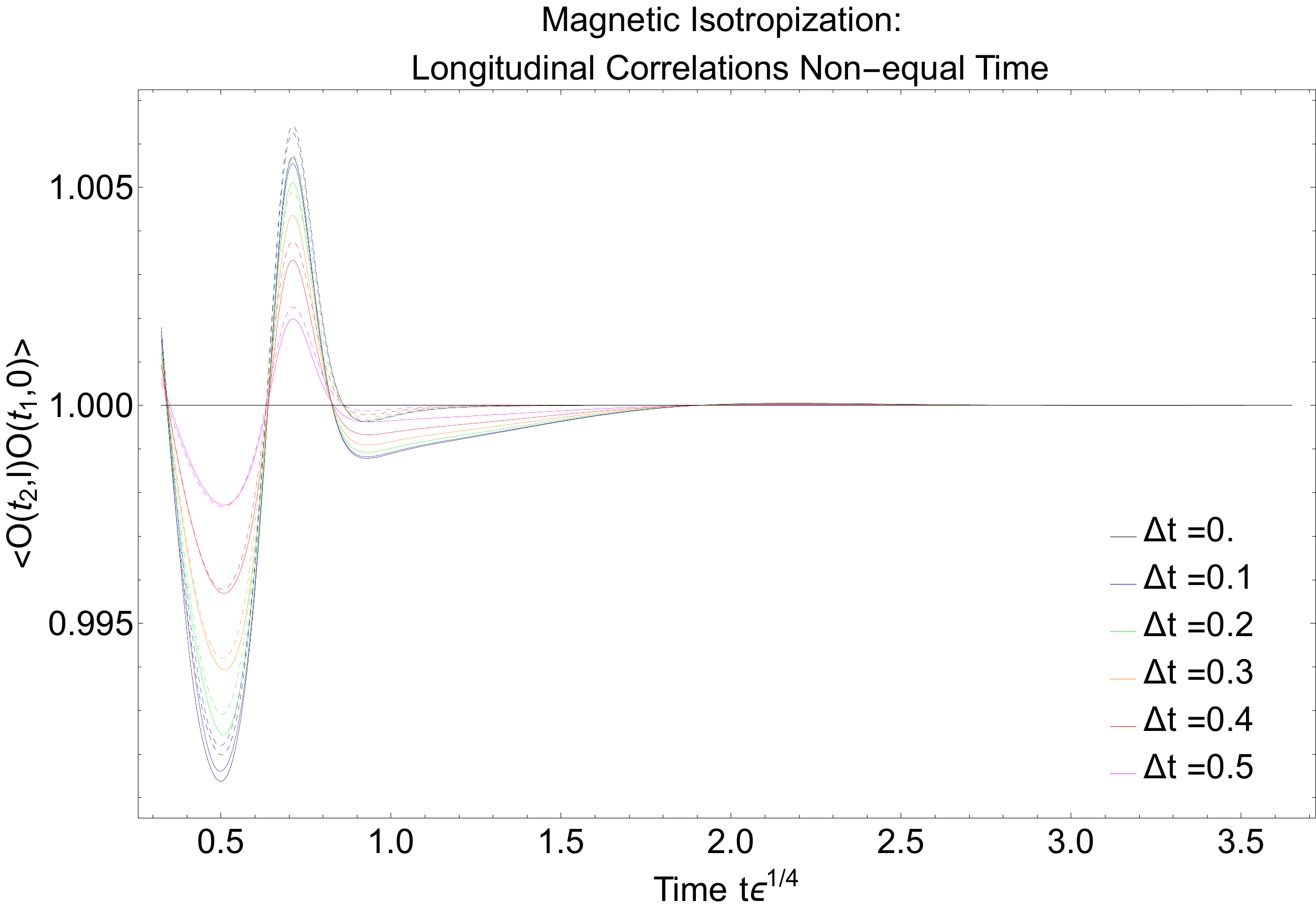}
  \end{center}
\end{subfigure}
\caption{\label{fig:MagneticCase}
2-point correlation functions of an (uncharged) scalar operator in a neutral thermalizing fluid (dual to evolution towards a magnetic black brane geometry), $\rho=0$, $\mathcal{B}=1$, case (c) in Fig.~\ref{fig:introFigureSetup}.
\textit{Top Left:} Various length separations in the transverse direction at $\Delta t=0$. 
\textit{Top right:} Various length separations in the longitudinal direction at $\Delta t=0$. 
\textit{Bottom left:} Various time separations $\Delta t$, fixed transverse length separation $l=0.65$.
\textit{Bottom Right:} Various time separations $\Delta t$, fixed longitudinal length separation $\Delta l=0.65$.
The purely anisotropic case (a) is shown for comparison in the top graphs as {an} inset, and in the bottom graphs as dashed curves.}
\end{figure}
In Fig.~\ref{fig:MagneticCase} we display the results of our calculations of 2-point functions in an initially anisotropic, uncharged thermal fluid subjected to an external magnetic field, which relaxes towards its anisotropic equilibrium state, as sketched in Fig.~\ref{fig:introFigureSetup} (c).\footnote{This is dual to the computation of geodesics in the time-dependent neutral magnetic black brane background.}   
\begin{figure}[H]
  \begin{subfigure}[b]{.5\linewidth}
  \begin{center}   
\includegraphics[width=2.9in]{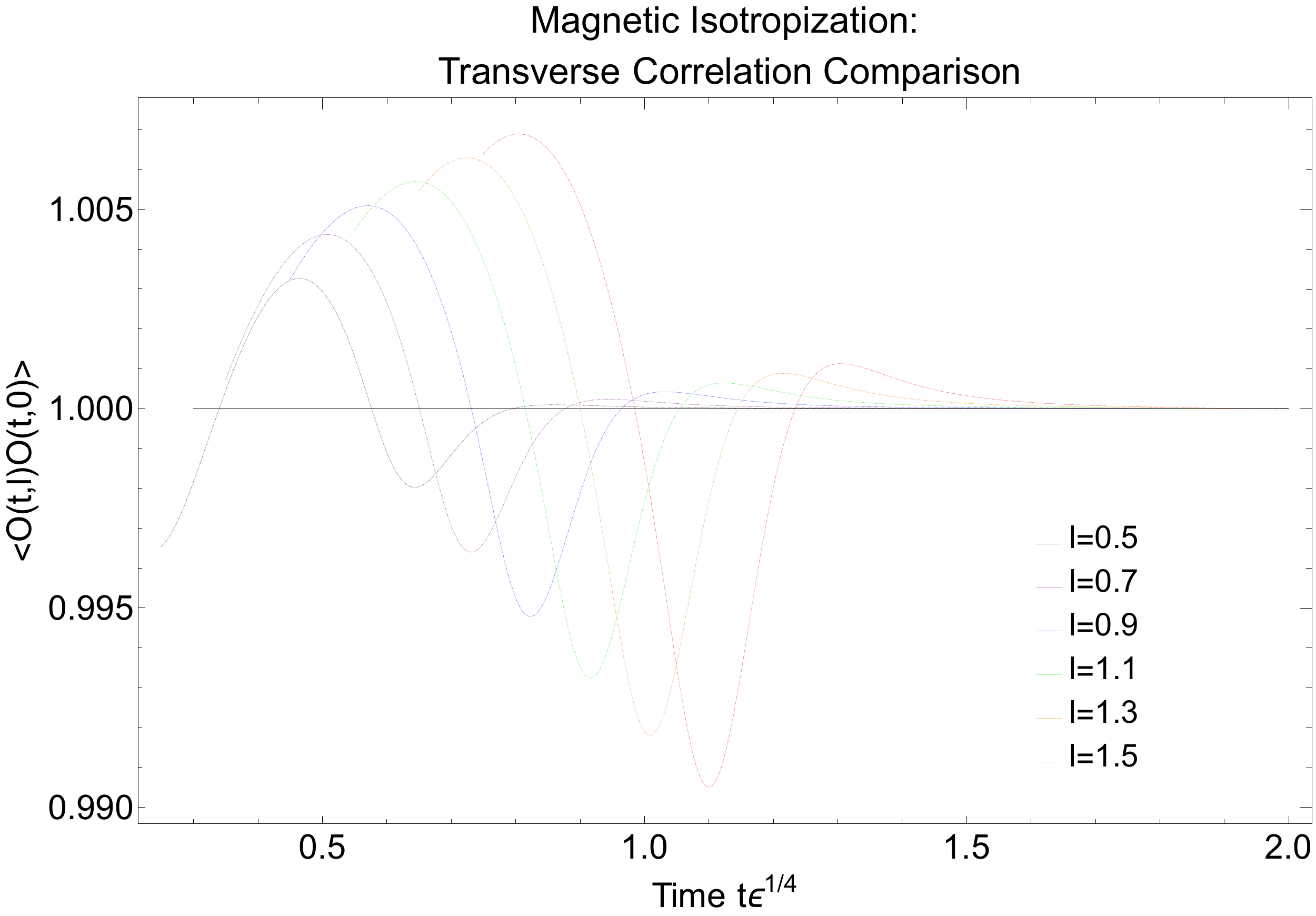}
  \end{center}
\end{subfigure}
 \begin{subfigure}[b]{.5\linewidth}
  \begin{center}   
\includegraphics[width=2.9in]{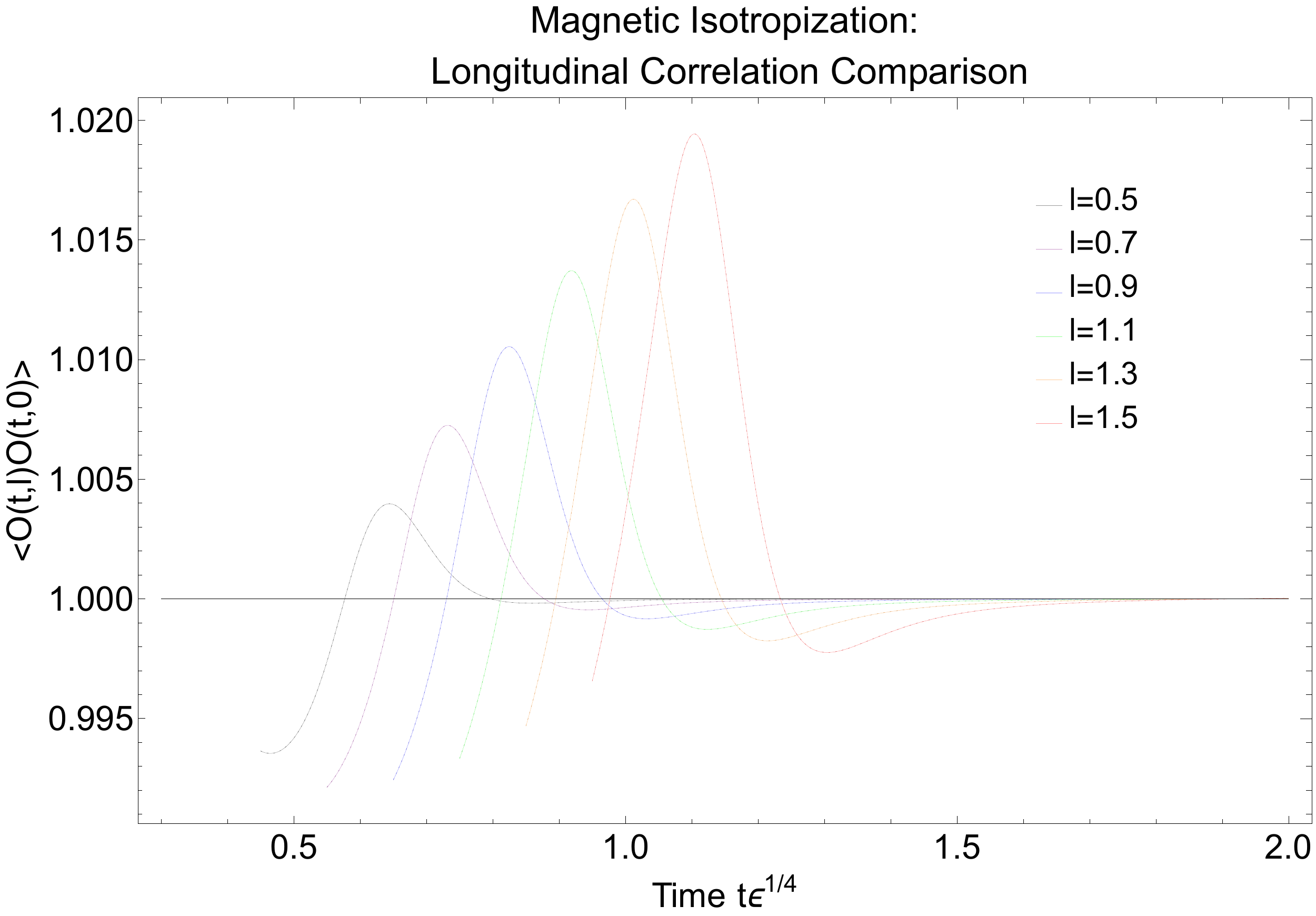}
  \end{center}
\end{subfigure}
\caption{\label{fig:MarginalMagField} 
{\it Small magnetic field correlators virtually identical to vanishing magnetic field case.} 2-point correlation functions of an (uncharged) scalar operator in a neutral thermalizing fluid (dual to the evolution towards a magnetic black brane geometry), $\rho=0$, $\mathcal{B}=0.1$. \textit{Left:} Various length separations in the transverse direction. \textit{Right:} Various length separations in the longitudinal direction. The purely anisotropic case (a) is shown for comparison as dashed curves. }
\end{figure}
In this case we find the most significant deviation from what was seen in the previous sections. Generally, the magnetic field pushes the 2-point functions away from equilibrium (compared to the case at $\mathcal B=0$). As a consequence for small magnetic field $\mathcal{B}=1$, the time to thermalization is significantly longer in both types of length separation; see table~\ref{tab:thermalizationTimes}. 
The background, however, 
takes approximately the same time to thermalize in both cases, $\mathcal{B}=1$ or 0, as in the uncharged case without magnetic field. 
The charged and uncharged fluids have 1-point functions which quickly come to their equilibrium values on time scales of $t \epsilon^{1/4}\approx 0.7$. 
This is also true for the magnetic case with $\mathcal{B}=1$. But both 1-point and 2-point functions thermalize faster and faster with increasing the magnetic field towards $\mathcal{B}=3$.\footnote{1-point functions thermalize at times
$t \epsilon^{1/4}\approx 0.6$ and 2-point functions at $t \epsilon^{1/4}\approx 1.4$ (compared to $t \epsilon^{1/4}\approx 1.9$.} 
In addition, for times between $0.8$ and $2.5$ where the 1-point functions are changing only slightly for $\mathcal{B}=1$, the 2-point functions are still subject to large changes. As a second consequence of the push away from equilibrium, the early equilibrium points (at $t=t_{cor,n}$) disappear, as the correlator curves now do not intersect the equilibrium line anymore. Large length separations are pushed away from equilibrium further in all cases with nonzero $\mathcal{B}$. This early-time push away from equilibrium is even more pronounced at larger $\mathcal{B}$ values where even correlations over short length scales are dragged by the magnetic field and no longer intersect the equilibrium line.

 \begin{figure}
  \begin{subfigure}[b]{.5\linewidth}
  \begin{center}   
\includegraphics[width=2.9in]{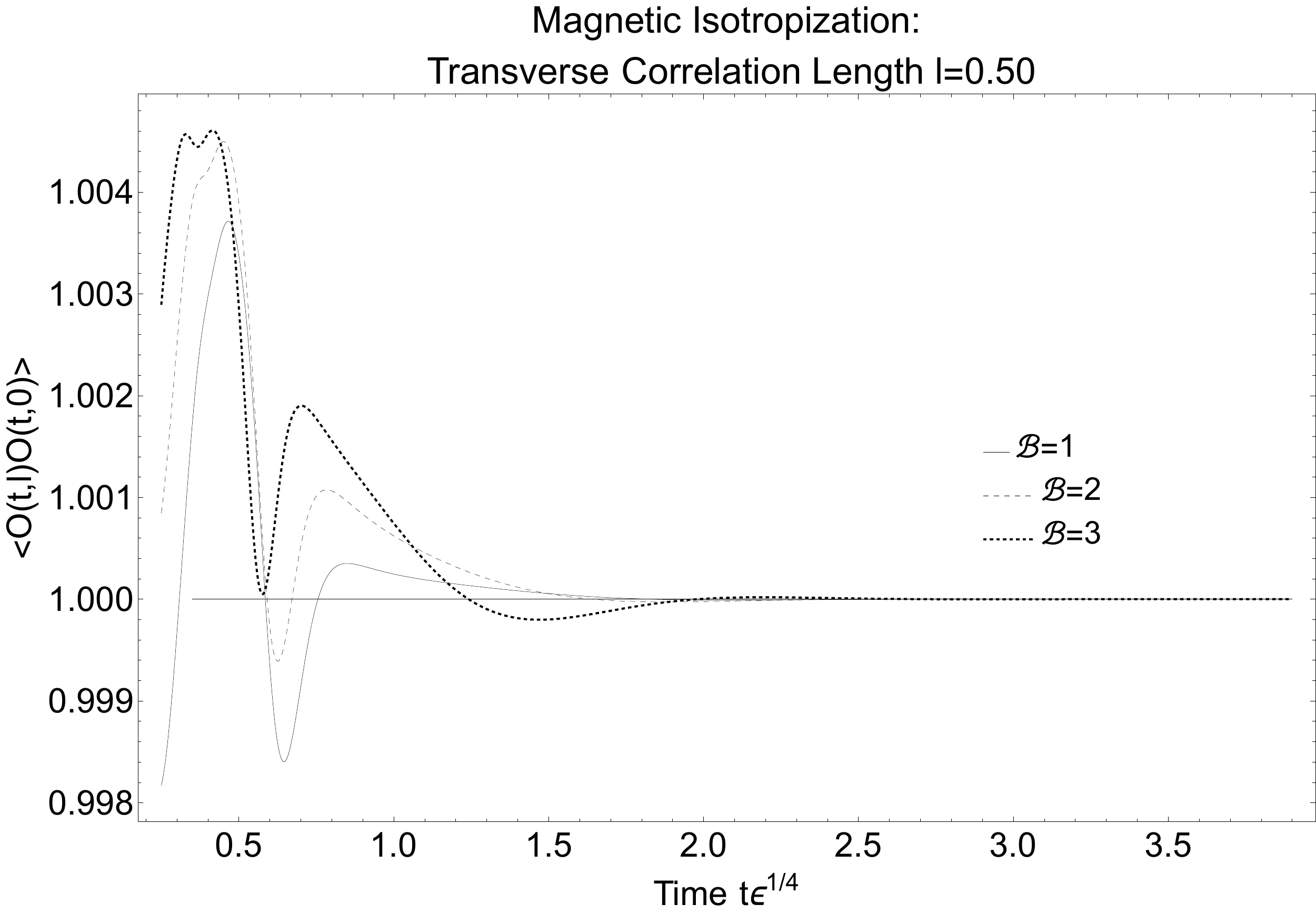}
  \end{center}
\end{subfigure}
   \begin{subfigure}[b]{.5\linewidth}
  \begin{center}   
\includegraphics[width=2.9in]{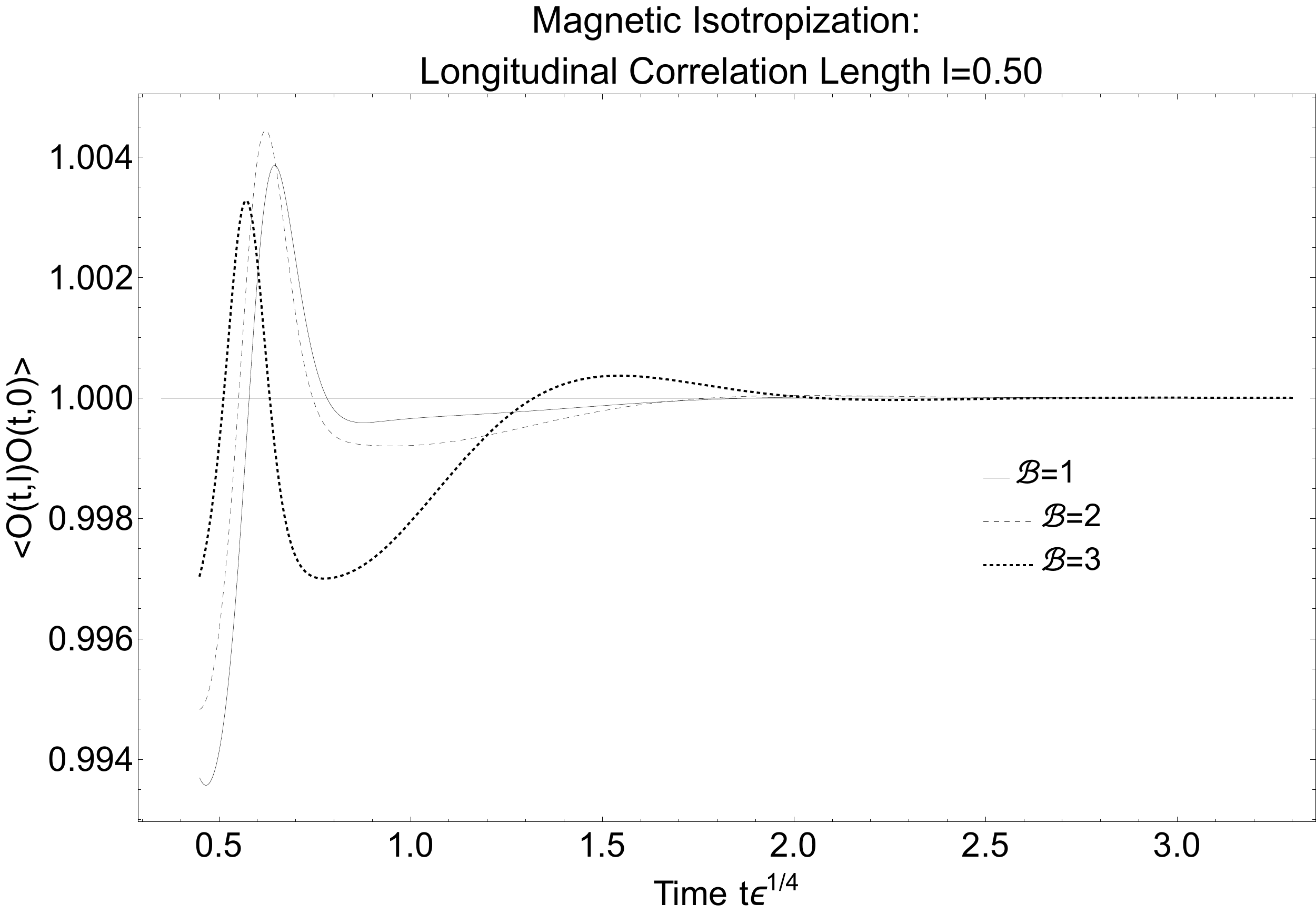}
  \end{center}
\end{subfigure}
  \begin{subfigure}[b]{.5\linewidth}
  \begin{center}   
\includegraphics[width=2.9in]{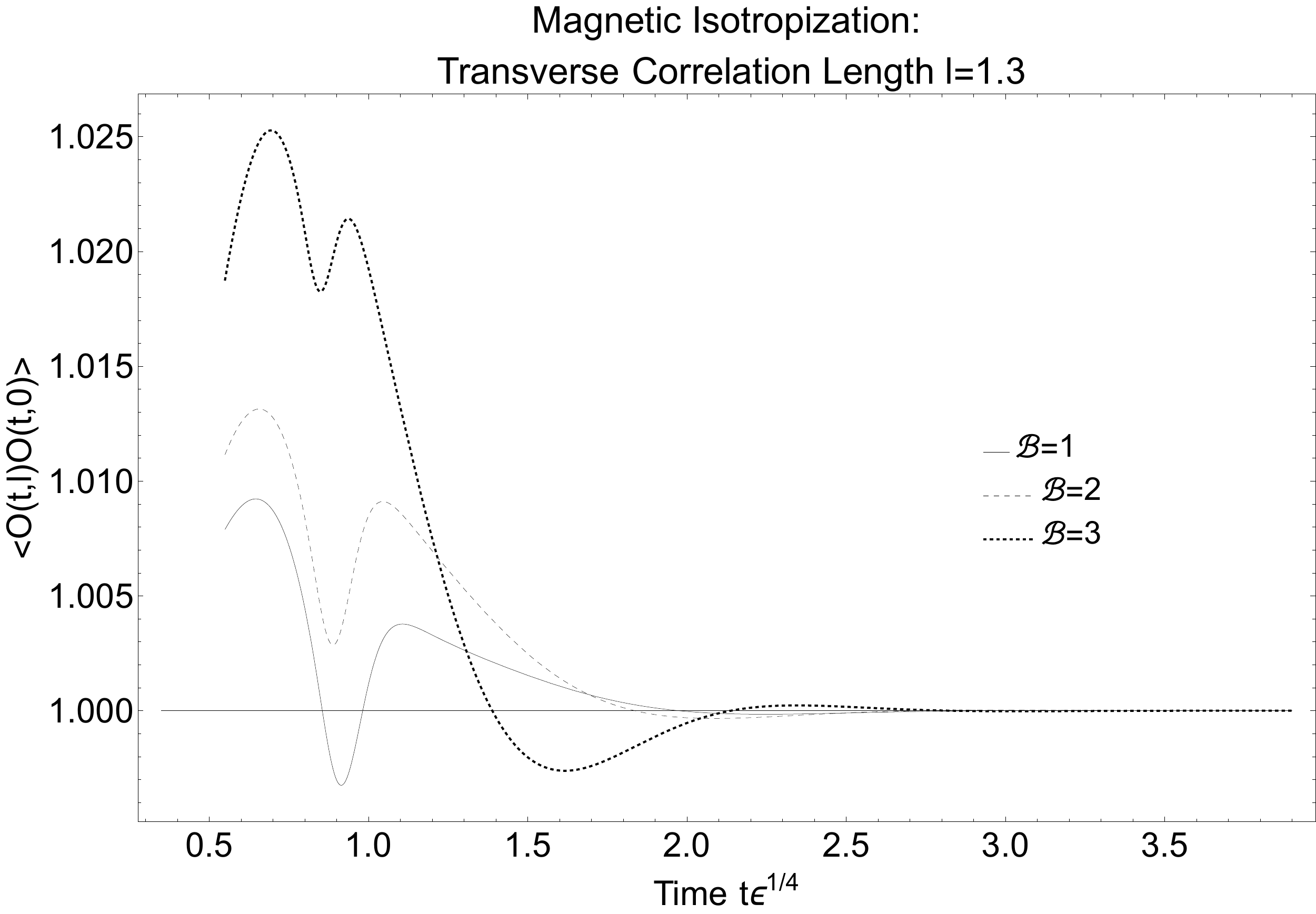}
  \end{center}
\end{subfigure}
  \begin{subfigure}[b]{.5\linewidth}
  \begin{center}   
\includegraphics[width=2.9in]{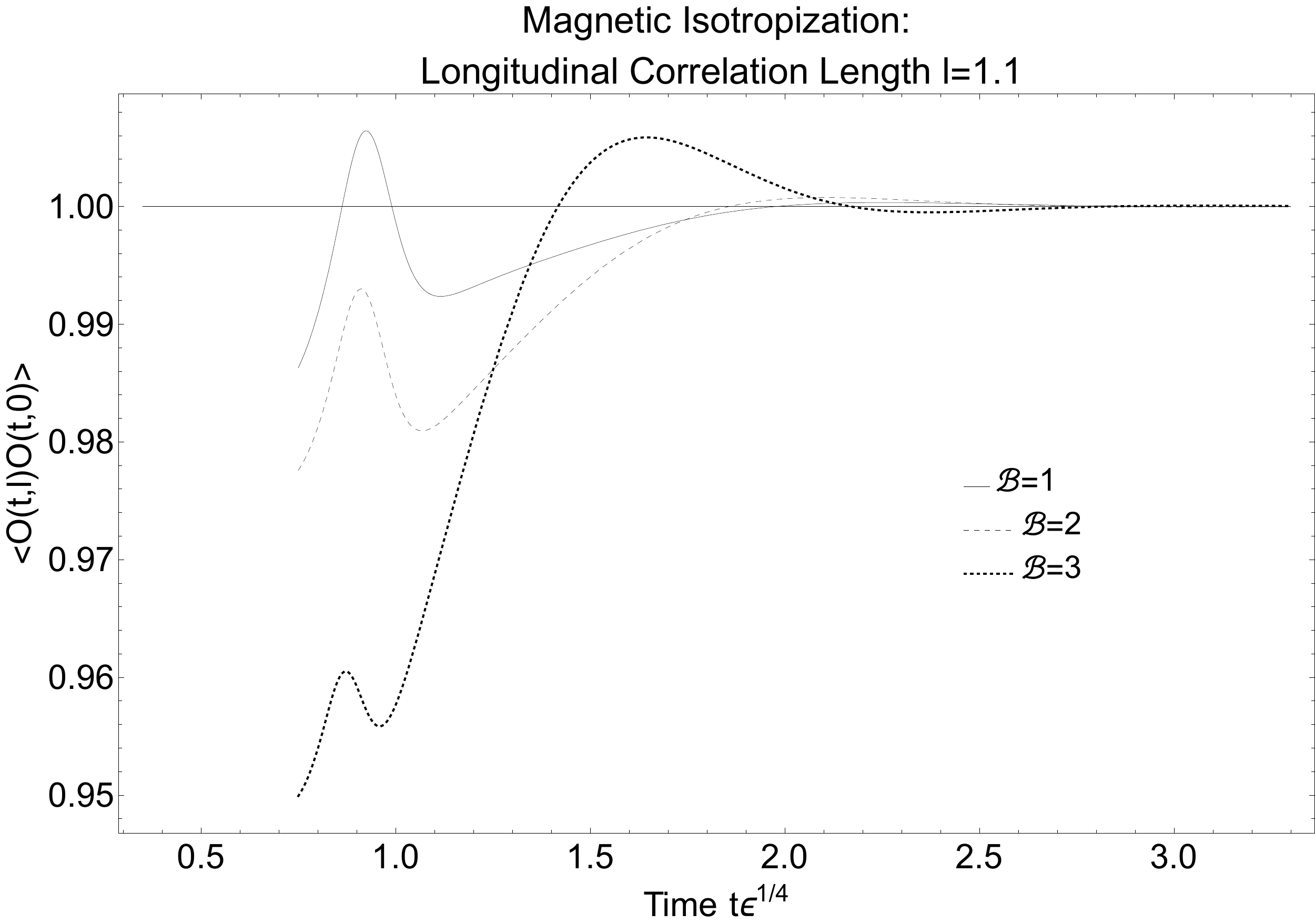}
  \end{center}
\end{subfigure}
\caption{\label{fig:CompareB} 2-point correlation functions of an (uncharged) scalar operator in a neutral thermalizing fluid (dual to the evolution towards a magnetic black brane geometry), $\rho=0$, $\mathcal{B}=1,2,3$. \textit{Top Left:} Length separation of $l=0.50$ in the transverse direction. \textit{Top Right:} Length separation of $l=0.50$ in the longitudinal direction. \textit{Bottom Left:} Length separation of $l=1.1$ in the transverse direction. \textit{Bottom Right:} Length separation of $l=1.1$ in the longitudinal direction.  }
\end{figure}

We additionally find a new feature at later times in the top two graphics of Fig.~\ref{fig:MagneticCase}. There appears to be a time $t\epsilon^{1/4}\approx 2.0$ at which the correlations attract to the same behavior independent of their length separation $l$, resembling the nodes previously found only in the non-equal time correlators of the neutral and charged cases. Looking at the inset graph of these top two graphs in Fig.~\ref{fig:MagneticCase} for comparison, we see such behavior is not displayed in the simple case of anisotropic thermalization. There we see each correlation function shift horizontally as a function of the length separation. This length-independent attraction observed in the magnetic case is particularly interesting because the peaks observed at earlier times $t\epsilon^{1/4}\approx 0.7$ do shift to later times with increasing length $l$. Yet the approach to equilibrium past this time of $t=\epsilon^{1/4}\approx 2.0$ of all the lengths probed appears to attract to a common behavior. Our observation is supported by our analysis in Sec.~\ref{sec:ComparisonofCases} where we find thermalization time of the correlation functions to be nearly independent of length as we increase the strength of the magnetic field. 

In order to investigate this behavior, and to show that the behavior changes smoothly with the parameters of the system, we have considered charge density and magnetic field values much smaller than their respective energy density scales as displayed in the appendix in Fig.~\ref{fig:ChecksSmallBsmallChar}. We find in this limit little if any deviation from the case of relaxation towards a Schwarzschild black brane. This indicates that the transition of the correlations to the
form displayed in Fig.~\ref{fig:MagneticCase} is a smooth function of magnetic field strength. This smooth behavior is further confirmed in Fig.~\ref{fig:MarginalMagField} and Fig.~\ref{fig:CompareB} for values of the magnetic field approximately two orders of magnitude larger, {i.e.}~at $\mathcal{B}=1, \, 2, \, 3$. The common feature is that the peak to peak amplitude of the curves grows with the magnetic field. This effect is obscured by the curves also being pushed more above or below the equilibrium line at early times at larger magnetic fields.

In the non-equal time correlators shown in the bottom two graphs of Fig.~\ref{fig:MagneticCase} we again observe that the magnetic field pushes the curves away from equilibrium. 
We again observe the attraction at $v\epsilon^{1/4}\approx 2.0$, which occurs in the equal time correlators discussed above. 
Remarkably, in the transverse non-equal time correlators, we observe that the earlier time node occurs at approximately the same time for both $\mathcal{B}=1$ or $0$, namely around $t\epsilon^{1/4}\approx0.62$. However, the node at slightly later time is shifted to occur earlier when $\mathcal{B}=1$. Meanwhile, in the longitudinal equal time correlators both nodes are shifted to earlier times; the fate of the longitudinal $\mathcal{B}=0$ node at $t \epsilon^{1/4}\approx 0.35$, see bottom right graph in Fig.~\ref{fig:MagneticCase}, is uncertain as our $\mathcal{B}=1$ data does not reach far enough back in time. 
Testing our hypothesis that such nodes should occur in 2-point functions a time of $l/2$ after the background oscillated through its equilibrium values, we find that our initial hypothesis does not explain the location of the node in the case of a fluid subjected to an external magnetic field. The results of this analysis are displayed in table~\ref{tab:EQ_Magnetic_Blackbrane_transverse} and table~\ref{tab:EQ_Magnetic_Blackbrane_Longitudinal}. Two important features of this table can be picked out. First the relative error does not display a length scale at which our system can be described by free streaming. All errors are above 15$\%$ and indicate that either non-local correlations or interaction with the strongly correlated fluid occurs. Furthermore while the relative error of the third node decreases with an increasing length scale the relative percent error of the fourth and fifth nodes increase with an increasing length scale. This indicates that larger scales are further from equilibrium. The red curve displayed in Fig.~\ref{fig:MagneticCase} is an example of this. Second, there are nodes in the background which are not present in the two point functions, these are entered in the table as Does Not Exist (D.N.E). These points further indicate the presence of interaction with the strongly correlated medium and the possibility of non-local correlations. 

\begin{table}
\begin{center}
\begin{tabular}{|c|c|c|c|c|c|}
\multicolumn{6}{c}{Transverse separation }\\
\hline
$n$   & $t_{eq,n}$ & $l\epsilon^{1/4}$& Predicted $t_{cor,n}$ & Numerical $t_{cor,n}$ & Relative percent error   \\
    \hline
\multirow{6}{*}{3 } &\multirow{6}{*}{0.518 }& 0.5& 0.768 & 0.585 & 27.043 \\
&&0.7 &0.868 & 0.663 & 26.779 \\
 &&0.9 &0.968 & 0.754 & 24.92 \\
&& 1.1&1.068 & 0.855 & 22.166\\
 &&1.3 &1.168 & 0.966 & 18.985 \\
 & &1.5 &  1.268 & D.N.E & D.N.E \\
   \hline
\multirow{6}{*}{4 } &\multirow{6}{*}{0.716}
&0.5  & 0.966 & 0.755 & 24.46 \\
& &0.7 & 1.066 & 0.835 & 24.283 \\
& &0.9 & 1.166 & 0.910 & 24.699 \\
& &1.1 & 1.266 & 0.982 & 25.255 \\
& &1.3 & 1.366 & 1.05 & 26.114 \\
& &1.5 & 1.466 & D.N.E & D.N.E \\
   \hline
 \multirow{6}{*}{5 } &\multirow{6}{*}{1.896}
&0.5  & 2.146 & 1.824 & 16.185 \\
& &0.7 &2.246 & 1.893 & 17.033 \\
& &0.9 &2.346 & 1.934 & 19.226 \\
& &1.1 &2.446 & 1.974 & 21.322 \\
& &1.3 & 2.546 & 2.018 & 23.138 \\
& &1.5 &  2.646 & 2.064 & 24.674 \\
   \hline
\end{tabular}
\end{center}
\caption{\label{tab:EQ_Magnetic_Blackbrane_transverse}
Example equilibrium times of the 1-point functions, $t_{eq,n}$, and 2-point functions, $t_{cor,n}$, in a thermalizing fluid (dual to the evolution towards a magnetic black brane geometry) with $\mathcal{B}=1$ and $\rho=0$. Entries of D.N.E stand for Does Not Exist, this can be seen in the red curves displayed in the top two graphics of Fig.~\ref{fig:MagneticCase}. The correlation functions of this length scale are lifted above the axis.}
\end{table}

\begin{table}
\begin{center}
\begin{tabular}{|c|c|c|c|c|c|}
\multicolumn{6}{c}{Longitudinal separation }\\
\hline
$n$   & $t_{eq,n}$ & $l\epsilon^{1/4}$& Predicted $t_{cor,n}$ & Numerical $t_{cor,n}$ & Relative percent error   \\
    \hline
\multirow{6}{*}{3 } &\multirow{6}{*}{0.518 }
& 0.5& 0.768 & 0.579 & 28.125 \\
&&0.7 & 0.868 & 0.660 & 27.224 \\
&& 0.9& 0.968 & 0.755 & 24.737 \\
&&1.1 & 1.068 & 0.863 & 21.322 \\
&& 1.3& 1.168 & 0.981 & 17.406 \\
 & &1.5 &  1.268 & D.N.E & D.N.E \\
   \hline
\multirow{6}{*}{4 } &\multirow{6}{*}{0.716}
&0.5  &  0.966 & 0.782 & 21.091 \\
&& 0.7& 1.066 & 0.844 & 23.268 \\
&&0.9 & 1.166 & 0.916 & 24.011 \\
&&1.1 & 1.266 & 0.99 & 24.44 \\
&&1.3 & 1.366 & 1.061 & 25.137\\
& &1.5 & 1.4659 & D.N.E & D.N.E \\
   \hline
 \multirow{6}{*}{5 } &\multirow{6}{*}{1.896}
&0.5  &  2.146 & 1.909 & 11.69 \\
&& 0.7& 2.246 & 1.915 & 15.879 \\
&&0.9 & 2.346 & 1.944 & 18.712 \\
&& 1.1& 2.446 & 1.983 & 20.867 \\
&& 1.3& 2.546 & 2.03 & 22.517 \\
&& 1.5& 2.646 & 2.083 & 23.776 \\
   \hline
\end{tabular}
\end{center}
\caption{\label{tab:EQ_Magnetic_Blackbrane_Longitudinal}
Example equilibrium times of the 1-point functions, $t_{eq,n}$, and 2-point functions, $t_{cor,n}$, in a thermalizing fluid (dual to the evolution towards a magnetic black brane geometry) with $\mathcal{B}=1$ and $\rho=0$. Entries of D.N.E stand for Does Not Exist, this can be seen in the red curves displayed in the top two graphics of Fig.~\ref{fig:MagneticCase}. The correlation functions of this length scale are lifted above the axis.}
\end{table}

Often it is said that the form of the vector potential does not influence the classical trajectory of particles. However just as the electric potential is the potential energy per unit charge so is the vector potential the momentum per unit charge~\cite{Konopinski1978}. This is seen clearly from the action with which we derive the Lorentz force law on a charged point particle, 
\begin{equation}
    S_{\text{Lorentz}}=\int \mathrm{d}\lambda \dot{x}^{\mu}A_{\mu}.
\end{equation}
To calculate the 2-point correlation function of a charged scalar operator we add to the action for a geodesic $S_{\text{Lorentz}}$, calculate the trajectories, and then explicitly compute the value of the total action. Of the backgrounds we consider only case (c) and case (d) have the ability to affect the value of the 2-point correlation function (case (b) is discussed in the appendix where it is argued that both charged and uncharged operator insertions will display the same 2-point correlations, as already mentioned in Sec.~\ref{sec:RNBlackBraneBG}). Let us consider two distinct choices for the introduction of a magnetic field in the $x^3$-direction, one which is $SO(2)$-symmetric $\mathcal{A}=(0,0,-y\mathcal{B}/2,x \mathcal{B}/2,0)$ and one which is not, namely $\mathcal{A}=(0,0,0,x \mathcal{B},0)$.

The calculation of the relaxing magnetic black brane is independent of this choice, the only quantity which enters the background Einstein equations is the field strength. In contrast to that, the action for the Lorentz force term required to calculate the trajectory of the classical charged particles in the bulk geometry is in fact sensitive to the choice of the gauge field. Writing down the action for the Lorentz force term for the rotationally invariant choice we find,
\begin{equation}
 S=\int \mathrm{d}\lambda \frac{\mathcal{B}}{2}(-y\dot{x}+x\dot{y}).
\end{equation}
Unfortunately trajectories through the bulk which obey $(-l/\sqrt{2},-l\sqrt{2})$ to $(l/\sqrt{2},l\sqrt{2})$ lead to solutions $x(\sigma)=y(\sigma)$. Clearly this leads to zero contribution to the total action $S=S_{\text{Geodesic}}+S_{\text{Lorentz}}$. This is due to the symmetry of the solution to the geodesic equation itself. Although this is already obvious analytically, we have explicitly verified the vanishing contribution to the 2-point function numerically as an additional check of our code. 

\begin{figure}[H]
    \begin{subfigure}[b]{.5\linewidth}
  \begin{center}   
    \includegraphics[width=2.75in]{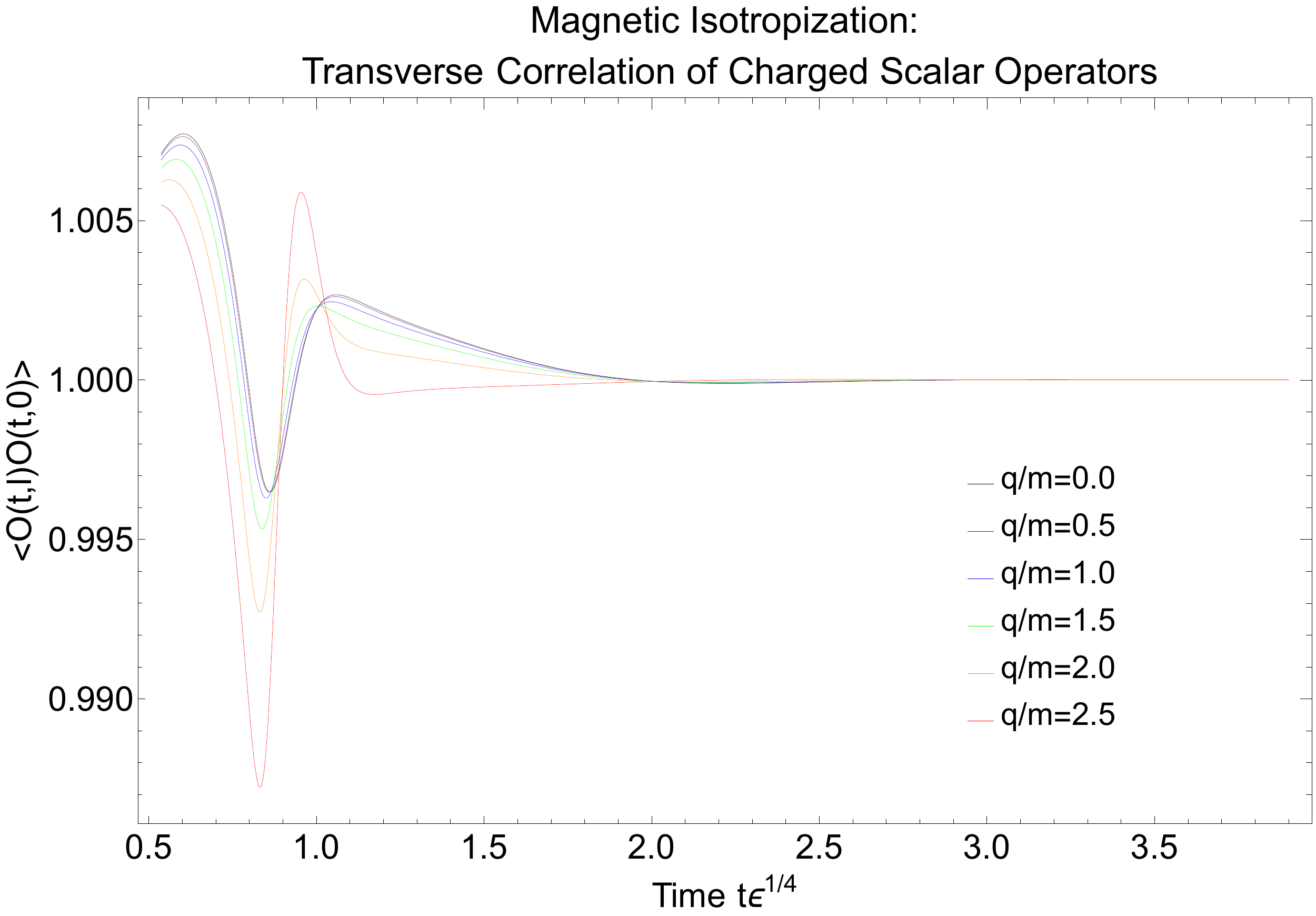}
    \end{center}
    \end{subfigure}
    \begin{subfigure}[b]{.5\linewidth}
  \begin{center}   
    \includegraphics[width=2.75in]{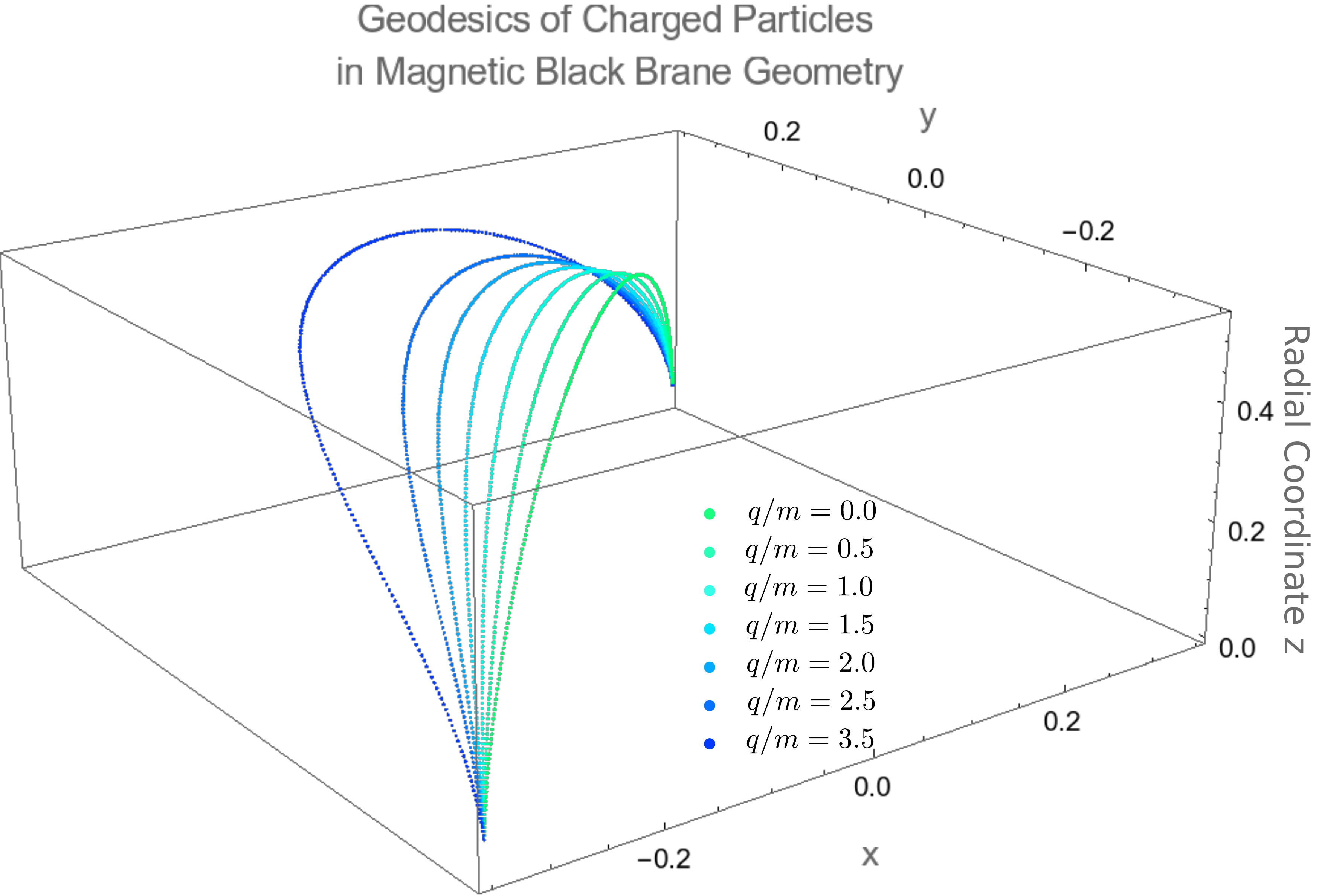}
    \end{center}
    \end{subfigure}
    \caption{\textit{Left:} 2-point correlation function of a charged scalar operator in a neutral thermalizing fluid (dual to the evolution towards a magnetic black brane geometry), $\rho=0$, $\mathcal{B}=1$. The correlation functions above are computed for $l=0.7\sqrt{2}$. \textit{Right:} The geodesics of a charged particle of various charge to mass ratios, $q/m$ displayed at a boundary time of $v=2.91$. These geodesics were used to calculate the correlation functions shown on the left. 
    \label{fig:ChargedOP}}
\end{figure}

Displayed in Fig.~\ref{fig:ChargedOP} is the result of using the gauge potential which is not $SO(2)$ symmetric. Here we find behavior which bears some resemblence to the case of operator insertions with spatial and temporal separation as shown in Fig.~\ref{fig:MagneticCase}. Though we do note that increasing charge to mass ratios display a shifting of the major features towards earlier times. The charge to mass ratio leads to decreasing thermalization times as we increase the charge to mass ratio.
\begin{table}[h]
    \centering
    \begin{tabular}{|c|c|c|c|c|c|c|c|}
    \hline
     $q/m$ & 0& 0.5 & 1.0& 1.5& 2.0& 2.5 & 3.5   \\
     \hline
        $t_{2pt.}$ &  2.382 & 2.381 & 2.378 & 2.365 & 2.314 & 1.836 & 1.3325 \\
        \hline
    \end{tabular}
    \caption{Example thermalization times of 2-point functions of charged scalar operators, $t_{2pt.}$, in a magnetic thermalizing fluid (dual to the evolution towards a magnetic black brane geometry) $\rho=0$, $\mathcal{B}=1$. The thermalization times are determined according to the measure defined in Eq.~\eqref{eq:thermal}. The length of separation considered is $l=0.7\sqrt{2}$   
    \label{tab:Charged_Operators}}
\end{table}

\subsection{Case (d): Isotropization towards a charged magnetic black brane}
\begin{figure}[H]
  \begin{subfigure}[b]{.5\linewidth}
  \begin{center}   
\includegraphics[width=2.9in]{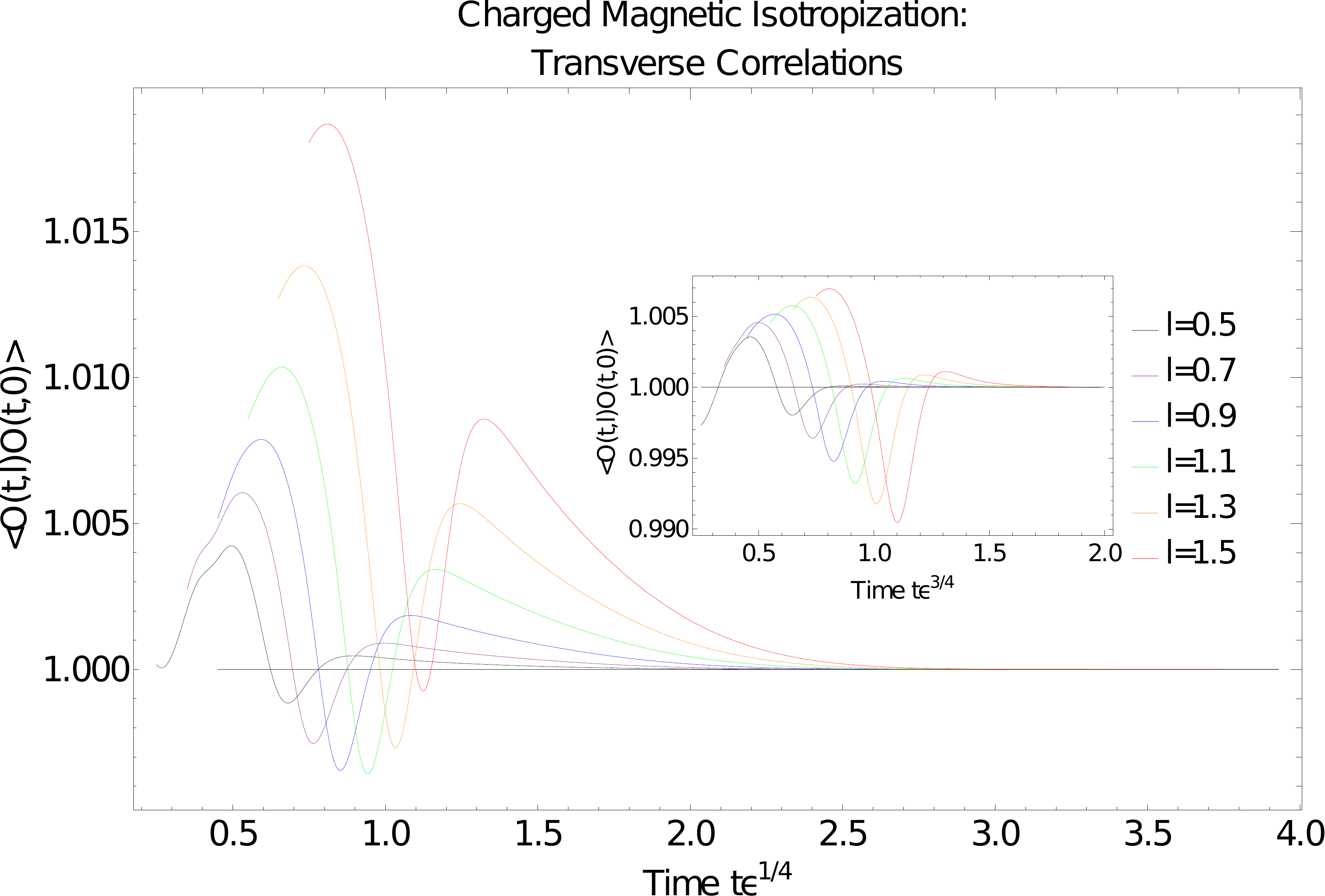}
  \end{center}
\end{subfigure}
 \begin{subfigure}[b]{.5\linewidth}
  \begin{center}   
\includegraphics[width=2.9in]{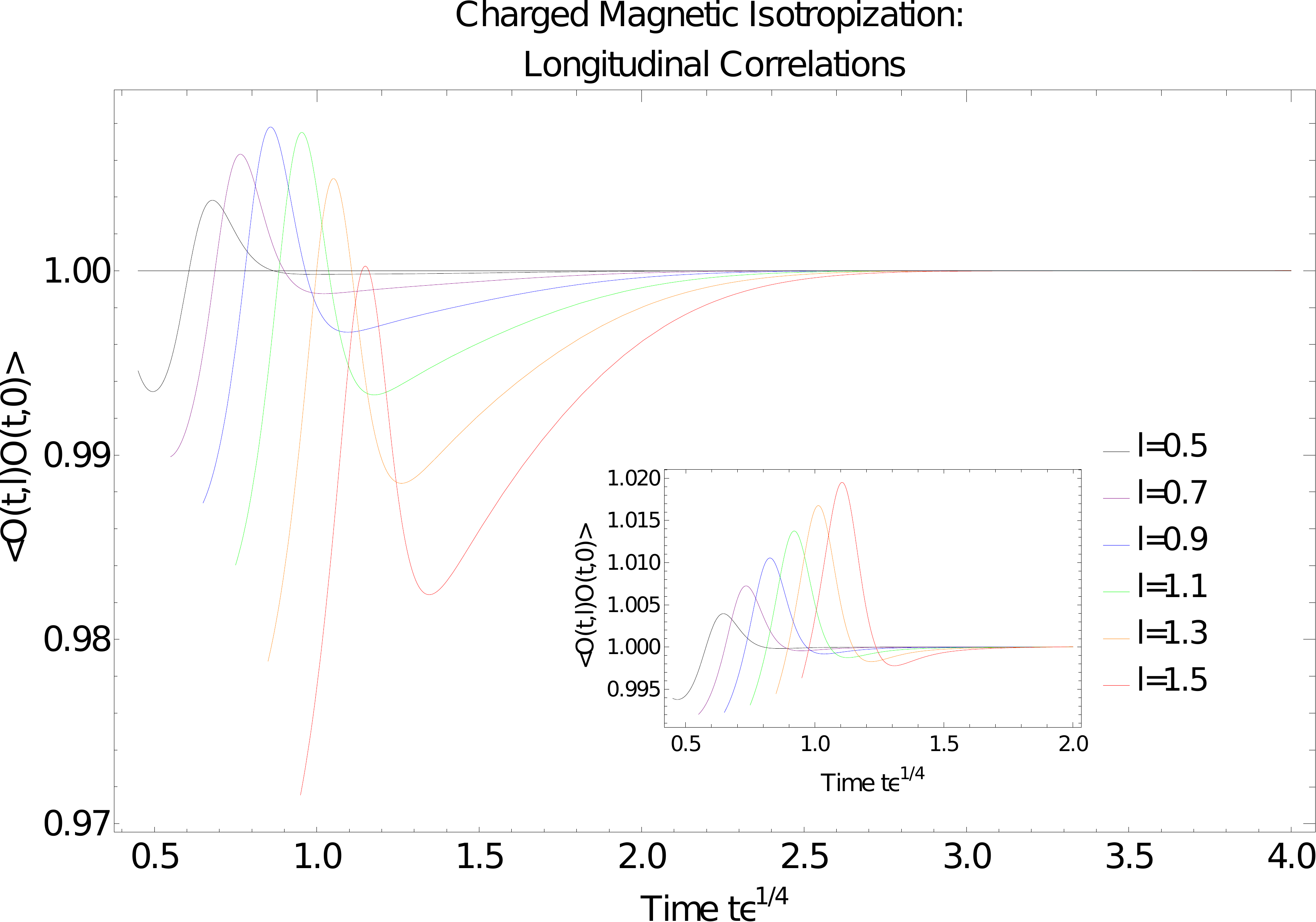}
  \end{center}
\end{subfigure}
 \begin{subfigure}[b]{.5\linewidth}
  \begin{center}   
\includegraphics[width=2.9in]{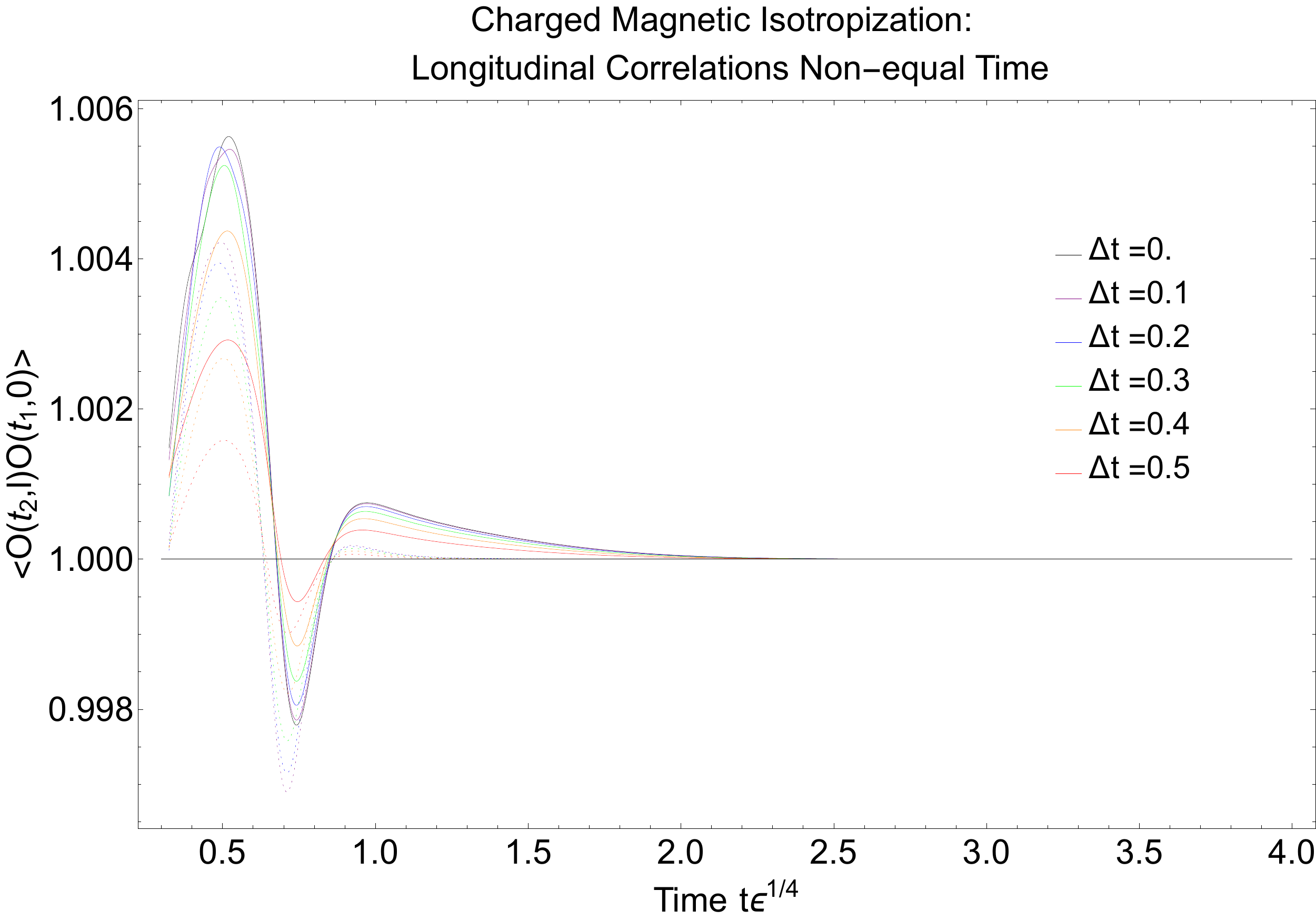}
  \end{center}
\end{subfigure}
 \begin{subfigure}[b]{.5\linewidth}
  \begin{center}   
\includegraphics[width=2.9in]{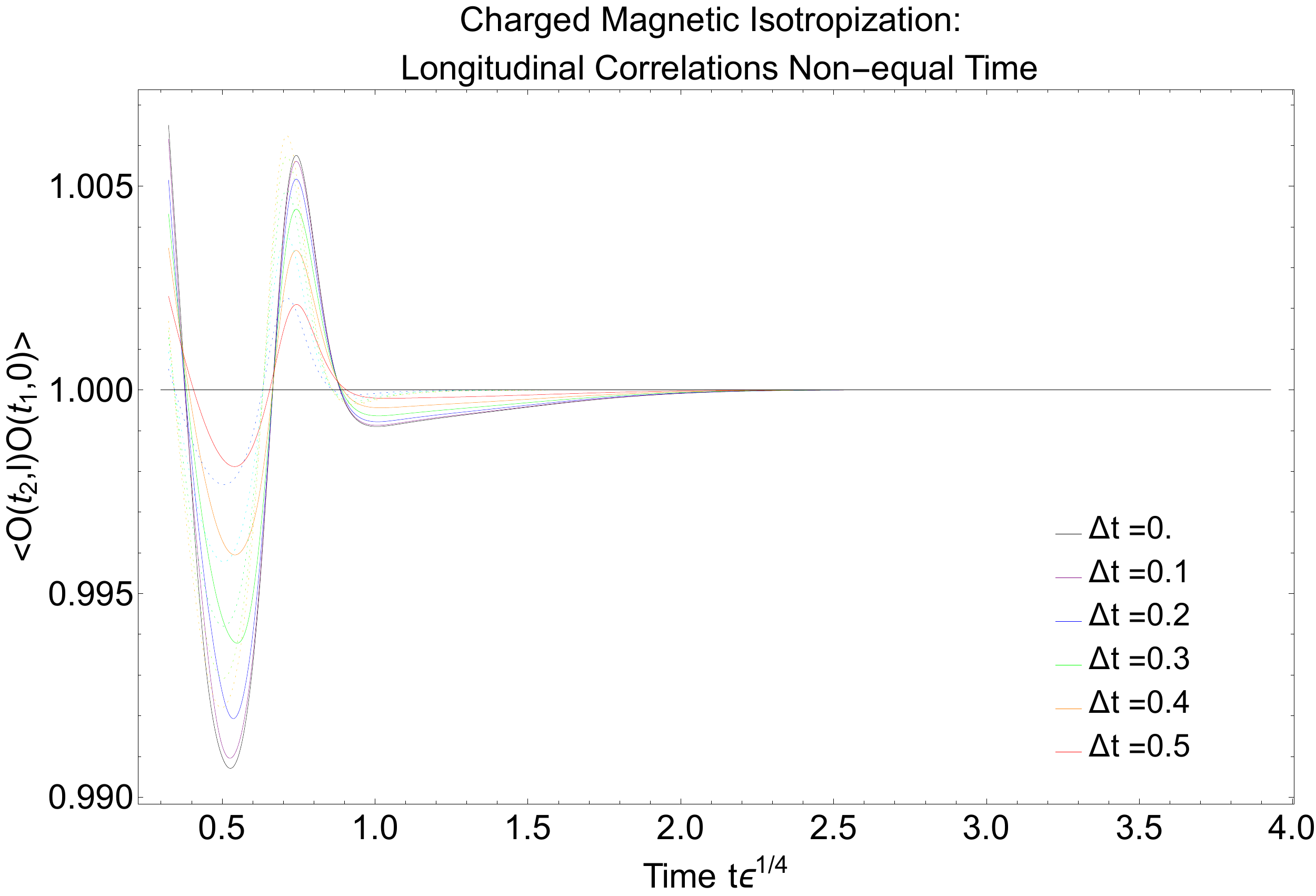}
  \end{center}
\end{subfigure}
\caption{\label{fig:ChargedMagneticCase}2-point correlation functions of an (uncharged) scalar operator in a neutral thermalizing fluid (dual to evolution towards a charged magnetic black brane geometry), $\rho=0.78\rho_{e}$, $\mathcal{B}=1$, case (c) in Fig.~\ref{fig:introFigureSetup}.
\textit{Top Left:} Various length separations in the transverse direction at $\Delta t=0$. 
\textit{Top right:} Various length separations in the longitudinal direction at $\Delta t=0$. 
\textit{Bottom left:} Various time separations $\Delta t$, fixed transverse length separation $l=0.65$.
\textit{Bottom Right:} Various time separations $\Delta t$, fixed longitudinal length separation $\Delta l=0.65$. The purely anisotropic case (a) is shown for comparison in the top graphs as {an} inset, and in the bottom graphs as dashed curves.
}
\end{figure}

In Fig.~\ref{fig:ChargedMagneticCase} we display the results of our calculations of 2-point functions in an initially anisotropic, charged thermal fluid subjected to an external magnetic field, which relaxes towards its anisotropic equilibrium state, as sketched in Fig.~\ref{fig:introFigureSetup} (d).\footnote{This is dual to the computation of geodesics in the time-dependent charged magnetic black brane background.}  
The initial effect of the charge is to push the 2-point correlators further away from equilibrium at small magnetic field $\mathcal{B}=1$. With increasing magnetic field, this is not the case any more. Overall, the background charge has similar effects at $\mathcal{B}\neq 0$ as it had at $\mathcal{B}=0$. 

 \begin{figure}
  \begin{subfigure}[b]{.5\linewidth}
  \begin{center}   
\includegraphics[width=2.9in]{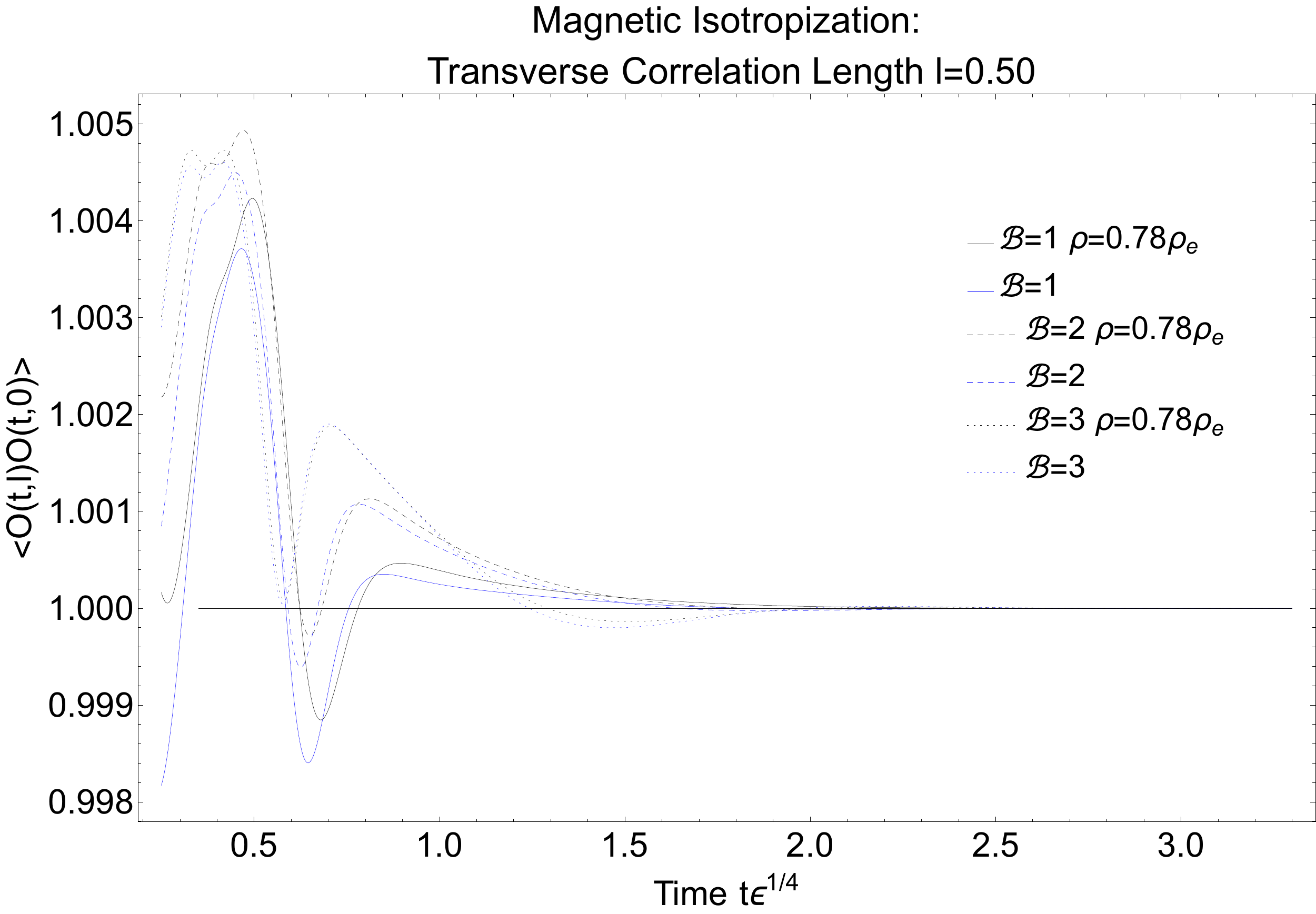}
  \end{center}
\end{subfigure}
   \begin{subfigure}[b]{.5\linewidth}
  \begin{center}   
\includegraphics[width=2.9in]{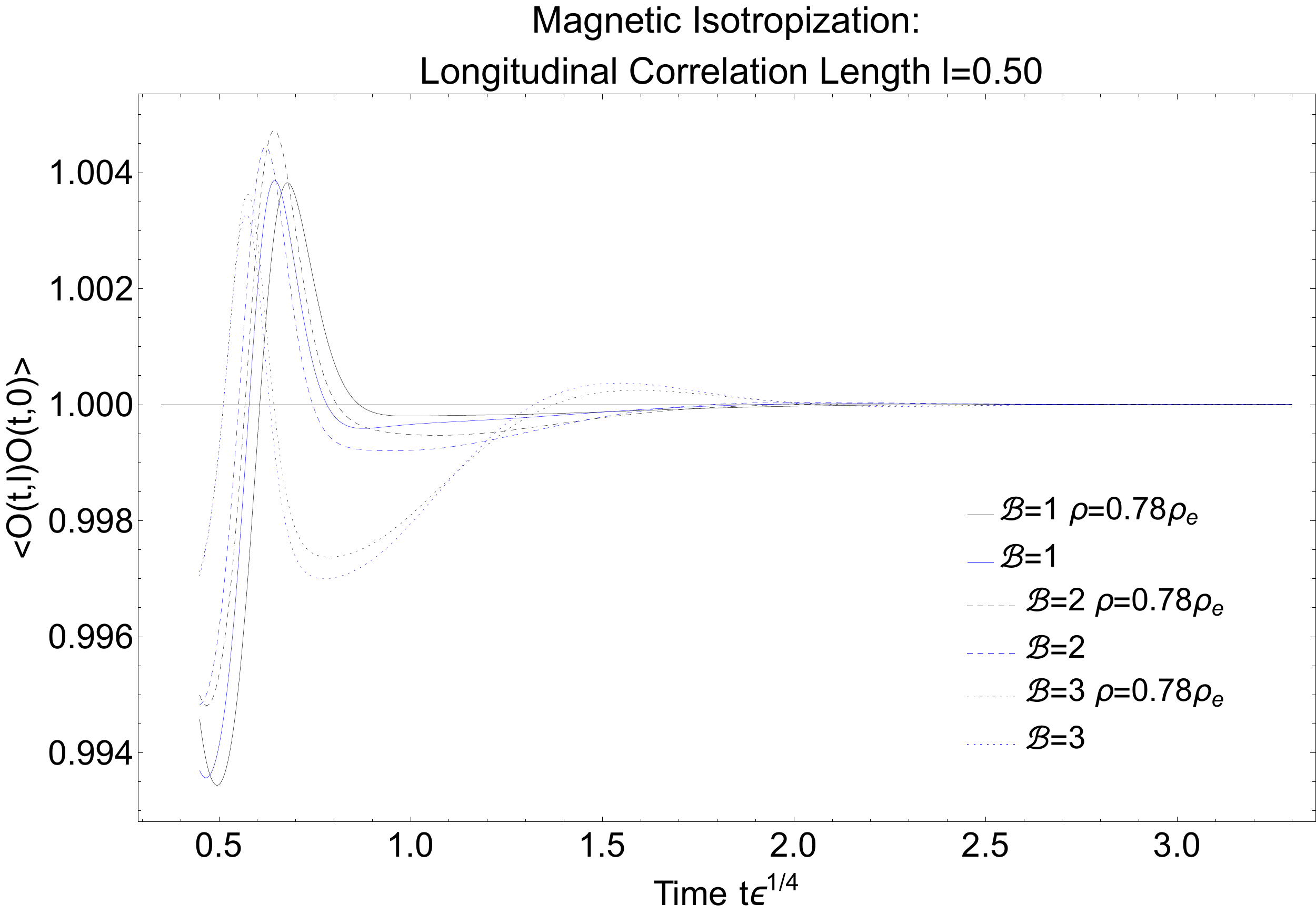}
  \end{center}
\end{subfigure}
  \begin{subfigure}[b]{.5\linewidth}
  \begin{center}   
\includegraphics[width=2.9in]{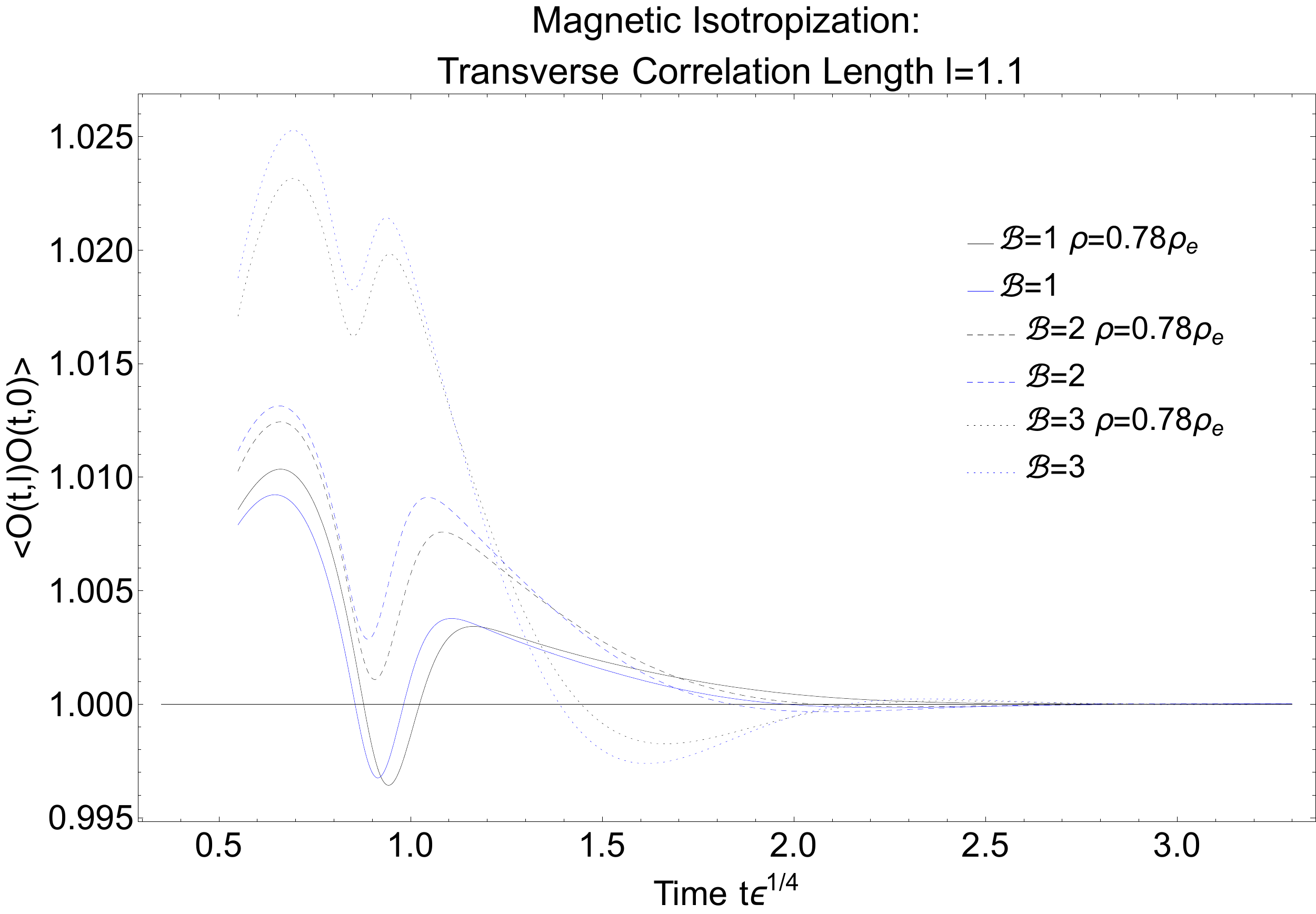}
  \end{center}
\end{subfigure}
  \begin{subfigure}[b]{.5\linewidth}
  \begin{center}   
\includegraphics[width=2.9in]{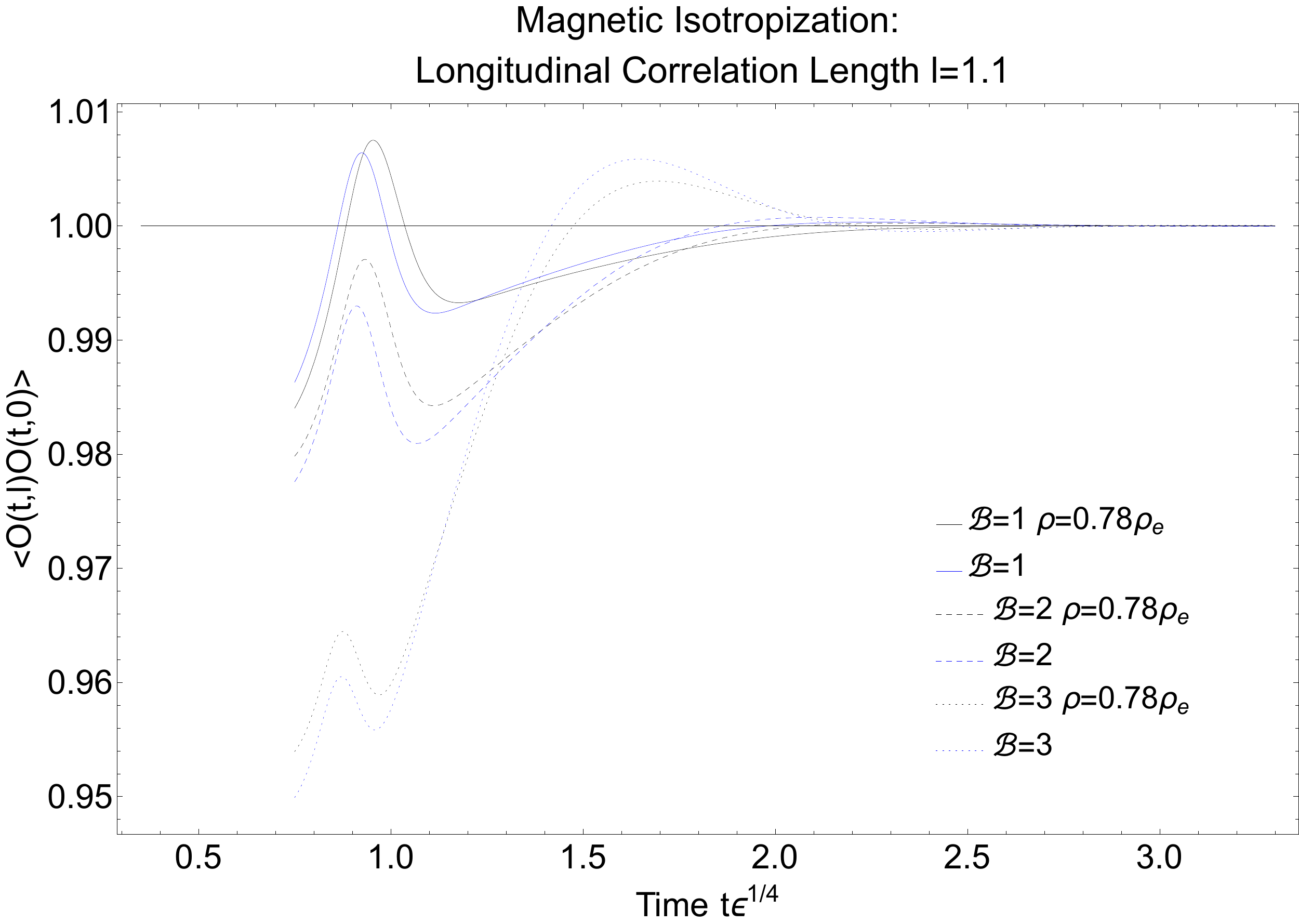}
  \end{center}
\end{subfigure}
\caption{\label{fig:CompareBWithCharge} {\it Separation length dependence at various magnetic fields.} 2-point correlation functions of an (uncharged) scalar operator in a charged magnetic thermalizing fluid (dual to the evolution towards a charged magnetic black brane geometry), $\rho=0.78 \rho_e$, $\mathcal{B}=1,2,3$. \textit{Top Left:} Length separation of $l=0.50$ in the transverse direction. \textit{Top Right:} Length separation of $l=0.50$ in the longitudinal direction. \textit{Bottom Left:} Length separation of $l=1.1$ in the transverse direction. \textit{Bottom Right:} Length separation of $l=1.1$ in the longitudinal direction.  
}
\end{figure}

In Fig.~\ref{fig:CompareBWithCharge}, we compare the cases $\mathcal{B}=1,\, 2,\, 3$ at length separations $l=0.50$ (top row figures) and $l=1.1$ (bottom row figures). We find that for the transverse correlations the amplitude is increased, peaks and other features are shifted to earlier times at increasing magnetic field values when the charge vanishes. However, in all correlators there appears to be a competition between initial anisotropy, magnetic field, and charge (once it is nonzero). Amplitudes and peaks do not change monotonously with the magnetic field anymore. This trend is also observed in the thermalization times, collected in table~\ref{tab:thermalizationTimes}.

\subsection{Comparison of cases}
\label{sec:ComparisonofCases}
\begin{table}[h]
\centering
\begin{tabular}{|l|c!{\vrule width 2pt}c|c|c|}
\hline
 &  $t_{1 pt.}$ &\multicolumn{3}{c|}{ $t_{2 pt.}$} \\
 \hline
case &  & $l=0.5$ & $l=1.1$ & $l=1.5$ \\
\hline
(a) $\rho=0,\,\mathcal{B}=0$ & 0.6869 &0.9624 &1.3942 & 1.6346\\
\hline
(b) $\rho\neq 0,\,\mathcal{B}=0$ & 0.7297 & 0.9917 & 1.4360 & 1.6656 \\
\hline
(c) $\rho=0,\,\mathcal{B}=1$ &0.6815 &1.5425 & 1.9180 & 2.0433 \\
\hline
(c) $\rho=0,\,\mathcal{B}=2$ &0.6403 & 1.5823 & 1.8229 & 1.9345 \\
\hline
(c) $\rho=0,\,\mathcal{B}=3$ &0.5537& 1.2811 & 1.4007 & 1.5108 \\
\hline
(d) $\rho\neq 0,\,\mathcal{B}=1$ &  0.7746 & 1.6526 & 2.3034 & 2.5742 \\
\hline
(d) $\rho\neq 0,\,\mathcal{B}=2$ &0.6803 & 1.6547 & 2.0043 & 2.130 \\
\hline
(d) $\rho\neq 0,\,\mathcal{B}=3$ &  0.5609 & 1.3232 & 1.4555 & 1.5716 \\
\hline
\hline
\end{tabular}
\caption{\label{tab:thermalizationTimes}
\textit{Thermalization times.} The thermalization times of 1-point functions, $t_{1pt.}$, 2-point functions, $t_{2pt.}$, are determined according to the measure defined in Eq.~\eqref{eq:thermal}. When nonzero, the charge density takes on the value $\rho=0.78\rho_{e}$, that is 78\% of the extremal charge density, $\rho_e$, of the Reissner-Nordstr\"om black brane. 
}
\end{table}

\begin{figure}[H]
  \begin{subfigure}[b]{.5\linewidth}
  \begin{center}   
\includegraphics[width=2.9in]{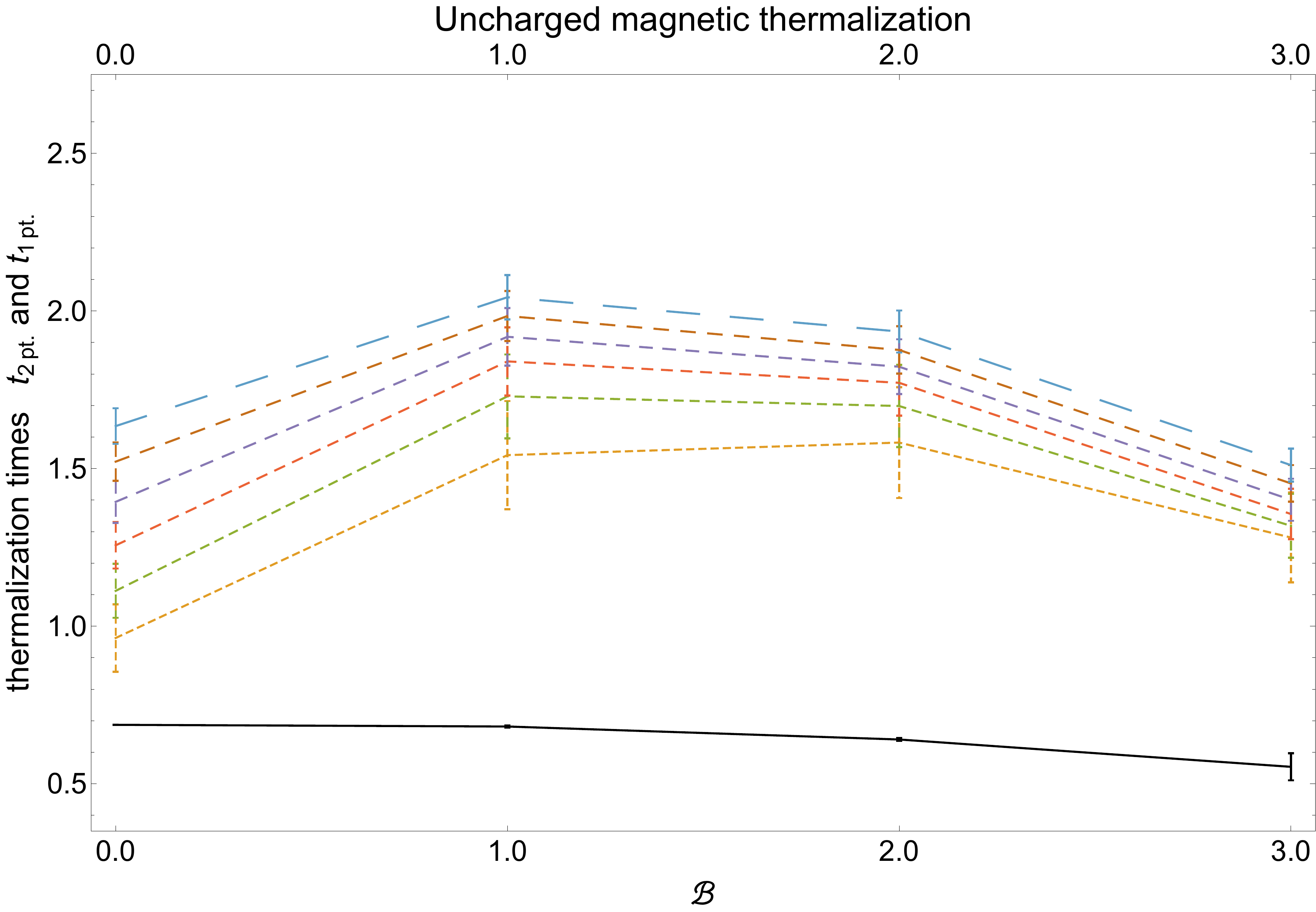}
  \end{center}
\end{subfigure}
 \begin{subfigure}[b]{.5\linewidth}
  \begin{center}   
\includegraphics[width=2.9in]{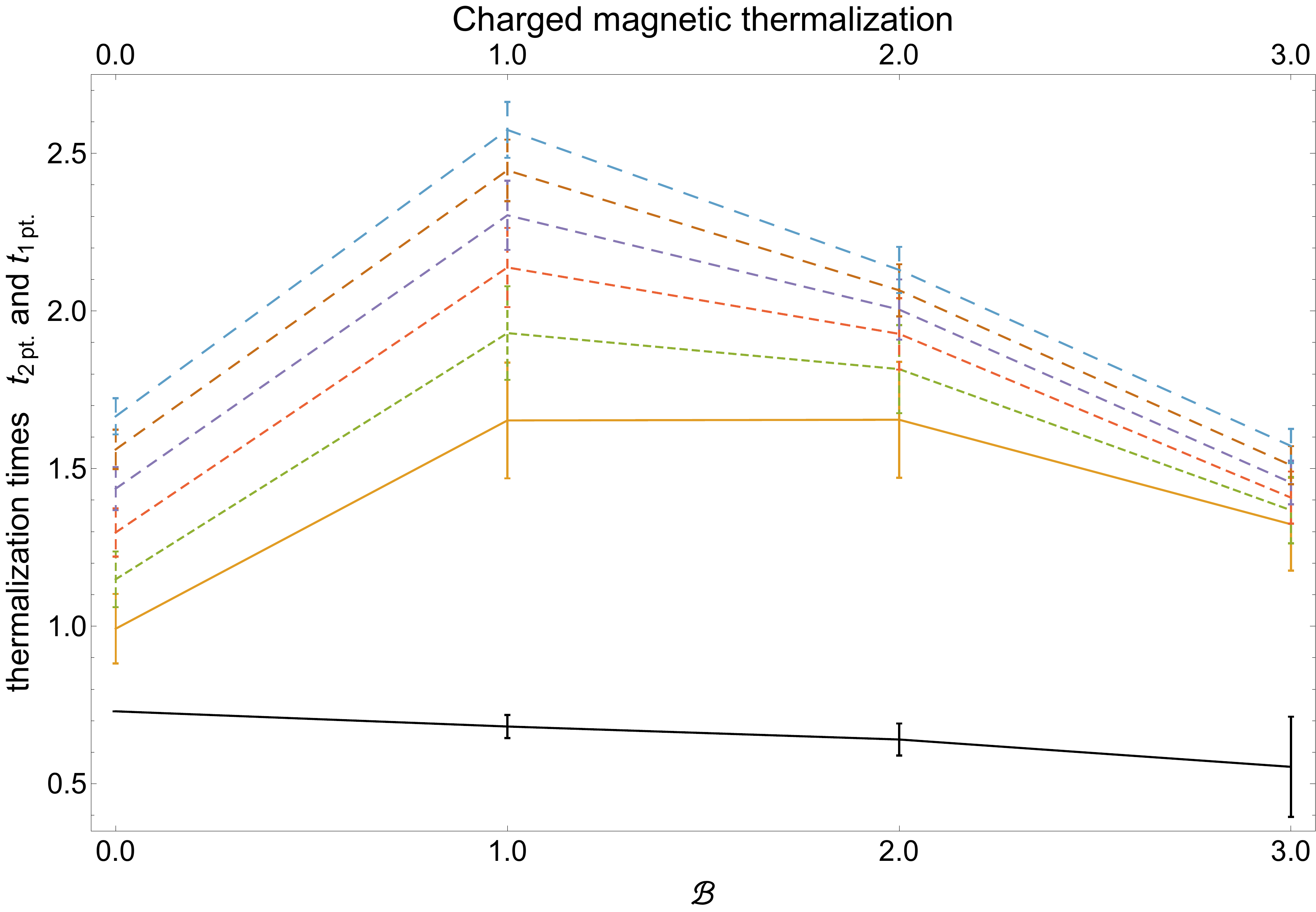}
  \end{center}
\end{subfigure}
\caption{\label{fig:comparingThermalizationTimes} 
{\it Thermalization times as a function of the magnetic field.} Shown here are 2-point function thermalization times $t_{2pt.}$ (dashed lines) and the 1-point function thermalization times $t_{1pt.}$ (solid line) in a thermalizing fluid. The dashed lines from bottom to top correspond to length separations $l=0.5, \, 0.7,\, 0.9, \, 1.1,\, 1.3,\, 1.5$. Error bars indicate an estimate of {our numerical error.} 
\textit{Left:} Uncharged fluid, $\rho=0$, $\mathcal{B}=0,\, 1,\, 2,\, 3$.  \textit{Right:} Charged fluid, $\rho=0.78 \rho_e$, $\mathcal{B}=0,\, 1,\, 2,\, 3$. Data points have been joined by straight lines to guide the eye.}
\end{figure}

\begin{figure}[H]
  \begin{subfigure}[b]{.5\linewidth}
  \begin{center}   
\includegraphics[width=2.9in]{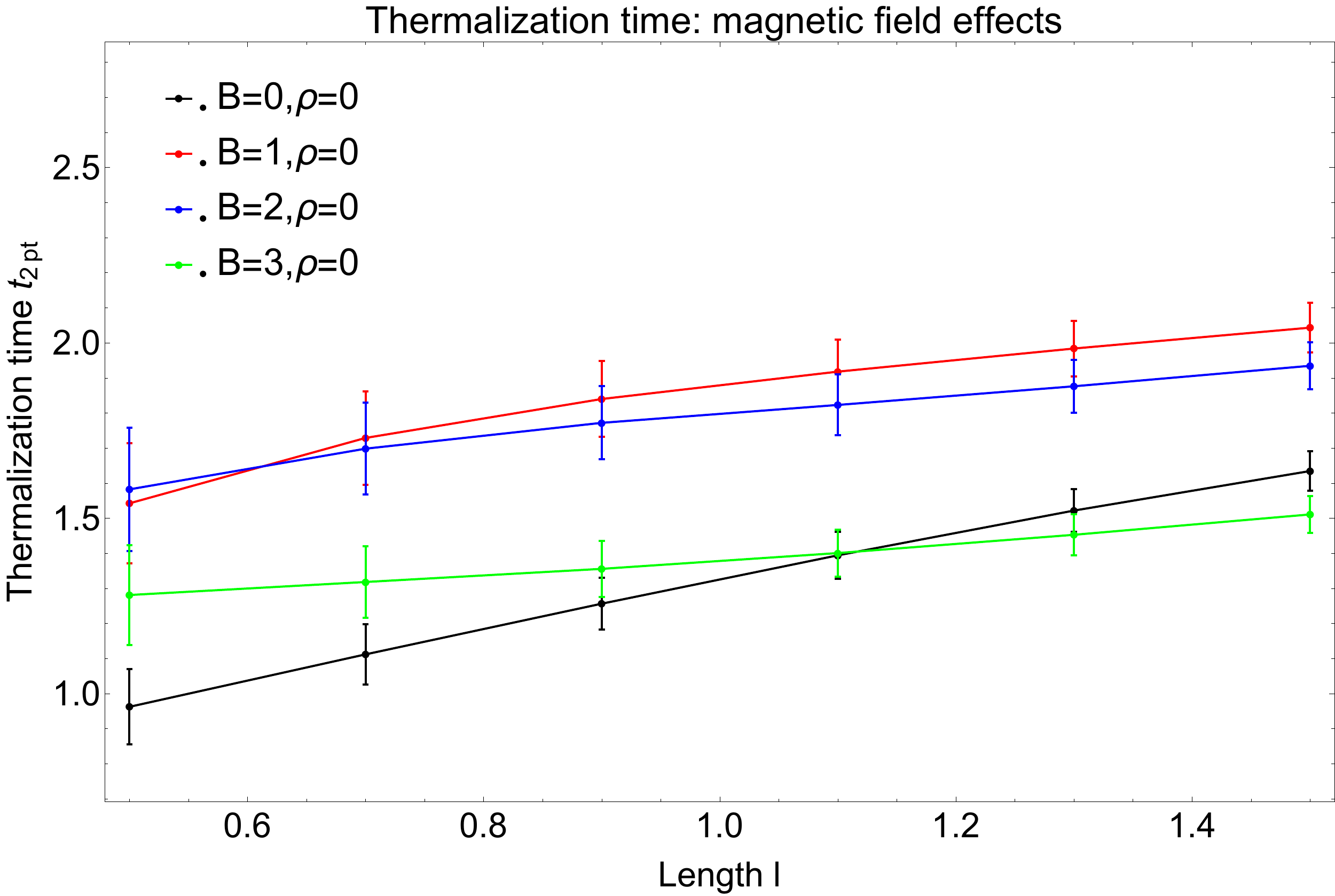}
  \end{center}
\end{subfigure}
 \begin{subfigure}[b]{.5\linewidth}
  \begin{center}   
\includegraphics[width=2.9in]{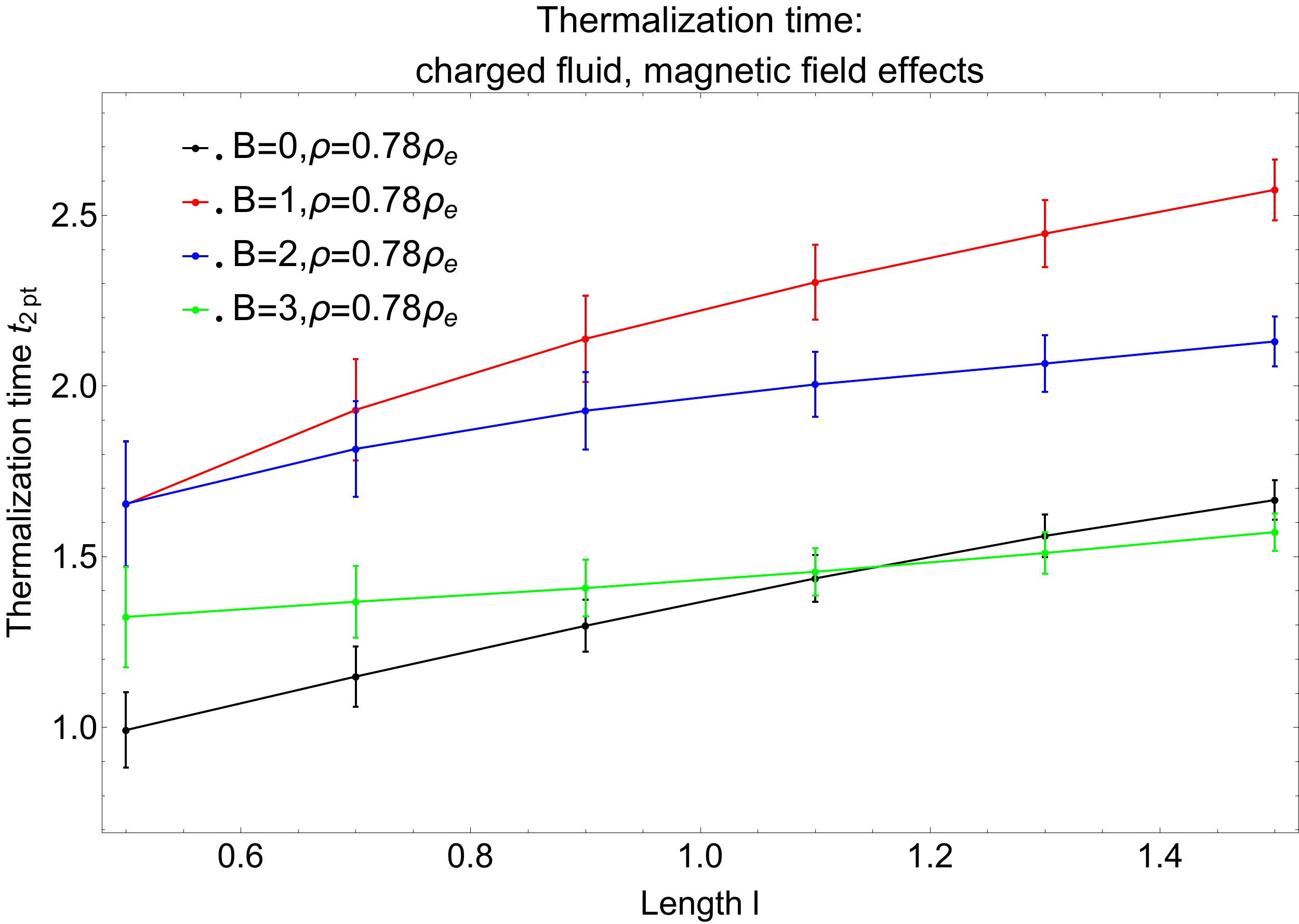}
  \end{center}
\end{subfigure}
\caption{\label{fig:comparingThermalizationTimes_Length_Dependence} 
{\it Thermalization times as a function of the length.} Shown here are 2-point function thermalization times $t_{2pt.}$ in a thermalizing fluid. 
\textit{Left:} Uncharged fluid, $\rho=0$, $\mathcal{B}=0,\, 1,\, 2,\, 3$.  \textit{Right:} Charged fluid, $\rho=0.78 \rho_e$, $\mathcal{B}=0,\, 1,\, 2,\, 3$. Data points have been joined by straight lines to guide the eye.}
\end{figure}

Table~\ref{tab:thermalizationTimes} shows approximate thermalization times which we observe for our four different cases in the 1-point versus the 2-point functions (with length separations $l=0.5,\, 1.1,\,1.5$). This data is also visualized in Fig.~\ref{fig:comparingThermalizationTimes} (including three additional length separations $l=0.7,\, 0.9,\,1.3$) and in Fig.~\ref{fig:comparingThermalizationTimes_Length_Dependence}. In order to provide a meaningful comparison, the initial anisotropy is kept fixed, as discussed in Sec.~\ref{sec:backgrounds}. As mentioned above, our working definition for thermalization time is the time after which the correlator stays within $1\%$ of its equilibrium value according to Eq.~\eqref{eq:thermal}. We include error bars to indicate a rough estimate of confidence in each data point. The geodesic approximation is good provided the dimension of the operator is large. However the results of calculation should get worse as we increase the size of $z_{UV}$ but should get better provided the relative difference between the length l and $z_{UV}$ is large. Hence we propose to use $z_{UV}/(l-z_{UV})$ as a measure of the relative error in the thermalization time at each length scale. Clearly as $l\rightarrow z_{UV}$ our measure is indeterminate and as $l\gg z_{UV}$ $z_{uv}/(l-z_{UV})\rightarrow 0$. 

The overall observation is that the 2-point functions take significantly longer to thermalize than the 1-point functions, see Fig.~\ref{fig:comparingThermalizationTimes}. 
Both, 1-point and 2-point functions show similar behavior as the magnetic field is increased from $\mathcal{B}=0$ to $3$: first the thermalization times increase towards $\mathcal{B}=1$, then decrease (or stay roughly constant) towards $\mathcal{B}=2$, then decrease again towards our largest magnetic field value  $\mathcal{B}=3$. This trend is even more pronounced in the charged backgrounds, see right plot in Fig.~\ref{fig:comparingThermalizationTimes}. 
At vanishing magnetic field, different length separations thermalize at very different times. For example, $l=0.5$ thermalizes at $t_{2pt.}=0.9624$, while the greater separation $l=1.5$ thermalizes much later at $t_{2pt.}=1.6346$. However, when the magnetic field is large, this behavior drastically changes. At large $\mathcal{B}=3$, all length separations thermalize after approximately the same time $t_{2pt.}\approx 1.4$. There appears to be a kind of saturation effect, at large magnetic fields. The large magnetic field scale seems to dominate the thermalization of the 2-point functions and renders the differences in length separations irrelevant. This saturation effect at large magnetic field becomes even more apparent in Fig.~\ref{fig:comparingThermalizationTimes_Length_Dependence}, where the green curve ($\mathcal{B}=3$) has a small slope as a function of length, indicating that the 2-point function thermalization times are virtually independent from the length separation between the operators. In all of the investigated cases we find thermalization to occur first at shorter length scales, then later at longer length scales (top-down thermalization). In none of our cases we observe a bottom-up thermalization~\cite{Baier:2000sb}.  

It is interesting to note the competition among scales displayed in our data. This competition shows between charged versus uncharged cases when we consider fixed magnetic field values. For example, at $\mathcal{B}=0$ we find that the presence of charge increases the thermalization time in the 1-point function from  $0.6869$ to $0.7297$. The 2-point functions show this same trend, as seen in table~\ref{tab:thermalizationTimes}. However, at a magnetic field value of $\mathcal{B}=1$ we find that the presence of charge increases significantly the thermalization time in the 1-point function from $0.6815$ to $0.7746$. This trend also manifests itself in the corresponding 2-point functions, in particular for $l=1.5$ the thermalization of the 2-point functions change from $2.0433$ to $2.5742$. Another example for the competition between the scales set by the initial anisotropy, the magnetic field, and the charge density, are the thermalization times for uncharged fluid at $\mathcal{B}=1,\, 2$. Considering $l=0.5$, we observe that the increased magnetic field leads to a slight increase in the thermalization time from $1.5425$ to $1.5823$. This is the behavior most of the correlators show. However, when considering $l=1.1$ and $l=1.5$, thermalization times are decreased when going from $\mathcal{B}=1$ to $2$. This effect of decreasing thermalization time as a function of the magnetic field at long length scales is even more dramatic in the pressence of charge. Another example for the competition of scales seems visible in Fig.~\ref{fig:comparingThermalizationTimes_Length_Dependence}, where 2-point thermalization time is a linear function of length separation at vanishing magnetic field. At intermediate magnetic field strength, $\mathcal{B}=1-2$, the functional dependence shows nonlinear behavior, i.e. curvature of the graphs. At large magnetic field, $\mathcal{B}=3$, there seems to be a saturation happening such that the thermalization time is becoming independent of the length separation, as already mentioned above.

\section{Discussion \& Summary}\label{sec:discussion}
\begin{figure}[H]
  \begin{center}   
\includegraphics[width=3.5in]{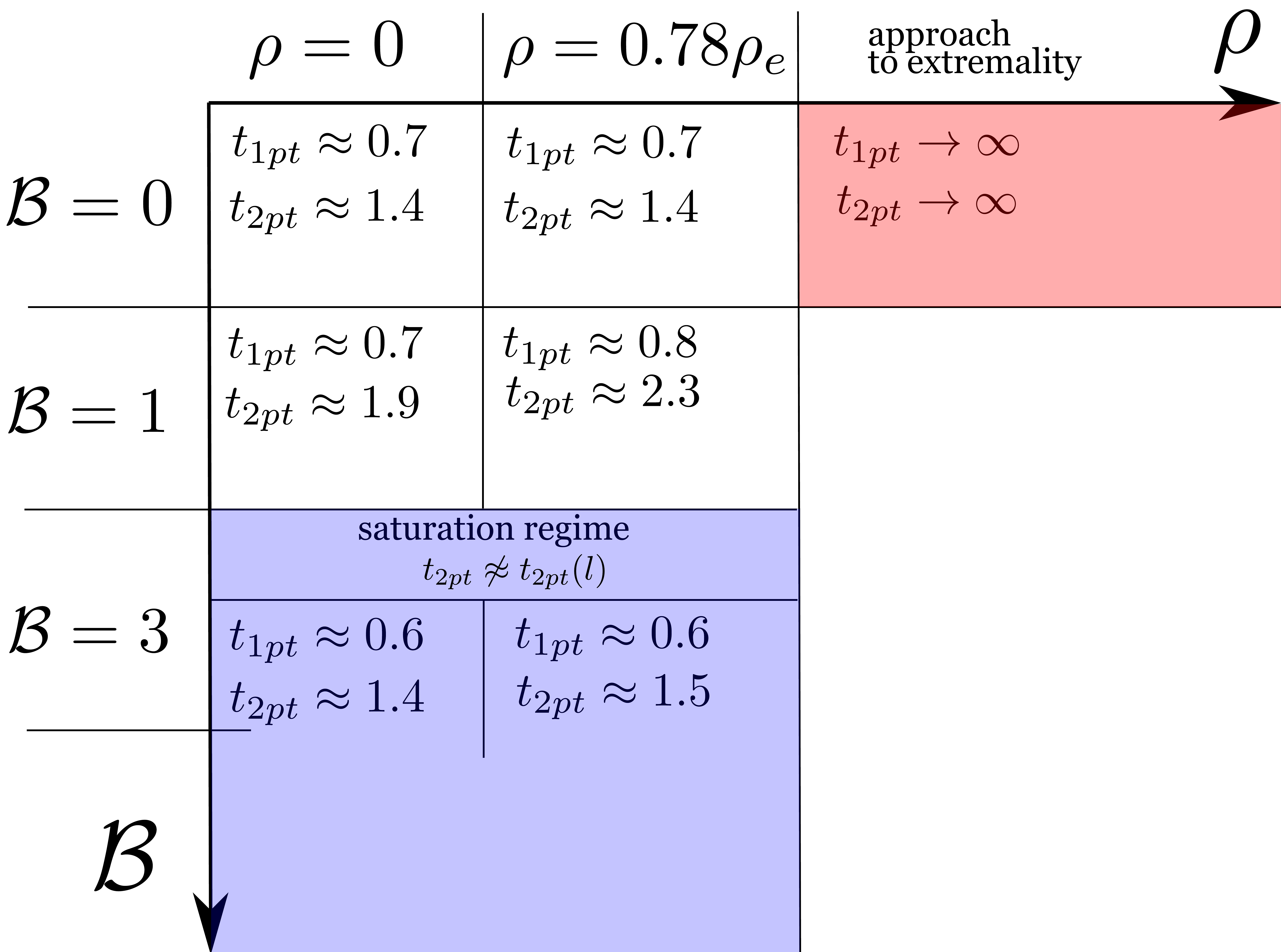}
  \end{center}
\caption{\label{fig:Phase_Portrait} 
 A ``phase portrait'' of the thermalization times of both the 1-point function and 2-point functions. We display this portrait at $l=1.1$. The actual dependence of the thermalization time on the length scale and magnetic field can be seen in Fig.~\ref{fig:comparingThermalizationTimes} and Fig.~\ref{fig:comparingThermalizationTimes_Length_Dependence}. 
 }
\end{figure}

In this work, we have studied the thermalization behavior of a thermal fluid, initially prepared in an anisotropic state far from equilibrium. We work within $\mathcal{N}=4$ Super-Yang-Mills theory at strong coupling and in the presence of charge density, $\rho$, as well as a strong external magnetic field, $\mathcal{B}$. Most notably, we have calculated 2-point functions as non-local probes. We compare their behavior over time to the behavior of 1-point functions as local probes. 2-point functions with transverse or longitudinal separation length $l$, or time separation $\Delta t$ have been studied in aforementioned cases illustrated in Fig.~\ref{fig:introFigureSetup}: (a)~$\rho=0,\,\mathcal{B}=0$, (b)~$\rho\neq 0,\,\mathcal{B}=0$, (c)~$\rho=0,\,\mathcal{B}\neq 0$, (d)~$\rho\neq 0,\,\mathcal{B}\neq 0$. The results are summarized in Fig.~\ref{fig:BBUnchargedProbe}, \ref{fig:RNUnchargedprobe}, \ref{fig:MagneticCase}, \ref{fig:ChargedMagneticCase}, respectively. As a new equilibrium result, we compute the charged magnetic black brane solutions to Einstein-Maxwell theory in equilibrium; see table~\ref{tab:lateTimeVsEquilibrium2} and Fig.~\ref{fig:Physically_Distinct_Solutions}.

A measure for thermalization has been defined in Eq.~\eqref{eq:thermal}, and thermalization times are determined for 2-point functions versus 1-point functions. The general trend is summarized in Fig.~\ref{fig:Phase_Portrait} for the various cases introduced in Fig.~\ref{fig:introFigureSetup}. See table~\ref{tab:thermalizationTimes}. The same data is visually summarized in Fig.~\ref{fig:comparingThermalizationTimes}. From those 
plots it is apparent that thermalization times of 2-point functions at length separations $l\ge 0.5$ are signficantly larger than those of the corresponding 1-point functions. This effect is enhanced by the magnetic field. As expected, the difference between 2-point and 1-point function thermalization times decreases with decreasing length separation $l\to 0$, because then the 2-point functions effectively become 1-point functions, probing the system locally. This behavior is observed for vanishing and small magnetic fields in Fig.~\ref{fig:comparingThermalizationTimes}. More striking, though, is the non-monotonous behavior of thermalization times with increasing magnetic field. All cases studied in this work are displayed in Fig.~\ref{fig:comparingThermalizationTimes}, showing an initial increase of thermalization time (from $\mathcal{B}=0$ to 1), then a decrease (from $\mathcal{B}=1$ to 2), and a final decrease (towards $\mathcal{B}=3$). We interpret this to indicate a competition of scales in the system. In the left plot of Fig.~\ref{fig:comparingThermalizationTimes}, at vanishing charge density, the two scales competing are the scale set by the initial anisotropy on one hand, and the magnetic field on the other. Comparing to the right side plot, the presence of charge density appears to enhance the increases and decreases of thermalization time in each regime. Remarkably, at large magnetic field values $\mathcal{B}\to 3$, all cases studied show that 2-point functions thermalize at approximately the same {\it saturation} time scale $t_{saturation}\approx 1.4$, regardless of their length separation $l$. In stark contrast to that, the 1-point functions thermalize much earlier at times earlier than $t_{1pt.}\approx 1$. Our interpretation of this apparent {\it saturation} is a combination of two effects 1) The appearance of attractor like solutions when the magnetic field is applied, 2) the dominance of the magnetic field scale as compared to all other scales for $\epsilon_B^{1/2}\ll B$. 

The main effect of the magnetic field is to push the 2-point correlator curves up above or down below the equilibrium value 1, see Fig.~\ref{fig:CompareB}. Note that features, such as peaks, in the curves are not shifted considerably along the time axis. At increasing magnetic field values, also the peak to peak amplitude of each curve increases. We interpret this to signal that a large magnetic field pushes the system far away from equilibrium at early times. While each of our systems is prepared in an anisotropic initial state, that initial anisotropy, for increasing magnetic field, is further and further away from the equilibrium anisotropy generated by the magnetic field at late times. However, although further from equilibrium, our measure (Eq.~\eqref{eq:thermal}) reveals it takes the system a shorter amount of time to reach the equilibrium state. This trend shows in the decreasing thermalization times for 1- and 2-point functions in table~\ref{tab:thermalizationTimes}.

The main effect of the charge density is to push all features, such as peaks or nodes, in the 2-point functions to slightly (roughly $1\% -25$\%) later times. This is also true for the 2-point function thermalization times, $t_{2pt.}$, which are always greater when the charge density is nonzero (compared to the same case at zero charge density). For example, table~\ref{tab:thermalizationTimes} reveals that the 2-point thermalization time at $\rho=0,\,\mathcal{B}=1$ for length separation $l=1.1$ is $t_{2pt.}=1.918$ (in units of energy density $\epsilon^{1/4}$), and increases to $2.3034$ as the charge density is increased to $\rho=0.78\rho_e$. This same statement of charge density pushing features to later times and increasing thermalization times is also true for all 1-point functions we have probed as well. 
Approaching the extremal charge value in absence of a magnetic field, the 2-point functions appear to thermalize at infinitely late times, as shown in Fig.~\ref{fig:nearExtremalLS}. Similarly for the 1-point functions. It would be interesting to study this regime further, but our current methods are limited to the charge densities displayed.

We find nodes in equal-time correlators, see {e.g.} the two bottom graphs in Fig.~\ref{fig:BBUnchargedProbe}. The location of these nodes changes with charge density and magnetic field, as seen {e.g.} in the bottom graphs in Fig.~\ref{fig:ChargedUncharged}. As argued in Sec.~\ref{sec:results}, we interpret this to indicate that there are strong (possibly non-local) medium effects in the fluid at early times, which seem to vanish at late times. More 1- and 2-point function features like these are discussed in detail in Sec.~\ref{sec:results}. 
 
Clearly, correlation functions are strongly affected by the presence of a magnetic field, as illustrated by Fig.~\ref{fig:CompareB} and Fig.~\ref{fig:CompareBWithCharge}. They are also showing qualitatively new features, even though the fluid itself (the dual gravitational background metric) is only mildly affected, e.g. by a 20 \% increase of the thermalization time compared to the case without magnetic field. One of these qualitatively new features are points which we will call {\it attractor points}, appearing in the correlators, see the top two plots of Fig.~\ref{fig:MagneticCase} at a time $t\epsilon^{1/4}\approx 2.0$ at which the 2-point functions attract to the same behavior independent of their length separation $l$.  These attractor points resemble but seem distinct from the nodes appearing in non-equal time correlators, discussed in Sec.~\ref{sec:case(a)}. 

The emerging universality in the thermalization time at large magnetic fields is interesting when considered in the context of the new field theoretic formulation of magnetohydrodynamics (MHD)~\cite{Grozdanov:2016tdf,Armas:2018atq,Armas:2018zbe}. In~\cite{Grozdanov:2016tdf} the authors studied the limit of large magnetic field. A symmetry enhancement from $SO(2) \to SO(2) \times SO(1,1)$ occurs, with the new $SO(1,1)$ reflecting boost invariance along the magnetic flux lines. In a holographic model, this limit was shown to imply non-dissipative physics and all first-order transport coefficients vanish~\cite{Grozdanov:2017kyl}; see also~\cite{Glorioso:2018kcp} for a follow-up study of the large magnetic field regime. We plan to follow up on this in future work with fully dynamic relaxations to the limiting large magnetic field solutions.

We have also studied 2-point functions of a {\it charged} scalar operator. At vanishing magnetic field, the background charge is uniformly distributed in the fluid at all times, hence there is no net effect on the 2-point function of a charged operator compared to that of an uncharged operator. However, there is an effect at nonzero magnetic field, the result is shown in Fig.~\ref{fig:ChargedOP}. For larger charge to mass ratios of the operator, features in the 2-point function are shifted to slightly earlier times and also thermalization appears to occur faster, (see table~\ref{tab:Charged_Operators}).
We also studied examples of initial anisotropy pulses with smaller amplitude. The overall effect is a general change in the slope of the thermalization time as a function of length.

Let us have a look out into future directions. In this work we have restricted our investigations to Einstein-Maxwell theory. However, when embedded consistently in type II B supergravity or superstring theory, then one is lead to include a Chern-Simons term with non-vanishing coupling. In the dual field theory, this introduces a chiral anomaly, which is known to affect the time-evolution near equilibrium considerably~\cite{Erdmenger:2008rm,Banerjee:2008th,Son:2009tf}. The effect of the Chern-Simons term far from equilibrium should reveal interesting behavior; for previous work in this direction see e.g.~\cite{Ammon:2016fru,Haack:2018ztx}. 
Our data indicates that our 1-point functions smoothly connect in their thermalization behavior to the 2-point functions at small length separation $l\to 0$. This is non-trivial,one might even say very surprising, because we compare two very different operators. The geodesic approximation forces us to consider 2-point functions of a scalar operator, whereas the 1-point functions we consider are those of a 2-tensor, namely the energy-momentum tensor $T^{\mu\nu}$ of the field theory. Therefore, it is going to be elucidating to actually calculate 2-point functions of the energy-momentum tensor with itself. For this purpose, a different method can be used.  Such correlation functions can be obtained as 1-point functions but in presence of a $\delta$-source~\cite{Wondrak:2017kgp,Ishii:2016wwa,Banerjee:2016ray,Wondrak:2019}. 
When developing an understanding of the field theory side, the comparison to effective descriptions would be useful. Phenomena such as prescaling and far from equilibrium hydrodynamics~\cite{Mazeliauskas:2018yef} could be searched for; the applicability of fluid dynamics far from equilibrium~\cite{Romatschke:2017vte} could be tested within our holographic model. 
It would furthermore be elucidating to consider different non-local observables, such as holographic entanglement entropy~\cite{Ryu:2006bv,Balasubramanian:2011ur}. One interesting possibility to be explored are geodesics/minimal surfaces penetrating the apparent horizon. We would need to know the full extended casual diagram of the isotropization process. If it is the case that geodesics which reach beyond the apparent horizon connect two asymptotically distinct AdS boundaries in our setup then those geodesics can be interpreted as field theory correlators in the real-time formalism~\cite{Aparicio:2011zy,Herzog:2002pc,Maldacena:2001kr}. 
Our current model may be pushed near/through a phase transition, see~\cite{Attems:2018gou} for such a study within a holographic model. This could provide a toy model for the beam energy scan~\cite{BEST} or similar experiments. It would be interesting to investigate the expected critical slowing down near the transition in real time. As a candidate transition one may mention the transition to the helical phase~\cite{Ammon:2016szz} driven by the Nakamura-Ooguri-Park instability~\cite{PhysRevD.81.044018}. However, this requires first to include the Chern-Simons term, mentioned above, into our system. 
As mentioned a few times throughout the text, holographic studies of dynamical electromagnetic fields on the boundary recently became feasible~\cite{Grozdanov:2017kyl,Grozdanov:2018fic} along with a modern view of magnetohydrodynamics~\cite{Grozdanov:2016tdf}. Extending our time-dependent setup to include dynamical electromagnetic fields on the boundary would be very interesting, as it will have implications for subjects such as the dynamical evolution of magnetic fields in heavy ion collisions (as well as plasma physics and cosmology).

Lastly, it remains ever tempting to try an analytically solvable formulation of holographic thermalization problems along the lines of~\cite{Horowitz:2013mia,Poole:2018koa}.

\acknowledgments
The authors thank Wilke van der Schee for very helpful discussions at the beginning of this project and for very useful comments on the manuscript, for which we also thank Saso Grozdanov. CC thanks Jackson Wu for many helpful discussions. MK thanks John Fuini and Laurence Yaffe for many helpful discussions, as well as the organizers and participants of the conference {\it HoloQuark2018} in Santiago de Compostela, Spain, for feedback on preliminary results of this work. This work was supported, in part, by the U. S. Department of Energy grant DE-SC-0012447.

\appendix

\section{Numerical accuracy and convergence}
\label{sec:appendixConvergence}

\begin{figure}[h]
\begin{subfigure}[b]{.49\linewidth}
\includegraphics[width=2.9in]{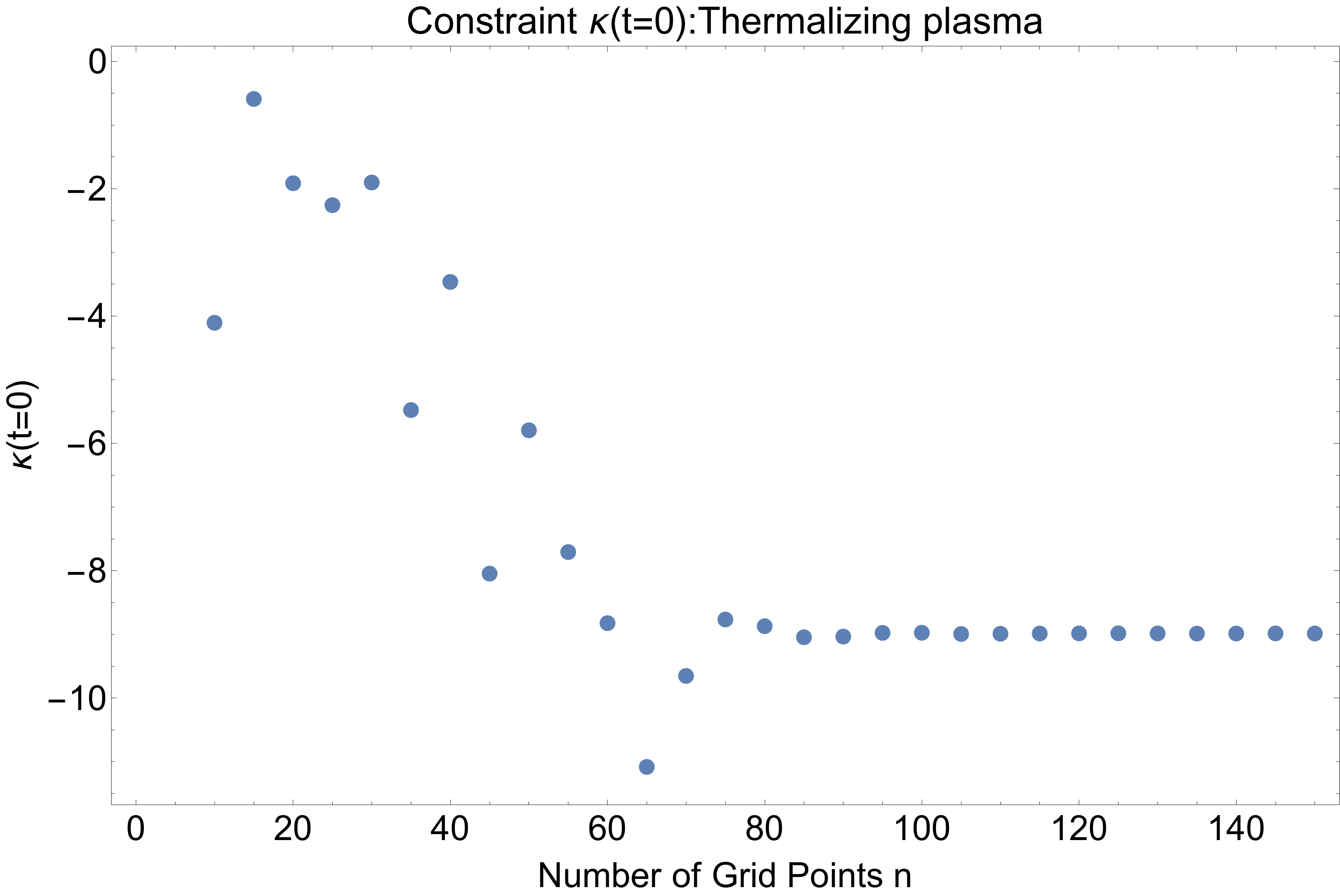}
\end{subfigure}
    \begin{subfigure}[b]{.49\linewidth}
\includegraphics[width=2.9in]{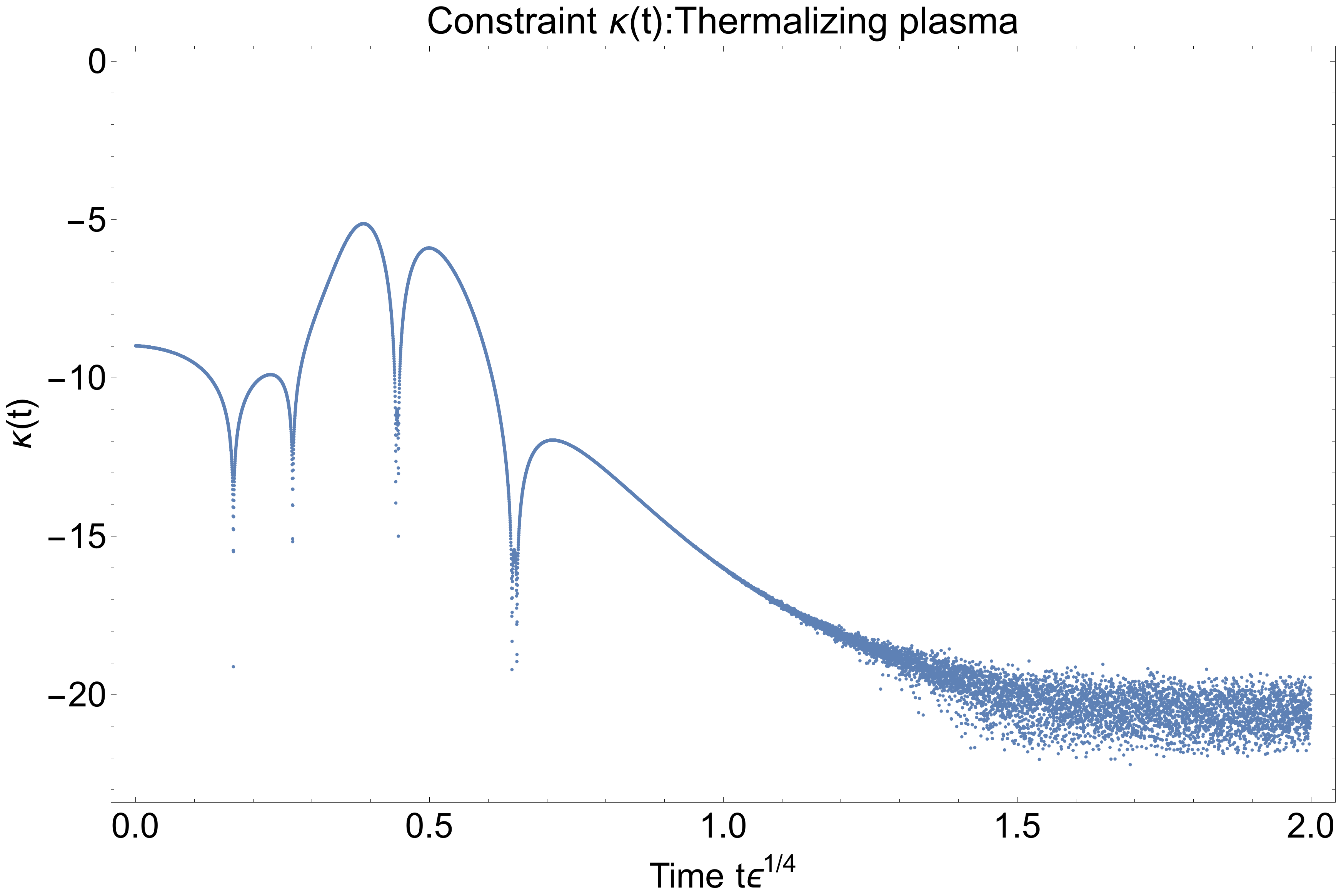}
\end{subfigure}

\begin{subfigure}[b]{.49\linewidth}
\includegraphics[width=2.9in]{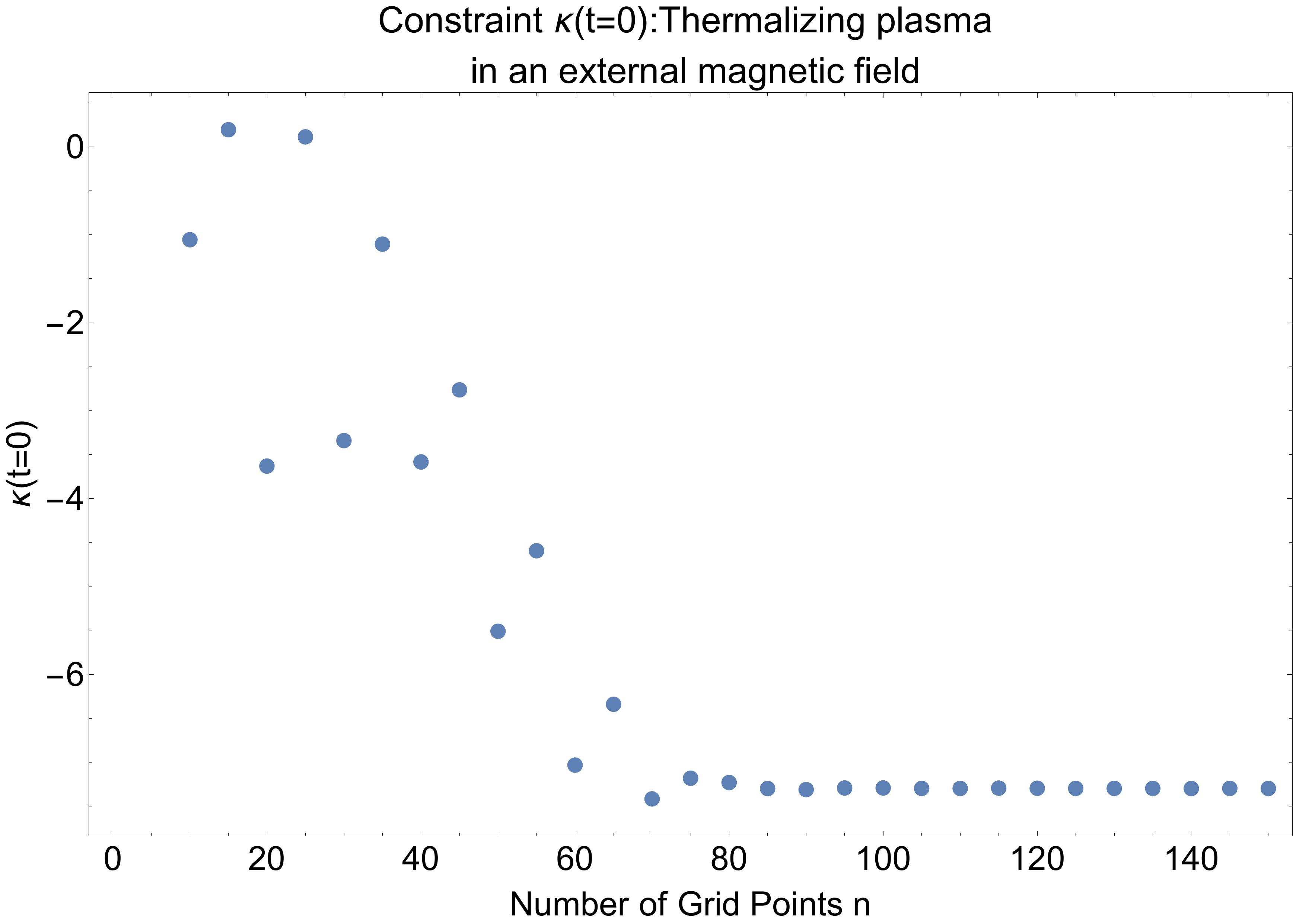}
\end{subfigure}
\begin{subfigure}[b]{.49\linewidth}
\includegraphics[width=2.9in]{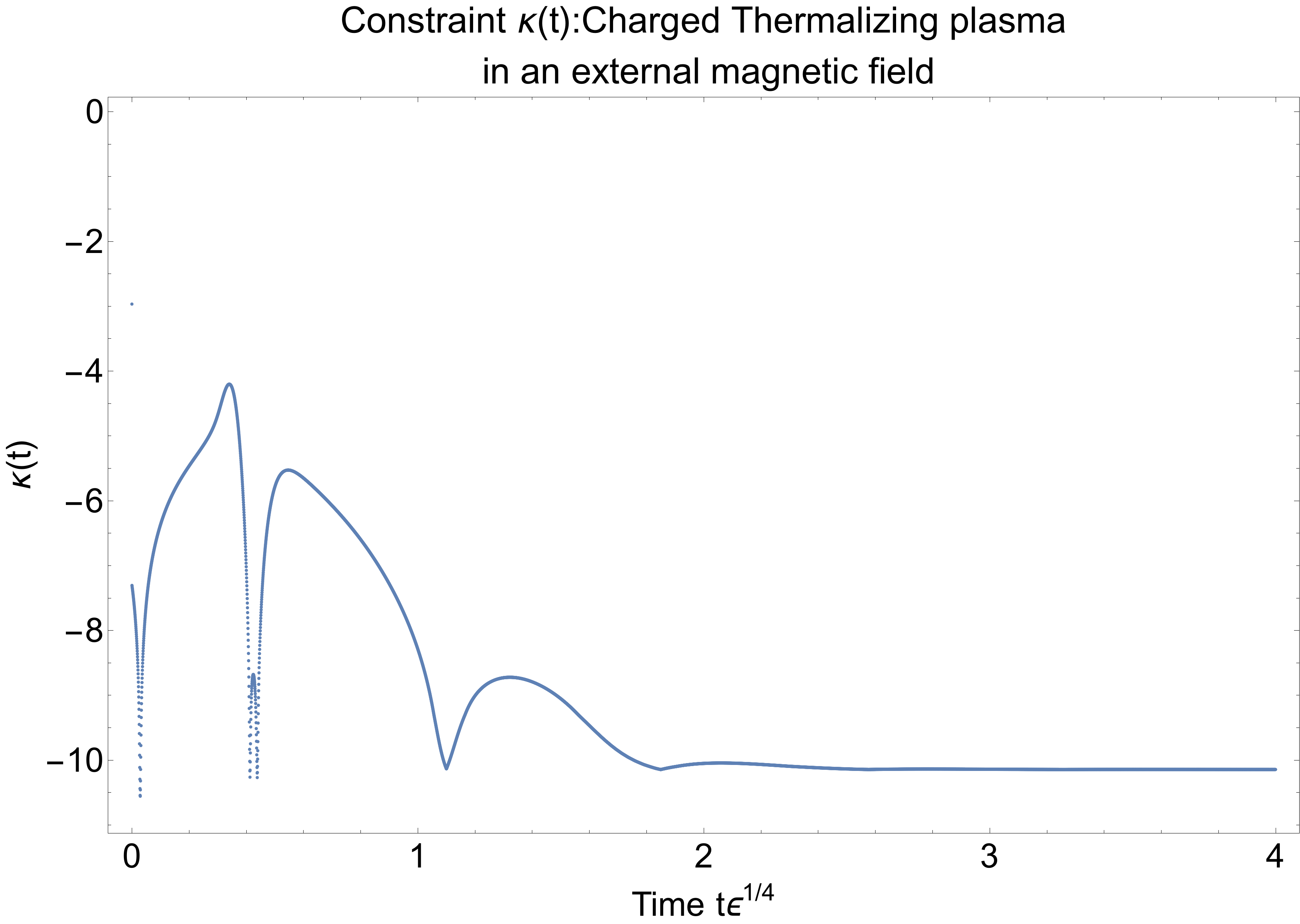}
\end{subfigure}
    \caption{
    \label{fig:neutralConstraint}Constraint violation measured by $\kappa(t)$ for the generation of the anisotropic geometries used in the calculation of geodesics. We display on the left the constraint on the initial time slice and on the right the value of the constraint throughout the evolution of the geometry. \textit{Top left:}$\rho=0,\mathcal{B}=0$. 
    \textit{Top right:}$\rho=0,\mathcal{B}=0$.
    \textit{Bottom left:}$\rho=0.78\rho_e,\mathcal{B}=3$..
    \textit{Bottom right:}$\rho=0.78\rho_e,\mathcal{B}=3$. }
\end{figure}

\begin{figure}[h]
    \begin{subfigure}[b]{.49\linewidth}
\includegraphics[width=2.9in]{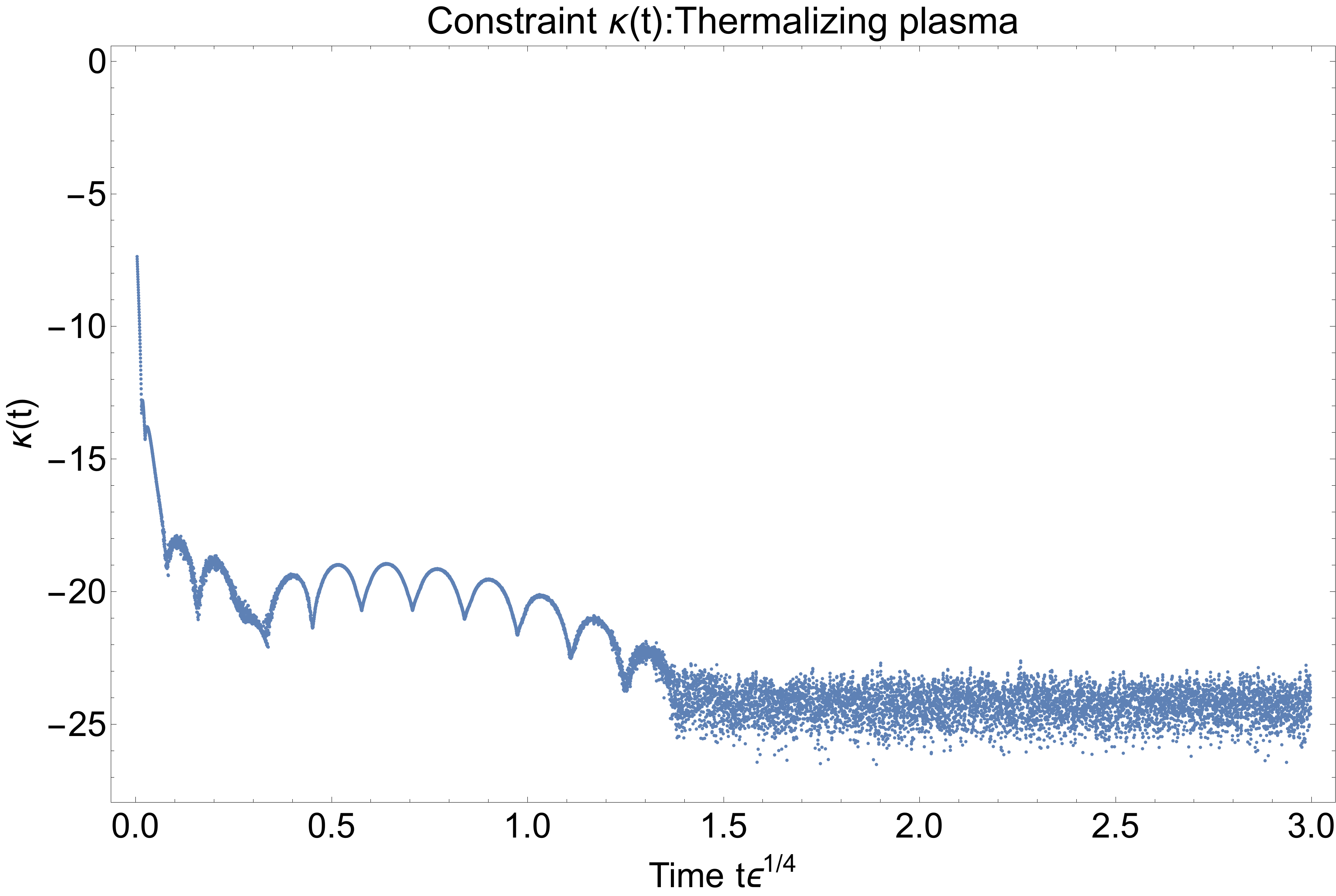}
\end{subfigure}
\begin{subfigure}[b]{.49\linewidth}
\includegraphics[width=2.9in]{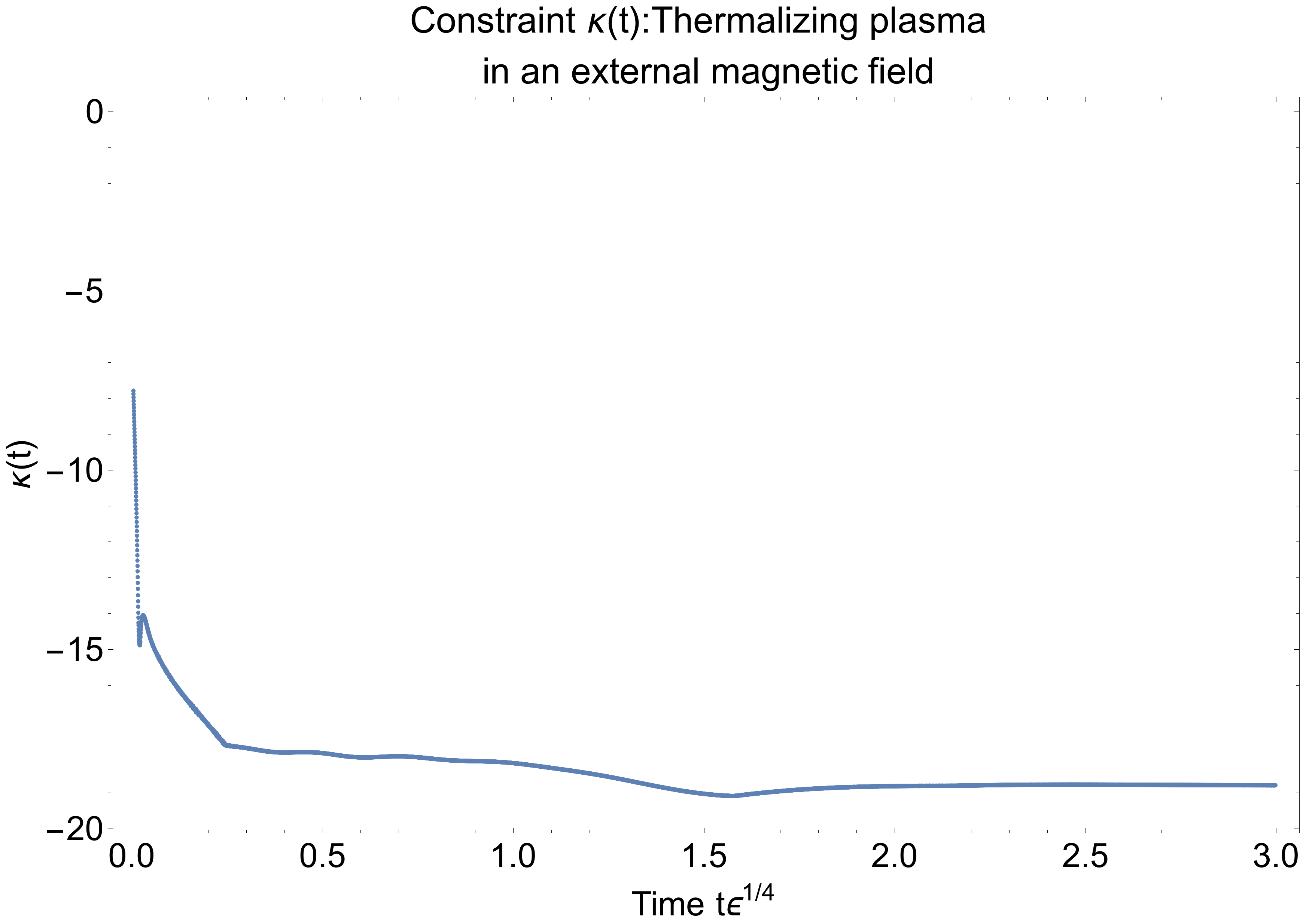}
\end{subfigure}
    \caption{
    \label{fig:Constraint_Constant_B_S}Constraint violation measured by $\kappa(t)$ for a constant initial anisotropy $B_s=\frac{8}{3}\epsilon_L$ \textit{Left:}$\rho=0,\mathcal{B}=0$.
    \textit{Right:}$\rho=0,\mathcal{B}=1$. }
\end{figure}

We provide a number of checks on our numerics. Our code to generate solutions to the Einstein equations monitors the residual value of an unused Einstein equation (~\eref{einsdotdot}),
\begin{equation}
    \kappa=\log{\left(\max_r \left. \left(\ddot{S}(v,r)- \frac{1}{2} A'(v,r) \dot{S}(v,r)+\frac{1}{2} \dot{B}(v,r)^2S(v,r))\right)\right|_{t}\right)}\label{eq:Constraint}
\end{equation}
The results of monitoring~\eref{eq:Constraint} are displayed in Fig.~\ref{fig:neutralConstraint}. We do not include the point at the boundary or the horizon in our constraint calculation. It is important to note that the residual displays information about the choice of initial anisotropy. This initial choice leads to different geometric evolution and the constraint is a reflection of the propagation of the different initial data. As an example consider instead a constant profile for the initial anisotropy $B_s=\frac{8}{3}\epsilon_L$ (see ~\cite{Heller:2013oxa} for comparison). We see that the residual of the unused Einstein equation quickly move to very small values and exhibit a series of oscillations before reaching a ``convergence floor'' where the value of the residual fluctuates rapidly at small values (see Fig.~\ref{fig:Constraint_Constant_B_S} left image). The pressence of the external magnetic field has additional effects on the constraint. In Fig.~\ref{fig:Constraint_Constant_B_S} (right image) we can see two effects, the reduction of the frequency of oscillations and the apparent disappearance of the ``convergence floor''. We see similar behavior in the constraint when working with the profile given in~\eref{eq:init_Profile}. In general what we see is that our code is well behaved, especially for constant initial anisotropy, and we have truly pushed our code to the limit in order to work with profiles of initial anisotropy as close as possible to those chosen in~\cite{Fuini:2015hba} in order to maintain a sense of continuity between their work and ours.

Our relaxation scheme depends on an ultra violet cutoff and the number of grid points chosen to subdivide the non-affine parameter $\sigma$. We display in table~\ref{tab:NeutralCut} and table~\ref{tab:CharMagCut} the results of the variation of the ultra-violet cutoff within a range of $z_{uv}\in[0.01,0.1]$ for two of the cases displayed in Sec.~\ref{sec:case(a)} and Sec.~\ref{sec:magneticBlackBraneResults}. We find slight cutoff variation in the data at earlier times in the evolution of the background. This is due to the large variation of the metric during this window of time. We have also worked with smaller, by an order magnitude anisotropy pulse ($\beta=.15$), finding very small $\mathcal{O}(10^{-6}$) cutoff dependence in the case of an isotropization towards a Schwarzschild metric. However we found this small amplitude difficult to work with in initial iterations of our code when both magnetic field and charges were present. This difficulty manifested in correlations which never reached equilibrium. This issue was alleviated by introducing larger ($\beta=1.5$) initial anisotropy. 

In table~\ref{tab:Grid_data} and table~\ref{tab:Grid_data_B_rho} we display the results of the variation of the number of grid points $N$ in the relaxation scheme for $N\in{50,100,175,250,500}$. We find that $N=175$ is sufficient for our work. We choose to work with 250 as a precaution as we found that the time difference to calculate with 250 versus 175 to be reasonably small.

\begin{table}[H]
    \centering
    \begin{tabular}{c|c|c|c|c|c|c|c}
    \hline 
 $z_{uv}$ & $t=0.7$ & $t=0.8875$ & $t=1.075$ & $t=1.2625$ & $t=1.45$ & $t=1.6375$ & $t=1.825$   \\
 \hline
0.01 & 0.999394 & 0.999584 & 1.00014 & 1.00003 & 1.00001 & 1. & 1. \\
0.02 &0.998744 & 0.999758 & 1.00013 & 1.00003 & 1.00001 & 1. & 1. \\
0.03 &0.998115 & 0.999894 & 1.00012 & 1.00003 & 1.00001 & 1. & 1. \\
 0.04 & 0.997544 & 1. & 1.00011 & 1.00002 & 1.00001 & 1. & 1. \\
0.05 & 0.997066 & 1.00008 & 1.0001 & 1.00002 & 1.00001 & 1. & 1. \\
0.06 & 0.996706 & 1.00014 & 1.00009 & 1.00002 & 1.00001 & 1. & 1. \\
0.07 & 0.996483 & 1.00018 & 1.00009 & 1.00002 & 1.00001 & 1. & 1. \\
0.08 & 0.9964 & 1.00021 & 1.00008 & 1.00002 & 1.00001 & 1. & 1. \\
0.09 & 0.99645 & 1.00022 & 1.00007 & 1.00002 & 1.00001 & 1. & 1. \\
0.1 &0.996614 & 1.00023 & 1.00007 & 1.00002 & 1.00001 & 1. & 1. \\
\end{tabular}
 \caption{\label{tab:NeutralCut} We vary our cutoff from .01 to .1 and find stable results of the correlation functions on the order of $10^{-5}$. These values were calculated for 2-point correlation functions of an (uncharged) scalar operator in a neutral thermalizing fluid (dual to the evolution towards a Schwarzschild  black  brane  geometry), $\rho= 0.0$, $B= 0.0$ and boundary conditions $l=0.7$. Our numerical background geometry was computed with a grid size of 150 points in the radial direction and a grid size of 250 points for our non-affine parameter.}
\end{table}

\begin{table}[H]
    \centering
    \begin{tabular}{c|c|c|c|c|c|c|c}
    \hline 
 $z_{uv}$ & $t=0.45$ & $t=0.85$ & $t=1.25$ & $t=1.65$ & $t=2.05$ & $t=2.45$ & $t=2.85$  \\
 \hline
0.01 & 1.00021 & 1.00008 & 1.0001 & 1.00005 & 1.00001 & 0.999995 & 0.999996 \\
0.02 & 1.00014 & 1.00002 & 1.00008 & 1.00003 & 0.999997 & 0.999989 & 0.999991  \\
0.03 & 1.00006 & 0.99997 & 1.00005 & 1.00002 & 0.999989 & 0.999983 & 0.999987  \\
0.04 & 0.999987 & 0.999917 & 1.00003 & 1.00001 & 0.999981 & 0.999978 & 0.999983 \\
0.05 & 0.999907 & 0.999864 & 1. & 0.999994 & 0.999973 & 0.999972 & 0.999979 \\
0.06 & 0.999825 & 0.999812 & 0.999977 & 0.999982 & 0.999965 & 0.999967 & 0.999975 \\
0.07 & 0.999741 & 0.999761 & 0.999954 & 0.99997 & 0.999958 & 0.999961 & 0.999971  \\
0.08 & 0.999656 & 0.999712 & 0.999931 & 0.999959 & 0.999951 & 0.999956 & 0.999968 \\
0.09 & 0.999571 & 0.999664 & 0.999909 & 0.999948 & 0.999944 & 0.999951 & 0.999964  \\
0.1 & 0.999485 & 0.99962 & 0.999888 & 0.999938 & 0.999938 & 0.999947 & 0.999961 
\end{tabular}
\caption{\label{tab:CharMagCut} We vary our cutoff from .01 to .1 and find stable results of the correlation functions on the order of $10^{-5}$. These values were calculated for 2-point correlation functions of an (uncharged) scalar operator in a charged thermalizing fluid (dual to the evolution towards a charged magnetic black brane geometry), $\rho= 0.78\rho_{e}$, $B= 1.0$ and boundary conditions $l=0.5$. Our numerical background geometry was computed with a grid size of 105 points in the radial direction and a grid size of 250 points for our non-affine parameter.}
\end{table}

\begin{table}[h]
    \centering
\begin{tabular}{c|c|c|c|c|c|c|c}
 N & $t=0.35$ & $t=0.6$ & $t=0.85$ & $t=1.1$ & $t=1.35$ & $t=1.6$ & $t=1.85$ \\
 \hline
 50 & 1.00131 & 1.00082 & 1.0004 & 1.0002 & 1.0001 & 1.00005 & 1.00002 \\
 100 & 1.0013 & 1.00081 & 1.0004 & 1.0002 & 1.0001 & 1.00005 & 1.00002 \\
 175 & 1.00129 & 1.0008 & 1.0004 & 1.0002 & 1.0001 & 1.00005 & 1.00002 \\
 250 & 1.00129 & 1.0008 & 1.0004 & 1.0002 & 1.0001 & 1.00005 & 1.00002 \\
 500 & 1.00129 & 1.0008 & 1.00039 & 1.0002 & 1.0001 & 1.00005 & 1.00002 \\
\end{tabular}
 \caption{A comparison of the value of the correlation function at various different grid sizes. This data was calculated for a transverse separation of $l=0.5$ in a neutral thermalizing fluid (dual to the relaxation towards a Schwarzschild black brane geometry) $\rho=0$, $\mathcal{B}=0$\label{tab:Grid_data}}
\end{table}

\begin{table}[h]
    \centering
\begin{tabular}{c|c|c|c|c|c|c|c|c}
N & $t=0.6$ & $t=1.05$ & $t=1.5$ & $t=1.95$& $t=2.4$ & $t=2.85$ & $t=3.3$ & $t=3.75$ \\
 \hline
 50 & 1.00082 & 1.00024 & 1.00007 & 1.00002 & 1. & 1. & 1. & 1. \\
 100 & 1.00081 & 1.00023 & 1.00007 & 1.00002 & 1. & 1. & 1. & 1. \\
 175 & 1.0008 & 1.00023 & 1.00007 & 1.00002 & 1. & 1. & 1. & 1. \\
 250 & 1.0008 & 1.00023 & 1.00007 & 1.00002 & 1. & 1. & 1. & 1. \\
 500 & 1.0008 & 1.00023 & 1.00007 & 1.00002 & 1. & 1. & 1. & 1. \\
\end{tabular}
 \caption{A comparison of the value of the correlation function at various different grid sizes. This data was calculated for a transverse separation of $l=0.5$ in a charged thermalizing fluid (dual to the relaxation towards a magnetic black brane geometry) $\rho=0.78\rho_{e}$, $\mathcal{B}=1.0$\label{tab:Grid_data_B_rho}}
\end{table}

\section{Charged operator correlators from charged probe geodesics}
\label{sec:chargedGeodesics}
We can also probe thermalization via charged scalar operators. In the dual gravitational system this corresponds to the geodesic equation with the addition of a Lorentz force term. However what we find is that for equal time correlations of charged scalar operators the thermalization of the correlator is uneffected by the charge of the operator! This is most easily seen via {the} action when considering a charged probe, the addition to the total action is,
\begin{equation}
\int{\exd\sigma A_{\mu}\dot{x}^{\mu}}=-\int{\exd\sigma\phi(v(\sigma),z(\sigma))\frac{\exd v(\sigma)}{\exd \sigma}}\label{eq:maxwellProblem}
\end{equation}
To see how this leaves us with no effect we can consider the shape of the curve for the time-like coordinate $v(\sigma)$. This coordinate solution is approximately quadratic and symmetric across $\sigma=0$. As a result the derivative, $g(\sigma)=d v/d\sigma$, is an odd function $g(-\sigma)=-g(\sigma)$. The rate of change of the timelike coordinate with respect to the non-affine parameter for negative values of the parameter is minus the value for positive values of the parameter. Furthermore the bulk radial coordinate also behaves nearly quadratically and is symmetric about the line $\sigma=0$. The final component is the potential $\phi$. The potential $\phi$ is nearly constant in time in our background and only displays a radial profile. As a result the potential in \eref{eq:maxwellProblem} is an even function while $\exd v/\exd\sigma$ is odd, as a result the integral vanishes! We then find no contribution to the length of the path leaving equal time correlations of charged operators in our thermalizing charged fluid identical to their uncharged counterparts. We have confirmed this simple analytical result numerically. 

In order to find any deviation in the correlations of charged operators we need to work with non-equal time correlators. In this case $\exd v/\exd\sigma$ is no longer an odd function of $\sigma$ and we have a chance at seeing the behavior which deviates from the correlations of uncharged scalar operators. However we have found that although the charge of the classical particle in the bulk does indeed lead to a longer proper length the overall contribution from the electromagnetic potential to this length is approximately same for each curve. This leads to an overall factor which is in the end is subtracted when normalizing the action, again leaving us with no effect to the normalized correlation functions.

\section{Additional checks}
\begin{figure}[H]
\begin{subfigure}[b]{.49\linewidth}
\includegraphics[width=2.9in]{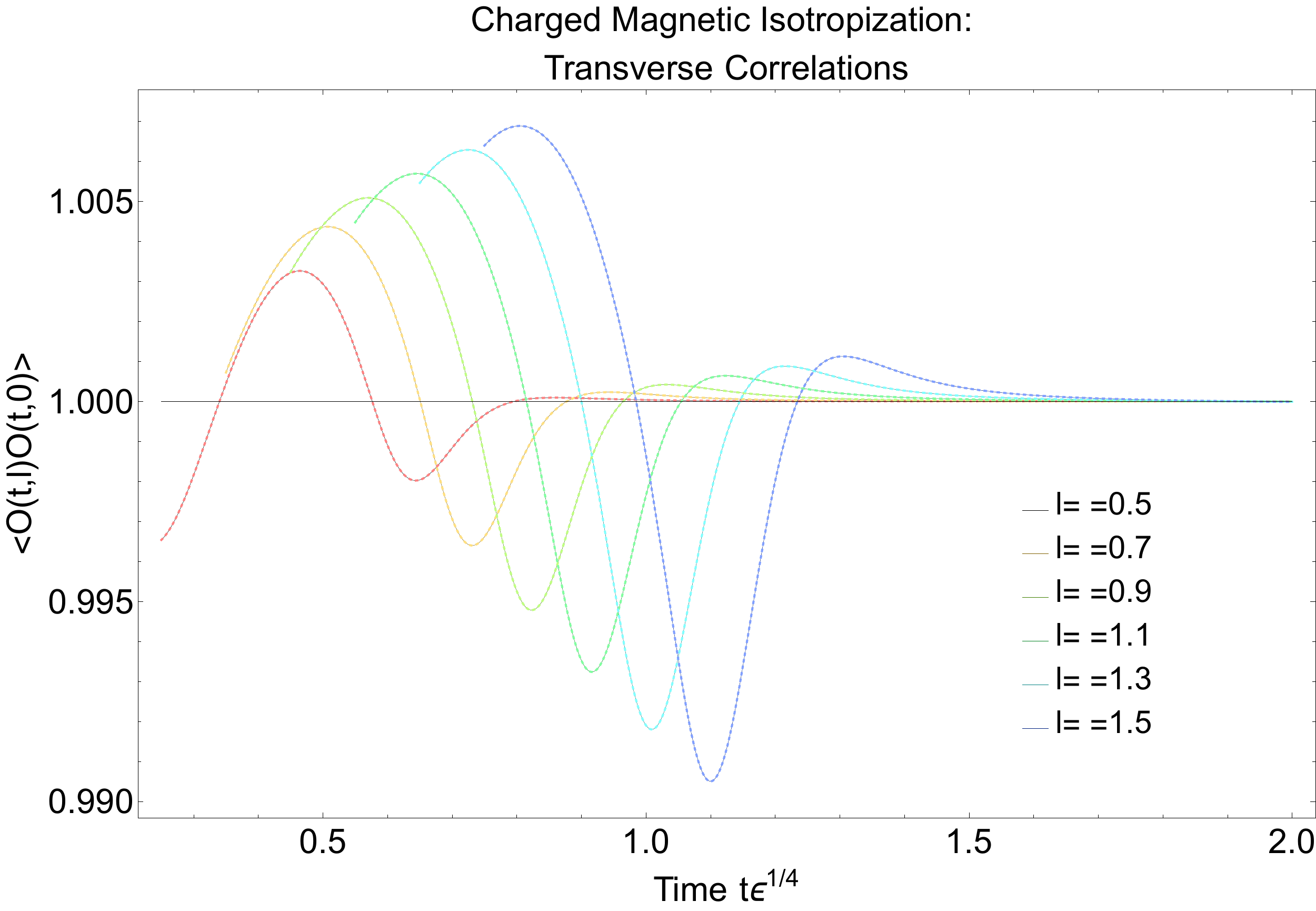}
\end{subfigure}
\begin{subfigure}[b]{.49\linewidth}
\includegraphics[width=2.9in]{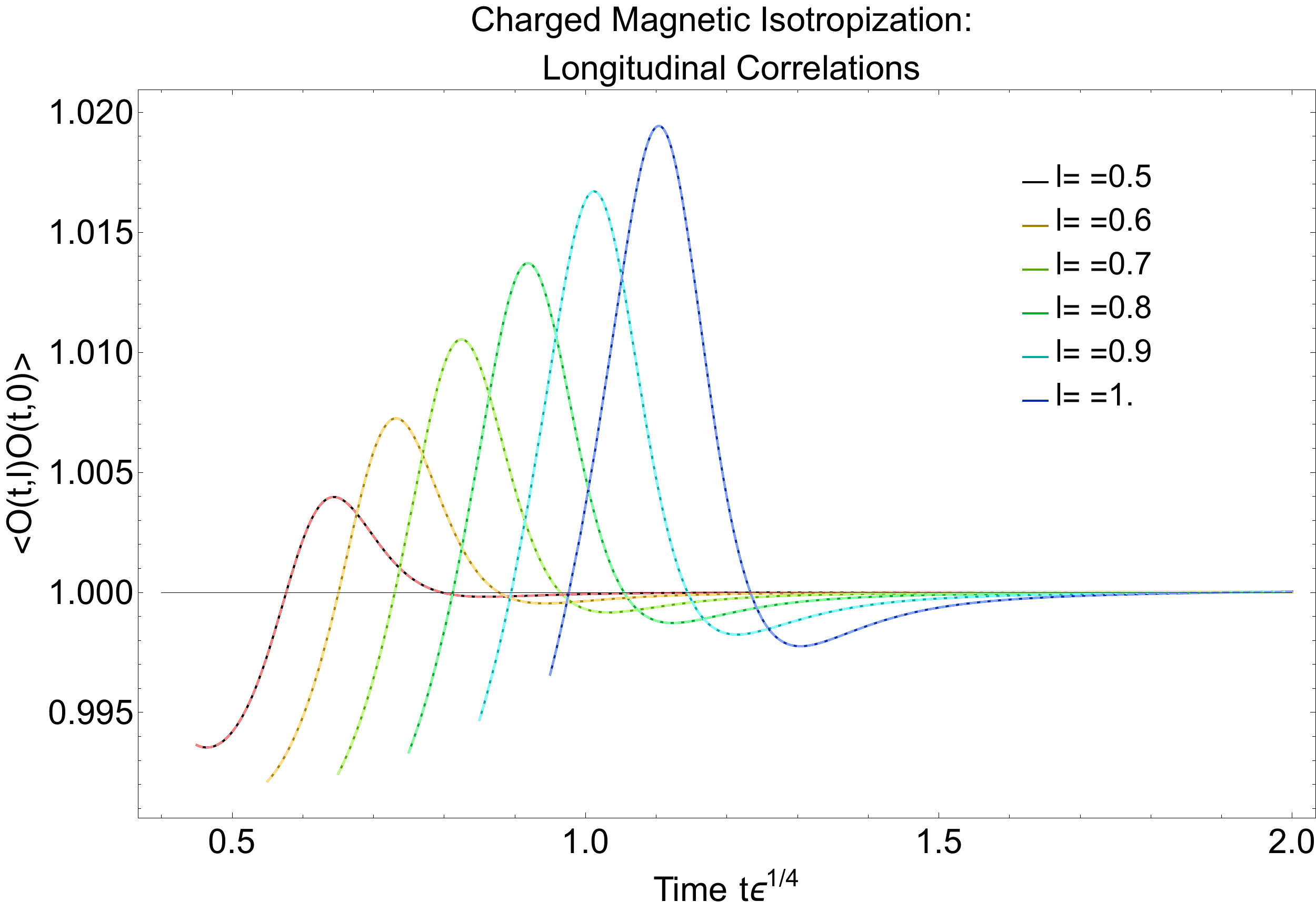}
\end{subfigure}
\caption{\label{fig:ChecksSmallBsmallChar} 2-point correlation functions of an (uncharged) scalar operator in charged thermalizing fluid (dual to the evolution towards a magnetic black brane geometry), $\rho=.001$, $\mathcal{B}=.0085$. \textit{Left:} Various length separations in the transverse direction. \textit{Right:} Various length separations in the longitudinal direction. The purely anisotropic case (a) is shown for comparison as dashed curves.}
\end{figure}

We confirm smooth transitions from zero magnetic to non-zero magnetic field by comparing the correlation functions for $\mathcal{B}=0$ and $\mathcal{B}^2\ll\epsilon_B$ we display this data in Sec.~\ref{sec:magneticBlackBraneResults} with a chosen value of $\mathcal{B}=.0085$. We repeat this exercise for the charged magnetic black brane by using the same magnetic field value while fixing the charge density to $\rho=.001$. We display this data in Fig.~\ref{fig:ChecksSmallBsmallChar}. 

\section{A different thermalization measure}
\label{sec:altMeasure}
\begin{figure}[H]
  \begin{subfigure}[b]{.5\linewidth}
  \begin{center}   
\includegraphics[width=2.9in]{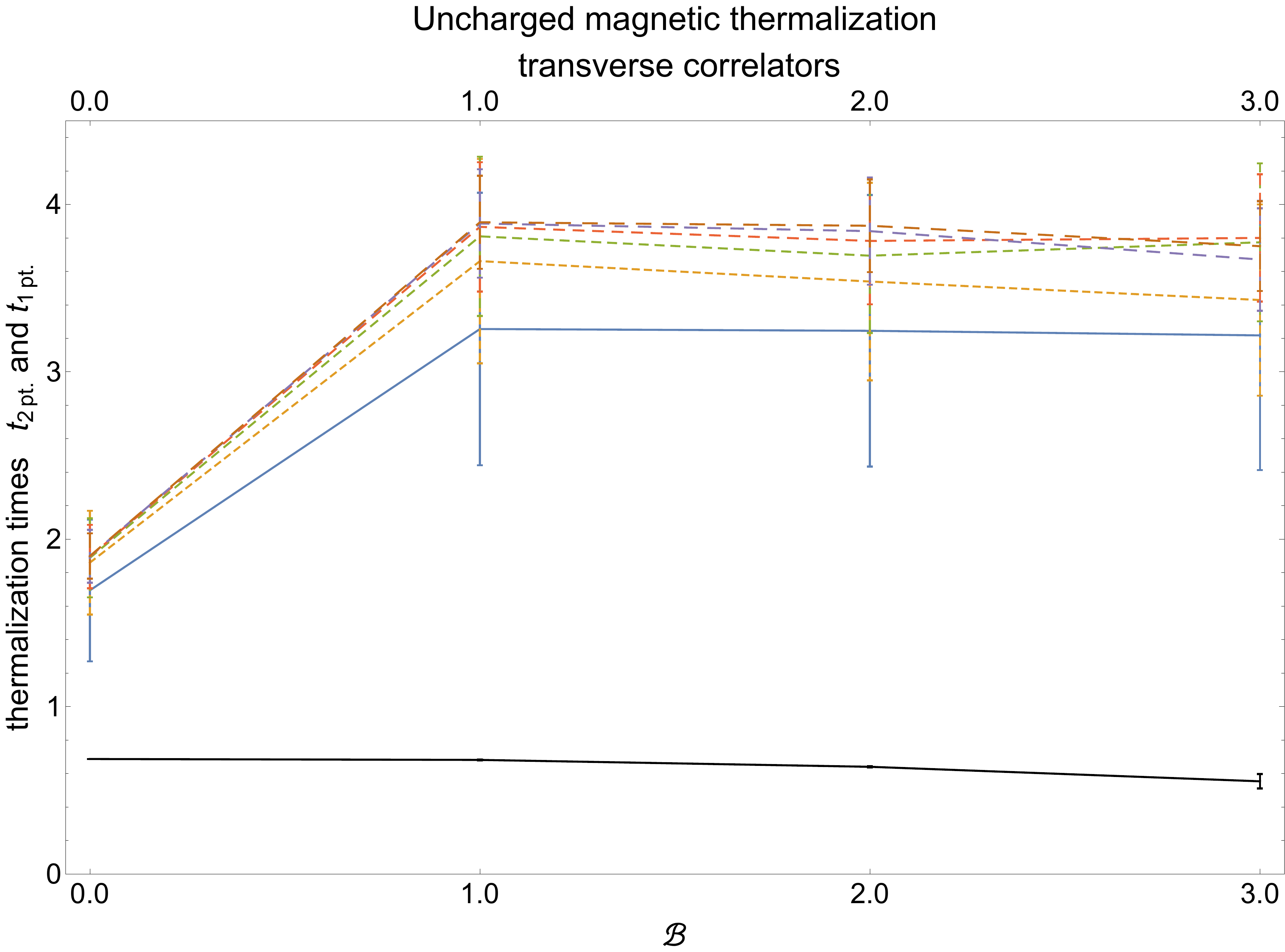}
  \end{center}
\end{subfigure}
 \begin{subfigure}[b]{.5\linewidth}
  \begin{center}   
\includegraphics[width=2.9in]{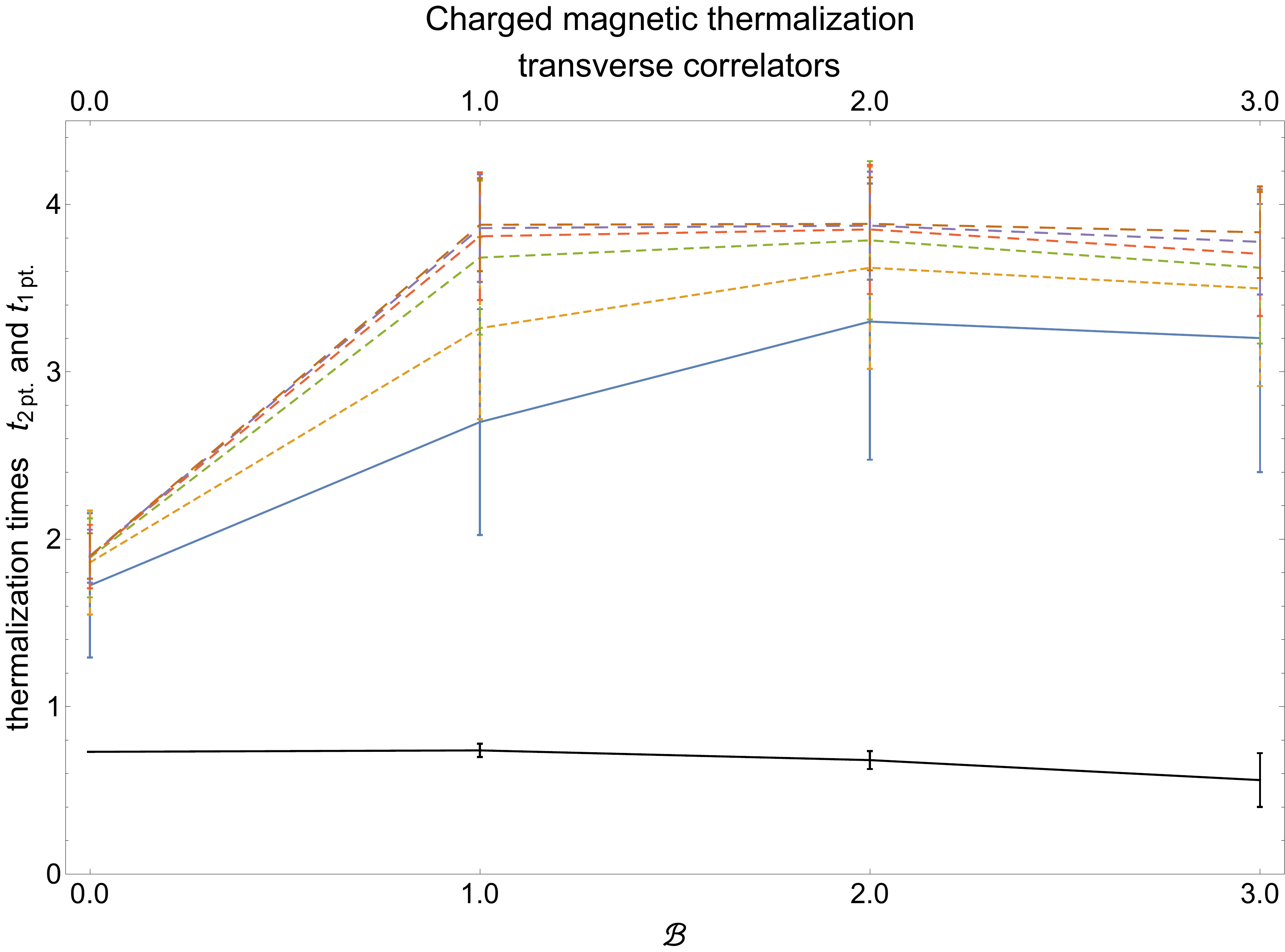}
  \end{center}
\end{subfigure}
\caption{\label{fig:comparingThermalizationTimes_ALT} 
{\it Thermalization times as a function of the magnetic field.} Shown here are the transverse 2-point function thermalization times $t_{2pt.}$ in a thermalizing fluid calculated with the alternative measure~\eqref{eq:thermal}. The dashed lines from bottom to top correspond to length separations $l=0.5, \, 0.7,\, 0.9, \, 1.1,\, 1.3,\, 1.5$. Error bars indicate an estimate of  
\textit{Left:} Uncharged fluid, $\rho=0$, $\mathcal{B}=0,\, 1,\, 2,\, 3$.  \textit{Right:} Charged fluid, $\rho=0.78 \rho_e$, $\mathcal{B}=0,\, 1,\, 2,\, 3$. Data points have been joined by straight lines to guide the eye.
}
\end{figure}

We have also considered the measure,
\begin{equation}
2\left|\frac{\langle\mathcal{O}(t,l)\mathcal{O}(t,0)\rangle-\langle\mathcal{O}(t=\infty,l)\mathcal{O}(t=\infty,0)\rangle}{\langle\mathcal{O}(t,l)\mathcal{O}(t,0)\rangle+\langle\mathcal{O}(t=\infty,l)\mathcal{O}(t=\infty,0)\rangle}\right|\leq 0.01 \tilde{A}\, .\label{eq:thermal2}
\end{equation} 
Most notably, this measure yields thermalization times for the transverse and longitudinal 1- and 2-point functions individually (while the measure in Eq.~\eqref{eq:thermal} only yields thermalization times for the difference of transverse and longitudinal quantities). Results for the correlations calculated in the transverse direction are shown in Fig.~\ref{fig:comparingThermalizationTimes_ALT}.
However we were unable to use this measure consistently for both 1-point and 2-point functions within the parameter region of this work. One of the purposes of this study is to compare the 1-point to 2-point functions hence this measure was not used to draw the main conclusions of the text. We can see in  Fig.~\ref{fig:comparingThermalizationTimes_ALT} that at each length the thermalization time saturates at large magnetic field. We can also see in terms of this measure the thermalization time is in general larger than that predicted by Eq.~\eqref{eq:thermal}. We attribute this increase to the fact that the difference of two functions can be within $1\%$ of the amplitude of the difference before each individual function is within $1\%$ of its amplitude.

\bibliographystyle{JHEP}
\bibliography{main}

\end{document}